\documentclass[
twoside,
openright,
titlepage,
numbers=noenddot,
footinclude,
cleardoublepage=empty,
paper=170mm:240mm,
fontsize=12pt
]{scrbook}

\usepackage[parts,linedheaders,dottedtoc,pdfspacing]{classicthesis}
\usepackage[
  inner=25mm,
  outer=20mm,
  top=25mm,
  bottom=30mm,
  bindingoffset=0mm
]{geometry}

\usepackage{amsmath,amssymb,amsthm}
\usepackage{rotating}
\usepackage{graphicx}  
\usepackage{wrapfig}
\usepackage{booktabs}
\usepackage[font={normalsize}]{caption}
\usepackage{subcaption}
\usepackage[ruled,vlined,linesnumbered]{algorithm2e}

\usepackage{wasysym}  
\usepackage{chngcntr}  
\counterwithout{figure}{chapter}
\counterwithout{table}{chapter}
\counterwithout{equation}{chapter}
\counterwithout{algocf}{chapter}
\usepackage[section]{placeins}  
\usepackage[
  backend=bibtex,
  style=numeric-comp,
  sorting=none,
  defernumbers=true,
  giveninits=true,          
  maxbibnames=10,           
  minbibnames=1,            
  maxcitenames=10,          
  mincitenames=1,           
  isbn=true,                
  doi=true,                 
  url=true,                
  eprint=true,              
]{biblatex}
\usepackage{longtable}
\newcommand{\ac}[1]{{\smaller #1}}

\usepackage{etoolbox}

\makeatletter
\pretocmd{\printbibliography}{%
  \renewcommand{\chaptermark}[1]{}%
}{}{}
\makeatother

\rohead{\mbox{\hfil{\headmark}}}
\lehead{\mbox{\headmark\hfil}}
\rofoot{\small\thepage}
\lefoot{\small\thepage}

\AtEveryCitekey{%
  \iffieldundef{doi}
    {}
    {\clearfield{url}\clearfield{urlyear}}%
}
\AtEveryBibitem{%
  \iffieldundef{doi}
    {}
    {\clearfield{url}\clearfield{urlyear}}%
}

\usepackage{hyperref}
\usepackage{cleveref}  

\let\cref\Cref
\Crefname{equation}{Equation}{Equations}
\crefname{algocf}{algorithm}{algorithms}
\Crefname{algocf}{Algorithm}{Algorithms}

\graphicspath{{figures/}}

\begin{document}

\frontmatter
\begin{titlepage}
\centering

{\Large PH.D. THESIS}

\vspace{1.5cm}

{\LARGE HoloHema: Digital Holographic\\
Hematology Analyzer} \\
\vspace{1cm}
{\large And HoloTile: Rapid and Speckle-suppressed\\
Computer-Generated Holography}

\vspace{4cm}

\begin{flushleft}
\textit{Author:}\\
Andreas Erik Gejl Madsen

\vspace{1cm}

\textit{Supervisors:}\\
Jesper Glückstad\\
Frank Nielsen\\
Peter Emil Larsen\\
Mohammad Esmail Aryaee Panah
\end{flushleft}

\vfill

\textit{A Thesis Submitted for the Degree of Doctor of Philosophy}\\
\vspace{0.5cm}
January 31, 2025

\end{titlepage}

\vspace*{\fill}
\begin{center}
© 2025\\
Andreas Erik Gejl Madsen\\
ALL RIGHTS RESERVED
\end{center}
\clearpage

\chapter*{Abstract}
\addcontentsline{toc}{chapter}{Abstract}

This industrial Ph.D. project, carried out in collaboration between Radiometer Medical ApS and SDU Centre for Photonics Engineering at the University of Southern Denmark, explored the use of digital holographic microscopy (DHM) for the purposes of differential white blood cell counts (dWBCs) in point-of-care (PoC) devices for acute care settings. Two DHM prototypes were developed; an initial lens-based system serving as the foundation for algorithm development, and experimental validation of the approach, achieving 89.6\% classification accuracy on a 3-part differential, and a subsequent lensless system for simplified design and increased field-of-view (FoV). Both prototypes employed convolutional neural networks (CNNs) for cell classification. With further optimizations, the lensless system achieved classification accuracies of 92.65\% and 89.44\% on the 3-part and 5-part differential, respectively. With the lensless system, the derivation of the monocyte distribution width (MDW), a biomarker for sepsis, was also demonstrated.

Additionally, pixel super-resolution and multi-wavelength DHM approaches were investigated to enhance the obtained cell information. Finally, a proof-of-principle physics-informed neural network (PINN) for holographic reconstruction was implemented, demonstrating the potential for machine learning (ML) reconstruction techniques.

In summary, this work represents an initial exploration of DHM for dWBC in PoC devices, laying the groundwork for future research.

\clearpage

\chapter*{Resumé}
\addcontentsline{toc}{chapter}{Resumé}

Dette industrielle Ph.D.-projekt, udført i samarbejde mellem \textit{Radiometer Medical ApS} og \textit{SDU Centre for Photonics Engineering} ved Syddansk Universitet, undersøgte brugen af digital holografisk mikroskopi (DHM) til differentialtælling af hvide blodlegemer (dWBC) i point-of-care (PoC) apparater til akut pleje.

Der blev udviklet to DHM-prototyper: et indledende linsebaseret system, der dannede grundlag for algoritmeudvikling og eksperimentel validering af metoden, og som opnåede 89, 6\% klassificeringsnøjagtighed på en 3-delt differentialtælling, samt et efterfølgende linseløst system med forenklet design og øget synsfelt (FoV). Begge metoder gjorde brug af convolutional neural networks (CNNs) til celleklassifikation.

Med yderligere optimeringer opnåede det linsefrie system klassificeringsnøjagtigheder på henholdsvis 92, 65\% og 89, 44\% på den 3-delte og 5-delte differentialtælling. Med det linsefrie system blev udledningen af monocyte distribution width (MDW), en biomarkør for sepsis, også demonstreret.

Derudover blev superopløsning og multi-bølgelængde DHM undersøgt for at forbedre den opnåede celleinformation. Endelig blev et proof-of-principle for et physics-informed neural network (PINN) til holografisk rekonstruktion implementeret, hvilket demonstrerede potentialet for rekonstruktion ved hjælp af machine learning (ML) teknikker.

Opsummerende repræsenterer dette arbejde en indledende udforskning af digital holografisk mikroskopi til differentialtællinger af hvide blodlegemer i PoC-apparater og lægger grundlaget for fremtidig forskning.

\clearpage

\chapter*{ACKNOWLEDGEMENTS}
\addcontentsline{toc}{chapter}{Acknowledgements}

The completion of this project can be attributed largely to the support of the people around me.

I would like to extend a massive thank you to my academic supervisor, Jesper Glückstad, for all the opportunities, guidance, and help he has given me. Our collaboration will surely continue to an even greater extent, towards ever greater goals.

To all employees at Radiometer Medical, thank you for making me feel welcome, despite the infrequency of my visits. I would like to extend my personal thanks to my industrial supervisors, Peter Emil Larsen, Mohammad Esmail Aryaee Panah, and Frank Nielsen for always making me feel welcome at Radiometer. It has been such a pleasure to work with you. In addition, an enormous amount of gratitude goes to Lene Eriksen and Ida Hollesen for their excellent work in the cell laboratory.

Finally, the support of my friends and family cannot be understated, and for that I am ever grateful.

\clearpage

\setcounter{tocdepth}{1}
\tableofcontents

\clearpage

\chapter*{PREFACE}
\addcontentsline{toc}{chapter}{Preface}

This thesis recounts the work performed during my industrial Ph.D. period in collaboration with Radiometer Medical ApS and the SDU Centre for Photonics Engineering at the University of Southern Denmark, supported by the Innovation Fund Denmark as part of their industrial Ph.D. programme.

As reflected in the table of contents of the previous pages, this thesis is divided into two distinct parts, each contributing to the incredibly broad field of holography.

\textit{Part I} explores the application of DHM for a specific challenge in hematology: performing rapid differential white blood cell counts with a simple, compact, and robust device. This investigation examines how the unique capabilities of DHM can be leveraged and implemented in a point-of-care unit. Such a device holds the potential to significantly reduce wait times for analysis in developing countries, acute care facilities, and operating rooms, where access to centralized laboratories may be limited.

\textit{Part II} shifts focus to the development of HoloTile, a novel approach within computer-generated holography (CGH), invented by Prof. Glückstad. This technology has far-reaching applications in numerous fields, including laser material processing, volumetric additive manufacturing, display technologies, and more.

While presented separately for the purpose of this thesis, DHM and CGH are deeply interconnected by the principles of holography. They are, in essence, two sides of the same coin. Much like capturing a photo and later replaying it by projection on a screen, or recording versus playing music, holography is a reversible discipline. While DHM captures and reconstructs the holographic information of a physical object, CGH takes the inverse path. Instead of recording a hologram with a sensor and reconstructing it numerically, holograms are created digitally. This digitally crafted hologram is then brought to life through physical reconstruction by the natural propagation of light, allowing us to effectively replay a synthetically generated wavefront, rather than one originating from a real-world object. This inherent duality is central to this project, as insights gained in one field is readily applicable to the other.

To avoid confusion for readers familiar with holography, it is important to define the specific scope of this thesis. Holography is a broad, multi-faceted field encompassing diverse applications. Here, I focus exclusively on optically thin holograms. This means that I will not address the complexities and phenomena associated with thick holograms and their applications.

Furthermore, the reconstruction of a hologram in this context refers specifically to the propagation of the complex wavefront encoded within to a series of axially separated, infinitely thin 2D planes.

\section*{Publications Contributing to This Work}

Several publications and manuscripts have been produced during my Ph.D. project, on which parts of this thesis is based. These publications listed here, and appended at the end of this thesis, provide further details on the specific aspects of my research:

\begin{enumerate}
\item \fullcite{madsen_holotile_2022}

\item \fullcite{madsen_-axis_2023}

\item \fullcite{gluckstad_new_2023}

\item \fullcite{gluckstad_gabor-type_2024}

\item \fullcite{gluckstad_holotile_2024-1}

\item \fullcite{madsen_holotile_2024-rgb}

\item \fullcite{madsen_axial_2025}
\end{enumerate}

\section*{Additional Publications During Ph.D. Period}

The following publications, while also produced during my Ph.D., are not directly included in this thesis. They include additional conference proceedings and journal articles, including a manuscript on the application of HoloTile for volumetric additive manufacturing. This work has been accepted for publication in Nature Communications (as indicated in a letter from the editor) and is currently awaiting processing. Therefore, it is referenced here in its preprint version, available on arXiv \cite{alvarez-castano_holographic_2024}.
\newpage
\begin{enumerate}
\item \fullcite{madsen_comparison_2022}

\item \fullcite{madsen_speckle-free_2022}

\item \fullcite{madsen_new_2022}

\item \fullcite{madsen_complex_2022}

\item \fullcite{madsen_review_2023}

\item \fullcite{madsen_holotile_2023}

\item \fullcite{gluckstad_comparing_2023}

\item \fullcite{gluckstad_holotile_2024}

\item \fullcite{gluckstad_holotile_2024-2}

\item \fullcite{alvarez-castano_holographic_2024-2}

\item \fullcite{madsen_generalized_2024}

\item \fullcite{madsen_digital_2024-1}

\item \fullcite{alvarez-castano_tomographic_2024}

\item \fullcite{alvarez-castano_holographic_2024}
\end{enumerate}

\clearpage

\chapter*{LIST OF ABBREVIATIONS}
\addcontentsline{toc}{chapter}{List of Abbreviations}

\renewcommand{\arraystretch}{1.25}
\begin{longtable}{ll}
2D & 2-Dimensional \\
3D & 3-Dimensional \\
ASC-PR & Amplitude and Sparsity Constrained Phase Retrieval \\
ASM & Angular Spectrum Method \\
CBC & Complete Blood Count \\
CGH & Computer-Generated Holography \\
CMOS & Complementary Metal–Oxide–Semiconductor \\
CNN & Convolutional Neural Network \\
CT & Computed Tomography \\
DHM & Digital Holographic Microscopy \\
DL & Deep Learning \\
DMD & Digital Micromirror Device \\
dWBC & Differential White Blood Cell Count \\
FC & Fully-Connected \\
FCN & Fully-Connected Network \\
FFT & Fast Fourier Transform \\
FNO & Fourier Neural Operator \\
FoV & Field-of-View \\
GAN & Generative Adversarial Network \\
GMM & Gaussian Mixture Model \\
GPU & Graphics Processing Unit \\
GS & Gerchberg-Saxton \\
IC & Integrated Circuit \\
ICU & Intensive Care Unit \\
LDA & Linear Discriminant Analysis \\
LED & Light Emitting Diode \\
MDW & Monocyte Distribution Width \\
ML & Machine Learning \\
NA & Numerical Aperture \\
NN & Neural Network \\
nRBC & Nucleated Red Blood Cell \\
PCA & Principal Component Analysis \\
PINN & Physics-Informed Neural Network \\
PLT & Platelet \\
PoC & Point-of-Care \\
PS & Polystyrene \\
PSF & Point-Spread Function \\
RBC & Red Blood Cell \\
RGB & Red-Green-Blue \\
ROI & Region of Interest \\
RMS & Root Mean Square \\
SLM & Spatial Light Modulator \\
SOTA & State-of-the-Art \\
T-VAM & Tomographic Volumetric Additive Manufacturing \\
ViT & Vision Transformer \\
VOI & Volume of Interest \\
VRAM & Video Random Access Memory \\
WBC & White Blood Cell \\
\end{longtable}
\renewcommand{\arraystretch}{1}

\mainmatter

\part{HOLOHEMA: DIGITAL HOLOGRAPHIC HEMATOLOGY ANALYZER}

\chapter{INTRODUCTION}
\label{ch:introduction}

Radiometer pioneered the world's first commercial blood gas analyzer in 1954, marking a significant milestone in medical diagnostics. Today, the company specializes in developing and manufacturing \ac{PoC} devices for acute care testing, including blood gas analysis, transcutaneous monitoring, and immunoassays. In recent years, there has been a growing interest in developing a device that can perform a \ac{dWBC} alongside blood gas measurements in a \ac{PoC} device. Given the importance of the \ac{dWBC} as a health indicator, the goal has been to create a compact, user-friendly analyzer for use in acute care facicilites, intensive care units (\ac{ICU}s), and operating rooms. This would enable healthcare professionals to quickly and accurately obtain \ac{dWBC}s without having to wait precious moments for processing in a centralized laboratory.

The current approach involves white blood cell (\ac{WBC}) classification using a bright-field microscope, with cells chemically stained to increase contrast and cell visibility. However, this project explores the potential of \ac{DHM} for \ac{dWBC} analysis. \ac{DHM} offers several significant advantages over conventional bright-field microscopy, such as amplitude and phase profile recovery, providing additional information for a cell classifier, large \ac{FoV}s, 3-dimensional (\ac{3D}) sample reconstruction, and a drastic simplification in mechanical complexity and size, possibly requiring neither an objective lens or a complex staining procedure to perform a white blood cell (\ac{WBC}) differential. As such, the objectives of the first part of this thesis are as follows:

\begin{enumerate}
\item To develop and construct lens-based and lensless holographic microscopes capable of reconstructing amplitude and phase images of \ac{WBC}s.

\item To create and implement an \ac{ML}-based \ac{WBC} classifier, capable of categorizing the most predominant \ac{WBC}s in the human body.

\item To investigate and apply experimental and computational techniques to further improve key parameters of the \ac{DHM} technique, including machine learning, super-resolution, and multi-wavelength holography.
\end{enumerate}

\section*{Contents of Part I: HoloHema: Digital Holographic Hematology Analyzer}

The following describes the contents of each chapter in Part I of this thesis:

\paragraph{Chapter 2} introduces the reader to the theoretical background of the thesis. The chapter includes an introduction to \ac{WBC}s, their function, and differentiation. Additionally, an overview of the inline holographic principle and its application to observing \ac{WBC}s is given. Finally, the chapter concludes with a description of image classification using machine learning.

\paragraph{Chapter 3} explores the possibility of using \ac{DHM} for \ac{WBC} classification. A best case scenario holographic microscope is constructed employing objective lens magnification for increased reconstruction fidelity. Furthermore an accompanying 3-part diff classifier is trained and tested on extracted \ac{WBC} reconstructions.

\paragraph{Chapter 4} investigates the possibility of removing the objective lens from the holographic microscope, while maintaining the ability to perform the 3-part diff. The lensless system allows for a simpler mechanical setup and a larger \ac{FoV}, at the cost of spatial resolution. Lastly the classification is expanded to the full 5-part diff. The classifier is trained using data from several more donors, and tested on blood from a previously unseen donor.

\paragraph{Chapter 5} accounts the implementation of a super-resolution algorithm, enabling the recording of features significantly smaller than the physical pixel size of the capturing image sensor.

\paragraph{Chapter 6} explores the use of multi-wavelength illumination for multi-colored sample reconstruction as a method to facilitate more extracted cell information for a classifier.

\paragraph{Chapter 7} explores the application of a state-of-the-art \ac{PINN} for self-supervised \ac{ML}-enabled hologram reconstruction. The model is trained on completely synthetic data and shows promising reconstruction results on previously unseen experimental holograms.

\chapter{BACKGROUND}
\label{ch:background}

This chapter serves as an introduction and overview of the key concepts that will be used extensively in this thesis, aiming to provide necessary context for subsequent chapters.

Specifically, Section 2.1 introduces white blood cells (\ac{WBC}s), their biological function, and diagnostic significance. This section also emphasizes the challenges involved in differentiation between various \ac{WBC} types — an important step in clinical decision making. Section 2.2 provides the background for holography and holographic microscopy. It is also discussed how its unique features, e.g., 3-dimensional (\ac{3D}) amplitude and phase imaging, may solve certain problems of conventional \ac{WBC} imaging and differentiation. Finally, Section 2.3 explores the machine learning (\ac{ML}) foundation for the \ac{WBC} classifiers introduced in Chapters 3 and 4, focusing on convolutional neural networks (\ac{CNN}s) and transfer learning.
\newpage
\section{White Blood Cells: Function and Significance}
\label{sec:wbc_function}

The differential white blood cell count (\ac{dWBC}), i.e., the count of each \ac{WBC} type in a unit volume, is an important health marker, providing a detailed overview over the hematological state of a patient. In this section, the \ac{WBC}s and their functions are briefly outlined, as well as the importance of their count in the human body. An overview is given over traditional \ac{WBC} classification methods, including their advantages and challenges in a point-of-care (\ac{PoC}) testing environment.

\subsection{Introduction to White Blood Cells}

White blood cells, also known as leukocytes, play a critical role in the defense system of the human body against infectious diseases. Their count and differentiation are essential diagnostic tools, providing valuable insights into the physiological conditions and pathological state of a patient \cite{blumenreich_white_1990}.

White blood cell counts are a fundamental component of the complete blood count (\ac{CBC}). An elevated or lowered \ac{WBC} count can be indicative of a range of underlying conditions, including infections, allergic reactions, and leukemia. Therefore, rapid and regular monitoring of \ac{WBC}s is essential for the early detection and diagnosis of serious health concerns.

Breaking down the \ac{WBC} count into its constituent cells, the count of each main \ac{WBC} type provides additional and more precise health-markers for a patient on which a physician can base a diagnosis \cite{tatsumi_practical_nodate, turgeon_linne_2016}.

\subsection{Differential White Blood Cell Counts}

In the human body there are five main types of \ac{WBC}s, making up approximately 1\% of all blood cells in a healthy adult. These being: \textit{monocytes}, \textit{lymphocytes}, \textit{neutrophils}, \textit{eosinophils}, and \textit{basophils}.

\begin{table}[htbp]
\centering
\caption{Overview of \ac{WBC}s and their function.}
\label{tab:wbc_overview}
\small
\begin{tabular}{lcl}
\hline
\textbf{\ac{WBC} type} & \textbf{Typical \% of leukocytes} & \textbf{Simplified function} \\
\hline
Monocytes & $2\%-10\%$ & Differentiate into macrophages \\
 & & and dendritic cells, which en- \\
 & & gulf pathogens and initiate an \\
 & & immune response. \\
Lymphocytes & $18\%-42\%$ & Produce antibodies, fight viral \\
 & & infections, regulate immune re- \\
 & & sponses. \\
Neutrophils & $40\%-70\%$ & Primarily combat bacterial in- \\
 & & fections by engulfing and de- \\
 & & stroying pathogens. \\
Eosinophils & $1\%-3\%$ & Fight parasitic infections, in- \\
 & & volved in allergic responses. \\
Basophils & $0.5\%-1\%$ & Involved in allergic responses, \\
 & & release histamine. \\
\hline
\end{tabular}
\end{table}

Each type exhibits distinct functional characteristics, as summarized, with extreme simplifications, in \cref{tab:wbc_overview} \cite{austin_differential_2021}, and the count of each cell type can provide valuable diagnostic information. For instance, an elevated neutrophil count is associated with bacterial infection and inflammation, while a lowered count may indicate tuberculosis, leukemia, or other autoimmune diseases.

Analyzers capable of automatic \ac{dWBC}s are typically found in several versions, depending on their degree of differentiation. For this thesis, the focus is on two versions; the three-part differential and the five-part differential.

\subsubsection*{The Three-Part Differential}

A three-part differential (``3-part diff'') provides a count and differentiation of the three main \ac{WBC} types: monocytes, lymphocytes, and neutrophils\footnote{In practice, the counts also include contributions from eosinophils and basophils. Their minute relative count is however considered negligible \cite{turgeon_linne_2016}.}. In the typical physician office laboratory, or acute care facilities, a 3-part diff provides sufficient information for a majority of clinical cases, being able to e.g., accurately distinguish between viral or bacterial infections.

\subsubsection*{The Five-Part Differential}

For larger hospitals and specialty laboratories, where a more detailed examination of the hematological state of a patient is necessary, a five-part differential (``5-part diff'') may be desirable. The minute populations of eosinophils and basophils make this differentiation significantly more difficult, and commercially available 5-diff analyzers can be several times more expensive than 3-diff devices.

\subsubsection*{Platelets}

An important factor for a \ac{dWBC} device is the platelet (\ac{PLT}) count in a blood sample. \ac{PLT}s (or thrombocytes) are approximately 2-3 µm in greatest diameter \cite{paulus_platelet_1975}, far smaller than any of the \ac{WBC}s. \ac{PLT}s play an important role in blood clotting, and their count can serve as a health marker for a number of autoimmune diseases, infections, and blood clot risk, among other conditions. Their size, along with their being nucleus-free and transparent, make them difficult to resolve and detect amidst the much larger red blood cells (\ac{RBC}s) and \ac{WBC}s.

\subsection{Traditional Methods for White Blood Cell Classification}

The \ac{dWBC} has been in use as a diagnostic tool since the early 20th century. Naturally, these counts were originally performed manually using optical microscopy. Although this method is still considered the gold standard to this day, electronic automated cell counting techniques have generally replaced the manual count. Electronic counting devices drastically reduce human error, analysis time, and labor-costs, as well as provide more statistically precise results due to the sheer number of cells that can be analyzed in a short time.

Modern analyzers combine several physical principles in order to make accurate classifications, typically based on electrical impedance (the Coulter principle \cite{don_coulter_2003, moldavan_photo-electric_1934}) and optical detection (flow cytometry \cite{mckinnon_flow_2018}). As cells are guided through a series of fluidic channels, numerous measurements are made, yielding individual cell information including size, nucleus morphology, granularity in the cytoplasm, and refractive index, on which the analyzer can base its final cell count \cite{turgeon_linne_2016}. While capable of achieving excellent results, these modern analyzers are often bulky and difficult to service due to the complexities of the examining system. They are therefore often placed in centralized laboratories serving multiple departments. For \ac{PoC} testing in acute care environments, rapidity of testing is of the utmost importance. The wait necessary for the analysis of a blood sample in a centralized laboratory may not be tolerable.

Advances have also been made in automatic \ac{dWBC}s with optical microscopy, especially following the increased popularity of \ac{ML} of the past decades. In particular, the application of image-based \ac{ML} classifiers on stained blood samples \cite{asghar_classification_2024, tavakoli_new_2021}. Although these can be made smaller, relatively speaking, making them more viable in a \ac{PoC} setting, they require significant sample manipulation through e.g., staining procedures. Staining, whether automated or not, increases both preparation time and chemical and mechanical complexity. In addition, the required optical magnification for accurate image classification limits the field-of-view (\ac{FoV}), and thus the parallel classification of many cells at once.

\subsection{Challenges and Consequences of Misclassification}

\subsubsection*{Relative Frequencies of \ac{WBC}s}

Performing accurate \ac{dWBC}s poses significant challenges, and continues to be an active field of research. First and foremost, the large discrepancy of relative proportions of the various \ac{WBC} types (e.g., basophil's 0.5\%-1\% to neutrophil's 40\%-70\%), makes it difficult to count the less frequent cell types with any statistical certainty \cite{kaushansky_williams_2021}, due to the limited samples. In addition, the difference in frequencies adds substantial risks of misclassification, which may lead to wildly incorrect counts for select cell types. For example, misclassifying a small percentage of neutrophils as basophils could dramatically inflate the basophil count, leading to an incorrect diagnosis. Conversely, misclassifying the few basophils present in a typical blood sample as neutrophils would have a negligible impact on the neutrophil count.

\subsubsection*{Rare Cell Types}

Secondly, any drawn blood sample will also contain a small, but not negligible, percentage of even rarer cell types including immature cells such nucleated red blood cells (\ac{nRBC}s), reticulocytes, and giant \ac{PLT}s \cite{schreier_blood_2020}. Even disregarding the diagnostic importance of these rare cells, their presence and similarity to any of the main cell types can also lead to misclassifications.

The consequences of misclassifying certain \ac{WBC}s as the wrong types, and thus presenting an incorrect count to a physician, are not insignificant. Critically, misclassifying lymphocytes as neutrophils, or vice versa, could lead to misdiagnosing a viral infection as bacterial, or the other way around, with significant implications for treatment. Therefore, it is important to determine both the risk of misclassification, and the impact a given misclassification could have on a \ac{dWBC}. This information, courteously provided by Radiometer Medical ApS and adapted for use in this thesis, is summarized in \cref{fig:wbc_misclassification_matrix} in Appendix A.

\subsubsection*{Transparent Samples and Staining}

Finally, and specifically important for this thesis, is the issue of staining in imaging-based analyzers such as bright-field microscopes. To enhance visibility and contrast of the nearly transparent \ac{WBC}s, it is common practice to stain the biological samples prior to differentiation by introducing dyes that make specific parts of a sample more opaque. The staining procedure can be quite extensive. It is common to add multiple dyes to the sample in order to stain different parts of the cell, e.g., plasma, nuclei, granules, etc., in different colors, thus introducing additional information on which a cell classifier can base its decision. While staining increases visibility of the cell structure, it also introduces additional complexity and time to the sample preparation. Additionally, the introduction of the dye may significantly interfere with the biological function of a living organism, and can in many cases be fatal for the organism.

Notably, as light passes through such a transparent sample, it undergoes a phase shift, rather than an amplitude attenuation, due to the refractive index — the optical density — of the sample. The relative phase shift of the light due to a cell is directly coupled to its complex morphology, both in its nucleus, cytoplasm, and cell wall. In conventional imaging configurations, this phase information is lost during recording, since common camera sensors and the human eye are only sensitive to light intensity.

However, if both amplitude and phase information of a cell are captured, such as is possible in holographic microscopy, it may enable the classification of cells using simpler staining procedures, or even without staining. This dual representation may allow a classifier to exploit inter-cellular differences in amplitude or phase, in order to differentiate them.

\section{Holography and Holographic Microscopy}
\label{sec:holography}

Digital holographic microscopy, and specifically inline digital holographic microscopy (\ac{DHM}), offers certain advantages that makes it an intriguing option for \ac{WBC} analysis and differentiation, especially in a \ac{PoC} testing environment. With its simple and cost-effective mechanical setup, a holographic microscope can be made compact, making it a viable solution for acute care facilities. In addition, with its \ac{3D} amplitude and phase imaging capabilities, it may enable \ac{WBC} classification without the use of staining.

In this section, the background for the application of inline \ac{DHM} for \ac{WBC} differentiation is explored. To that end, an overview of the principles behind holographic microscopy is included, the theory of which will be used extensively in the subsequent thesis chapters.

\subsection{Introduction to Holographic Microscopy}

In an attempt to improve the information regained from electron microscopes, Hungarian-British electrical engineer and physicist Dennis Gabor (1900-1979) experimented on capturing defocused images that, despite showing virtually no semblance to the sample in question, contained both amplitude and phase information of the entire modulated wavefront \cite{gabor_new_1948}. From this defocused image — the hologram\footnote{From Greek: holos, meaning whole and grafe, meaning write.} — a \ac{3D} image of the sample could be reconstructed. Although the technique was first imagined as a technique for electron microscopy, it has since gained most traction in optical applications.

The application of the holographic technique to microscopic objects has given rise to the enormous scientific field of \ac{DHM}, allowing the recovery of both amplitude and phase information from samples such as \ac{WBC}s. In recent years, \ac{DHM} has also been employed in water quality monitoring \cite{pitkaaho_digital_2016, gorocs_-line_2010}, detection of microplastics \cite{zhu_microplastic_2022, zhu_digital_2021} and air pollutants \cite{baker_lensfree_2023, basso_digital_2023}, \ac{3D} motion tracking of micro- and nanoparticles \cite{verpillat_digital_2011, riekeles_motion_2024}, cancer diagnostics \cite{yuan_digital_2021, gangadhar_deep_2023}, surface profilometry \cite{fernandez_development_2018, kim_phase_2022}, among other applications.

For \ac{PoC} testing and \ac{dWBC} specifically, inline \ac{DHM} exhibits certain characteristics that could be immensely desirable. With the ability to recover both the amplitude and phase of a sample, it may be possible to differentiate the \ac{WBC}s with limited or no staining, addressing a key limitation of conventional imaging-based methods. Furthermore, since the whole wavefront is recorded, the hologram captures \ac{3D} sample information, providing additional information on cell morphology for classification. Last, but not least, inline \ac{DHM} setups can be constructed with very few components, allowing for extremely compact, cost-effective, robust devices.

\subsection{Principles of Holography}

Holography allows for the recording and subsequent reconstruction of a temporally and spatially coherent wavefront by the capture of both its amplitude and phase information. Since all recording media, e.g., photographic film or digital camera sensors, are only sensitive to light intensity, phase information is typically lost upon capture. Therefore, the solution enabled by a holography is the conversion of the wavefront phase information into recordable interferometric intensity variations.

Such a wavefront may be the result of amplitude and/or phase modulation by an object in its propagation path. Recording this wavefront holographically also records these modulations. Consequently, by reconstructing the hologram, amplitude and phase information about the object, or objects, that caused them is also obtained.

The holographic process thus involves two key steps: recording and reconstruction. The following presents an overview of both, including the formation of the interferometric hologram, and the subsequent reconstruction.

\subsubsection*{Hologram Recording}

To describe the hologram recording process, care must be taken to first describe the representation of the objects. These objects, when placed in the propagation path of a coherent beam of light, modulate the wavefront of the beam according to their absorbance and relative refractive index. The objects in question are complex-valued and are defined to have a finite lateral extent, which can be modeled using an indicator function $\mathcal{X}$ \cite{gluckstad_slack-notes_2025}:

\begin{equation}
\mathcal{O}(x_o, y_o) = \mathcal{O} \mathcal{X}(\mathcal{O})
\label{eq:object-definition}
\end{equation}

where the indicator function $\mathcal{X}$ is defined as:

\begin{equation}
\mathcal{X}(\cdot) =
\begin{cases}
1 & \text{if } (x_o, y_o) \text{ within object area} \\
0 & \text{if } (x_o, y_o) \text{ outside object area}
\end{cases}
\end{equation}

Ensuring that the region outside the extent of the object is zero, while the region inside contains the complex values of the object\footnote{The formulation of \cref{eq:object-definition}, and the inclusion of the indicator function, is attributed to Professor Glückstad.}. 
The imparted modulation, e.g., amplitude attenuation and phase shift, on an incident wave is commonly expressed as a complex-valued 2-dimensional (\ac{2D}) object transmission function $t(x_o, y_o)$. 
The objects are thus considered to be optically thin, simplifying the following derivations. 
The complex object transmission function, $t(x_o, y_o)$, can be modeled by a background term and a term relating to the presence of the object. 
The background transmission term is simply 1, corresponding to complete transparency of the background plane. 
The object transmission term, however, must include the complex object and remove the region of the background term that is within the object area. Therefore, within the finite extent of the object, the background must be subtracted:
\begin{align}
t(x_o, y_o) &= 1 + \mathcal{X}(\mathcal{O}) \cdot (\mathcal{O} - 1) \\
&= 1 + t_o
\end{align}
$t_o(x_o, y_o)$ is therefore not a direct representation of the complex object. Instead, it relates to the change in complex transmittance due to the presence of a modulating object relative to the background transmittance \cite{goodman_introduction_2017, fournier_unconventional_2024}.

Illuminating a \ac{2D} plane with complex transmittance described by $t(x_o, y_o)$ by a monochromatic plane wave $\mathcal{R}_0$ results in the transmission of a wave expressed immediately following modulation as:

\begin{equation}
\mathcal{R}(x_o, y_o) = \mathcal{R}_0 t(x_o, y_o) = \mathcal{R}_0 + \mathcal{R}_0 t_0(x_o, y_o)
\end{equation}

The transmitted wave thus consists of both a uniform plane wave component and a scattered wave component due to the object transmittance term. This duality is crucial for the hologram recording process, as it allows for interference to occur between them. Since the interference depends on both the amplitude and phase of the scattered wave component, this information is contained within.

The resulting interference pattern is generated by the free-space propagation of the transmitted wave $\mathcal{R}(x_o, y_o)$ over a distance $z_r$, and subsequent intensity capture by a detector. Modeling this propagation by spatial convolution by the impulse response of free-space yields the field in the detector plane (described by spatial coordinates $x$ and $y$) can be expressed as:

\begin{equation}
\mathcal{A}(x, y) = \mathcal{R}(x_o, y_o) \circledast h(x, y, z_r)
\end{equation}

Where $h(x, y, z_r)$ denotes the propagation kernel of free-space to a distance $z_r$, e.g., the Fresnel kernel if applying a paraxial approximation:

\begin{equation}
h(x, y, z) = \frac{1}{i\lambda z} \exp\left[\frac{ik}{2z}(x^2 + y^2)\right]
\end{equation}

where $k$ is the wavenumber of the propagating wave, $\lambda$ is its wavelength, and $z$ is the propagation distance.

As any detector solely records the intensity of incident light, the captured pattern is given by the squared magnitude of the field $\mathcal{A}(x, y)$:

\begin{equation}
\mathcal{I}_h(x, y) = |\mathcal{A}(x, y)|^2 = |\mathcal{R}(x_o, y_o) \circledast h(x, y, z_r)|^2
\end{equation}

Expanding this intensity expression yields\footnote{Lateral coordinates have been omitted to accommodate the page layout.}:

\begin{align}
\mathcal{I}_h &= |\mathcal{R}_0(1 + t_o) \circledast h(z_r)|^2 \nonumber \\
&= |\mathcal{R}_0|^2 |(1 + t_o) \circledast h(z_r)|^2 \nonumber \\
&= |\mathcal{R}_0|^2 (1 + t_o \circledast h(z_r))(1 + t_o \circledast h(z_r))^* \nonumber \\
&= |\mathcal{R}_0|^2 \left[1 + |t_o \circledast h(z_r)|^2 + t_o \circledast h(z_r) + t_o^* \circledast h(-z_r)\right]
\label{eq:hologram-intensity}
\end{align}

where $*$ denotes the complex conjugate. Due to their equivalence, the complex conjugate of the free-space propagation kernel has been substituted by propagation in the reverse direction ($h^*(z_r) = h(-z_r)$).

\begin{figure}[t]
\centering
\captionsetup{format = plain}
\includegraphics[width=\linewidth]{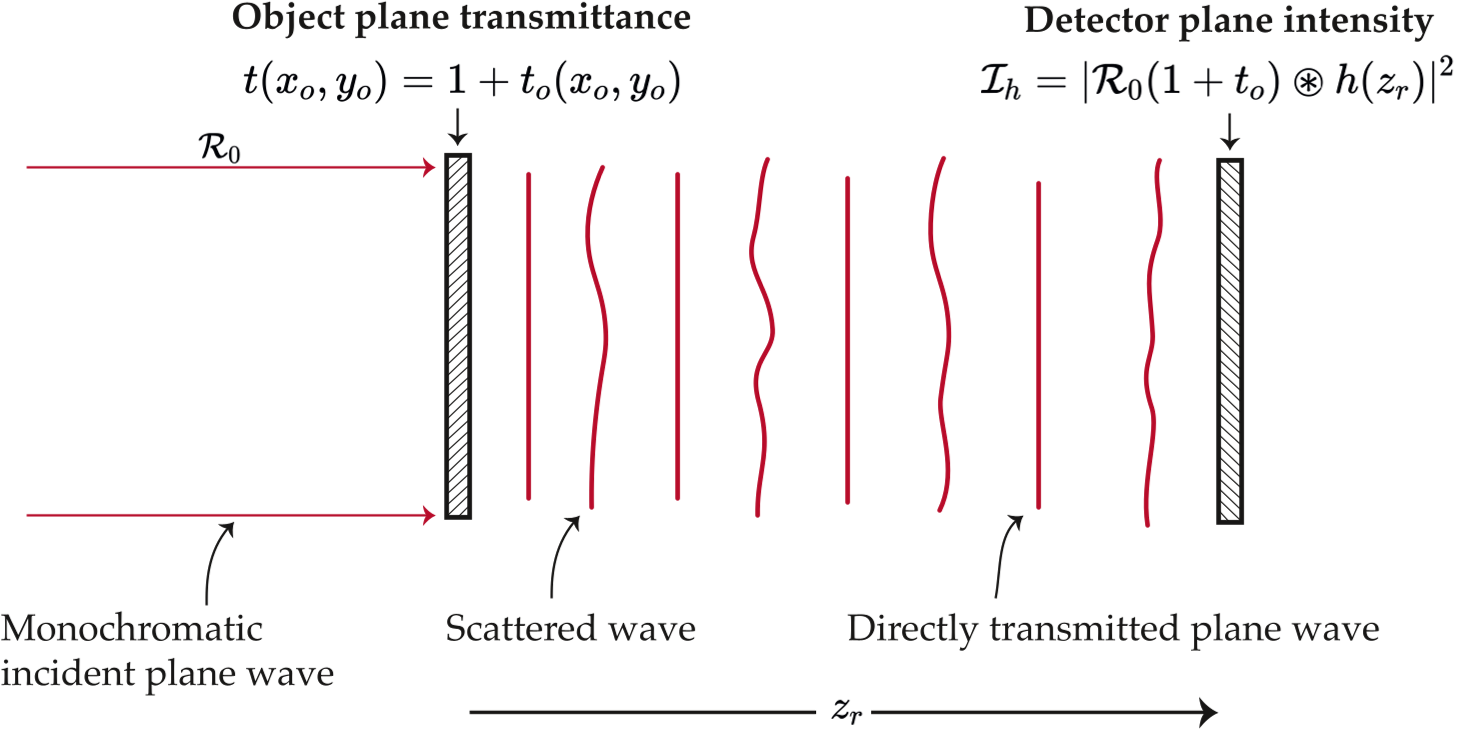}
\caption{Simplified illustration of the hologram recording process. A monochromatic plane wave, $|\mathcal{R}_0|$, illuminates the complex transmission function $t(x_o, y_o)$ in the object plane. The transmitted field consists of both a plane wave component due to the background transparency in the object plane, and a scattered wave component due to the modulation by the objects. Following free-space propagation of the transmitted field a distance $z_r$, the intensity of the field, $\mathcal{I}_h$, is captured by a detector.}
\label{fig:hologram_recording}
\end{figure}

The interference pattern described in \cref{eq:hologram-intensity} is the captured hologram, consisting of four individual terms. In the third term, the transmittance term relating to the object, $t_o$, is present in its entirety, including its amplitude and phase information. The last term is a conjugate image of the object transmittance term which, due to the reverse propagation, appears to have originated from behind the hologram. The formation of the captured hologram intensity of the object transmittance $t(x_o, y_o)$ is visualized in \cref{fig:hologram_recording}.

\subsubsection*{Holographic Reconstruction}

Holograms were traditionally recorded on photographic film/plates and developed as \ac{2D} transparencies with amplitude transmittance proportional to the exposure of the intensity hologram. This hologram transmittance can be expressed as:

\begin{equation}
t_h(x, y) = \beta|\mathcal{R}_0|^2 \left[1 + |t_o \circledast h(z_r)|^2 + t_o \circledast h(z_r) + t_o^* \circledast h(-z_r)\right]
\end{equation}

where $\beta$ is a proportionality constant representing the parameters of the recorded and developed media, e.g., photographic film or plates. By illuminating this hologram transmittance by a monochromatic plane wave equivalent to the recording beam, the transmitted wave consists of four terms, each corresponding to one of the terms in \cref{eq:hologram-intensity}. The first term corresponds to an attenuated plane wave passing directly through the hologram, without scattering. The second term corresponds to the defocused object intensity, and will be revisited later. The third term represents a component of the transmitted wave proportional to the object transmittance term, appearing as an image of the object located a distance of $z_r$ in front of the hologram; in its original object plane. Similarly, the fourth term corresponds to a conjugate image of the object, appearing as an image of the object, mirrored in the hologram, originating from a distance $z_r$ behind the hologram.

\begin{figure}[t]
\centering
\captionsetup{format = plain}
\includegraphics[width=\linewidth]{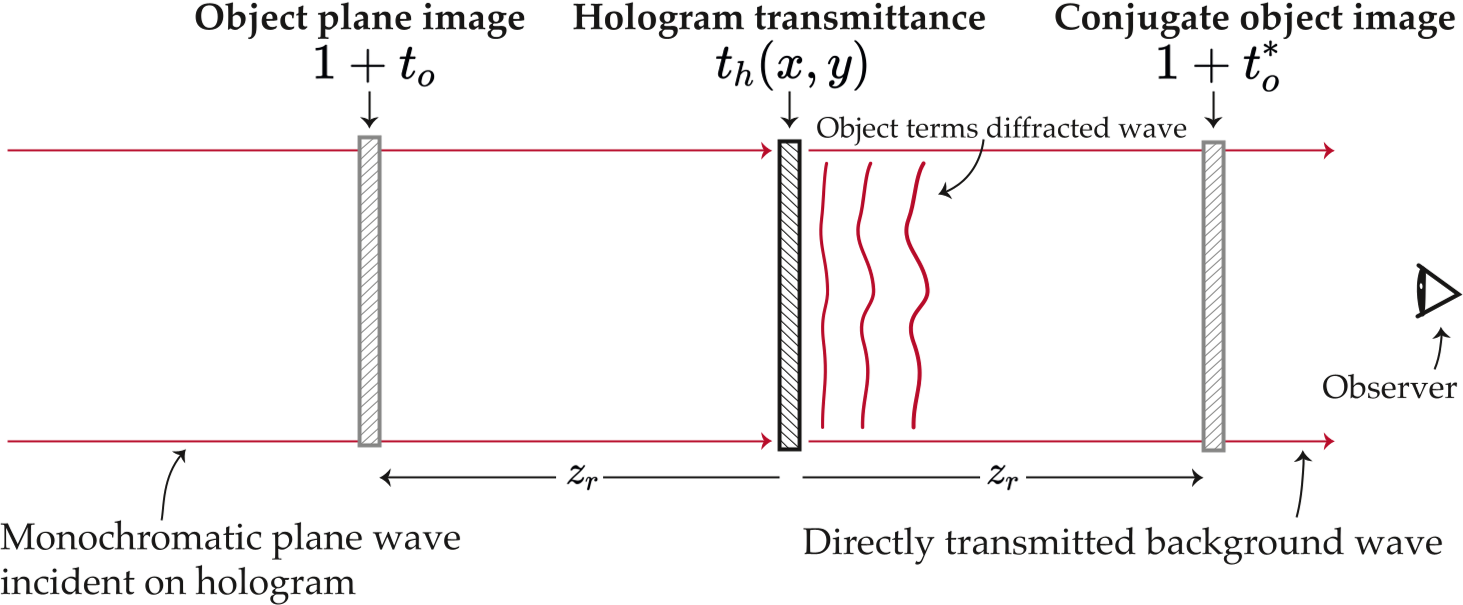}
\caption{Illustration of the formation of the twin-images in inline holography. A monochromatic plane wave illuminates the hologram which is expressed by the amplitude transmittance $t_h(x, y)$. The diffracted wave consists of terms corresponding to an image of the object transmittance term $t_o$, as well as a conjugate image, $t_o^*$. The images are accompanied by a constant background due to the background transparency of the hologram. To an observer, the two images appear as if located on either side of the hologram, separated by a distance of $2z_r$.}
\label{fig:twin_images}
\end{figure}

For an observer viewing the hologram, as depicted in \cref{fig:twin_images}, the reconstructed objects of the third and fourth term will appear to overlap exactly. Focusing on one of the two images brings the other out of focus, and vice versa. These images are conventionally called twin-images, and their inseparability was a major factor for the initial dismissal of inline holography.

Transitioning to the present day, the scientific use of holography has mainly become a digital endeavor. In digital holography, the hologram is recorded on a digital camera sensor, and the reconstruction is performed entirely numerically. However, the reconstruction process is still analogous to the reconstruction performed by Gabor and others using photochemical recording media.

Naturally, by capturing the interference patterns on a digital sensor, the stored holograms are now discrete, both in terms of bit depth (the amount of bits available to represent a transmittance value) and the finite sampling grid of the image sensor. To distinguish between the representation of digital images to their analog continuous counterparts, these will be denoted by their non-calligraphic symbols, i.e. the discrete sampling of the hologram $\mathcal{I}_h$ will be denoted as $I_h$.

To reconstruct the field in the original object plane, the recorded hologram is digitally illuminated by an equivalent monochrome plane wave, and the transmitted wave is numerically propagated to the object plane. The propagation corresponds to the convolution of the hologram transmittance with the discrete free-space propagation kernel to a distance $-z_r$. Due to computational efficiency, this impulse response convolution is most often performed in the Fourier domain by way of the convolution theorem \cite{goodman_introduction_2017}, thus employing a transfer function approach:

\begin{equation}
\mathcal{I}_h \circledast h(-z_r) \Rightarrow \text{FFT}^{-1}\{\text{FFT}\{I_h\} H(-z_r)\}
\end{equation}

where $\text{FFT}\{\cdot\}$ denotes the \ac{2D} fast Fourier transform, and $H = \text{FFT}\{h\}$ is the discrete transfer function of free-space propagation. Setting $R_0 = \beta = 1$ for simplicity, the complex-valued discrete reconstructed field in the object plane can then be expressed as:

\begin{equation}
C_r = \text{FFT}^{-1}\{\text{FFT}\{I_h\} H(-z_r)\}
\end{equation}
\begin{equation}
\approx 1 + |t_o \circledast h(z_r)|^2 \circledast h(-z_r) + t_o + t_o^* \circledast h(-2z_r)
\label{eq:reconstructed-field}
\end{equation}

where $h(-2z_r) = h(-z_r) \circledast h(-z_r)$. $C_r$ is approximate here due to the truncation of the finite extent of the sensor area, and the discrete sampling.

The reconstructed field in the object plane given in \cref{eq:reconstructed-field} thus consists of four terms, corresponding to the terms recorded in the hologram intensity. The first term is a constant background due to the directly transmitted plane waves\footnote{In practice, the reconstructed background is not constant, but oscillatory near the reconstruction edges due to diffraction by the finite extent of the image.}. The second term, which here will be denoted as the self-interference term, corresponds to the propagation of the defocused object term intensity. The self-interference term is often considered negligible in the literature. However, the effect of this superimposed complex component can significantly affect both the reconstructed amplitude and phase, as was examined in the publication in Appendix J \cite{gluckstad_gabor-type_2024}. The exact effect of the self-interference term and its dependence on the parameters of the holographic system deserves significantly more attention, but is beyond the scope of this work.

The third term describes the reconstruction of the object transmittance term, while the fourth describes the overlapping twin-image of the object term, defocused by a distance of $2z_r$. The defocused twin-image also severely impacts the accuracy of the reconstructed object transmittance by the addition of its own amplitude and phase. Consequently, distinguishing between the desired image of the object and its twin-image counterpart quickly becomes challenging, especially for biological samples where subtle differences in cell characteristics, e.g., subtle differences in refractive indices, may reveal important details about cell-structure and function.

\subsubsection*{Advanced Holographic Reconstruction}

In Madsen et al. \cite{madsen_-axis_2023}, the review paper published in relation to this PhD project, the landscape of state-of-the-art (\ac{SOTA}) hologram reconstruction methods is detailed. These reconstruction algorithms all have the common goal of efficiently reconstructing the original object, while attenuating the effects of the degrading terms as much as possible. In the review, the reconstruction methods are divided into two main categories; direct and iterative.

Direct methods obtain the object reconstruction in a single computational step, without the need for iterative refinement. This category also includes various \ac{ML}-based reconstruction and twin-image suppression approaches, as the processing can be performed in a single step following training of the given model. These \ac{ML}-based reconstruction methods have become increasingly popular, owing to the incredible computation speed which can be reached using a trained model on a graphics processing unit (\ac{GPU}). However, these methods typically require extensive training data and computational resources in order to facilitate training.

Iterative methods, on the other hand, utilize multiple cycles of computations to arrive at a reconstruction solution. As a direct measurement of the phase of the wavefront is impossible using intensity-sensitive detectors, the hologram in \cref{eq:hologram-intensity} encodes the phase of the incident wavefront indirectly through the fringes in the interference pattern.

Since this indirect measurement leads to the formation of the two conjugate object images – the twin images – it is pertinent to consider whether this lost phase information can be recovered algorithmically. If the phase of the wavefront in the detector plane could be recovered and combined with the measured amplitude ($A_h = \sqrt{I_h}$) to form a complex field, the reconstruction should contain the object transmittance information only.

The aim of such phase retrieval methods is then to recover the lost wavefront phase. Most of these methods employ some variation of the Gerchberg-Saxton (\ac{GS}) algorithm \cite{gerchberg_practical_1972}, Fienup algorithms \cite{fienup_phase_1982}, or inverse problem approaches \cite{madsen_-axis_2023,herve_alternation_2020,mallery_regularized_2019}. The iterations of the algorithms typically enforce certain constraints on the reconstructed object, e.g., positive absorption, and includes regularization terms in the cost function such as to encourage solutions exhibiting certain properties, e.g., sparsity. These methods provide a clearly defined algorithm with which the original object can be extracted. However, due to the iterative nature of the algorithms, computation time can be a limiting factor.

\subsection{Benefits of Inline Holographic Microscopy}

Given the inherent degradation of the reconstructions in inline holography due to the twin-image and self-interference terms, it may be wise to ask the question as to why this method is chosen for the given problem. Historically, the off-axis holography modality was preferred over inline methods due to the ability to filter the twin-image artifact. In off-axis configurations, a separate reference beam is positioned at an angle relative to the object beam, thereby introducing a carrier frequency, resulting in a hologram in which the twin terms are spatially separated in reconstruction.

Despite these advantages of off-axis holography, recent advancements in algorithmic design, \ac{ML}, and computing power have led to a revival of inline holography. With the newly acquired ability to effectively remove the twin-image and self-interference artifacts algorithmically, the inherent advantages of inline holography can finally be realized:

\subsubsection*{Simplified Optical Setup}

The introduction of a separate reference beam in off-axis holography, such that it is no longer colinear with the object beam, requires precisely aligned mirrors, lenses, prisms, beamsplitters, etc., to manipulate the beams. On the other hand, inline holograms can be recorded using simply an illumination source and a detector. This also means that a digital holographic microscope can be incredibly compact, increasing versatility drastically.

\subsubsection*{Interference Pattern Stability}

Due to the separate beams, an off-axis system is quite sensitive to vibrations during the recording process. Even minute movement between the two optical paths can disrupt the interference pattern and lead to blurry or unusable holograms. In inline holography, the reference and object beams are always colinear, and the interference pattern is not easily disturbed by vibrations.

\subsubsection*{Preservation of High-Frequency Information}

In order to separate the real and twin image in digital off-axis holography, usually a filtering of the spatial frequencies is required. Due to the addition of the carrier wave from the off-axis geometry, the two images are separated in the Fourier domain. Extracting only the real image in the spectrum inevitably discards some high-frequency information [48, p. 427]\cite{macovski_hologram_1970}, leading to reduced fidelity in the reconstruction. Digital inline holography, however, utilizes the entire captured spectrum of the hologram.

\subsection{\ac{DHM} Configurations for \ac{WBC} Differentiation}

\begin{figure}[t]
\centering
\captionsetup{format = plain}
\includegraphics[width=\linewidth]{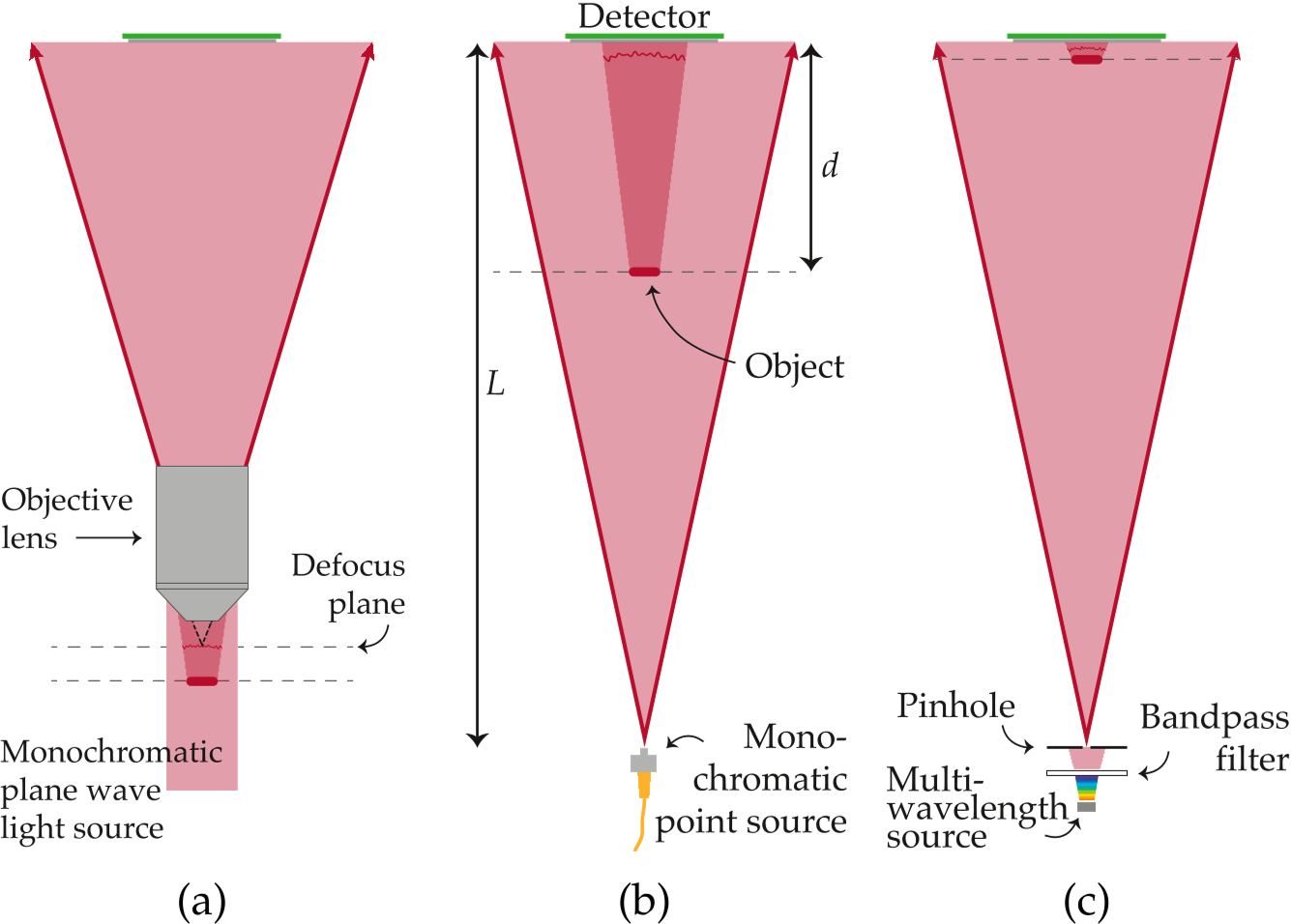}
\caption{Illustrations of the \ac{DHM} configurations used in this thesis. (a) The lens-based setup as described in Chapter 3. (b) The lensless setup as described in Chapter 4. (c) The large \ac{FoV}, unity magnification, lensless system as described in Chapters 5 and 6 for use with a pixel super-resolution algorithm and multicolor reconstructions.}
\label{fig:dhm_configurations}
\end{figure}

With the aforementioned advantages of inline \ac{DHM}, its application to \ac{WBC} differentiation appears promising. However, the holographic imaging of the \ac{WBC}s can be performed using several different experimental configurations, each with their own strengths and weaknesses. In \cref{fig:dhm_configurations}, the three configurations used in this project and thesis are depicted.

All versions rely on the propagation of the wavefront due to modulation by some object, and subsequent capture of the hologram on an intensity-sensitive detector, usually a complementary metal–oxide–semiconductor (\ac{CMOS}) camera sensor. In \cref{fig:dhm_configurations}a, an object lens is added between the sample and the sensor, responsible for magnifying the defocus plane prior to its capture. This results in a decreased effective pixel size of the camera sensor, and thus increased resolution. The specific magnification is determined by the numerical aperture (\ac{NA}) of the objective, and the distance between the objective and the camera sensor. While this configuration increases the resolution of the captured hologram, allowing reconstruction of smaller features in the object plane, it also impacts the mechanical complexity and compactness of the system.

In \cref{fig:dhm_configurations}b, the objective lens is removed, and the hologram is captured in a lensless fashion, akin to the illustrations in \cref{fig:hologram_recording}. Due to the divergent beam of the coherent light source, e.g., from a terminated fiber laser, it is still possible to achieve a degree of magnification of the hologram. In this case, the magnification factor is governed by the ratio of source-detector distance, and the object-detector distance via:

\begin{equation}
M = \frac{L}{L-d}
\label{eq:magnification-M}
\end{equation}

Both configurations in \cref{fig:dhm_configurations}a and \cref{fig:dhm_configurations}b achieve magnification by physical means. However, if optimizing for \ac{FoV}, for example to detect and differentiate as many \ac{WBC}s as possible in one single capture, it can be beneficial to lower the optical magnification significantly. While a large magnification factor increases resolution, it also limits the observable area of a sample. In the configuration in \cref{fig:dhm_configurations}c, the sample is placed immediately in front of the detector, yielding effectively unity magnification. Consequently, the sample \ac{FoV} is the entire area of the detector, and the spatial resolution is limited by the pixel size of the capturing camera sensor. In Chapter 5, this resolution limit is increased using a pixel super-resolution algorithm, with which high-resolution holograms are synthesized, while maintaining the large \ac{FoV}. This configuration is also used in Chapters 6 and 7, in which a multi-wavelength light source is used.

\subsection{Coherence Considerations}

A light source can exhibit both temporally and spatially coherence. A certain degree of coherence of both kinds are a prerequisite for holographic recording, as coherence directly facilitates the ability to create optical interference patterns.

Temporal coherence, often expressed in terms of the coherence time, $\tau_c$, or length, $L_c$, the distance over which a wave maintains a consistent phase relationship, is related to the monochromaticity of a light source. The spectral bandwidth of a source impacts the coherence length directly, with temporally coherent sources having a long coherence length. A source with a wide spectral bandwidth consists of a wide range of wavelengths. The recorded intensity of such a light source corresponds to the incoherent sum of the intensities of each wavelength component. Recording a hologram with a source that is temporally incoherent leads to a blurring of the hologram \cite{fournier_unconventional_2024, gopinathan_coherence_2008}. Spectral bandpass filters are commonly used to increase the monochromaticity of temporally incoherent light sources by limiting their spectral bandwidth, as illustrated in \cref{fig:dhm_configurations}c.

Spatial coherence describes the degree to which the phase of a wave remains uniform across different points in its wavefront. For most practical light sources, spatial coherence is inversely proportional to the size of the source. Therefore, to record holograms with clear interference fringes without blurring, the size of the source should be made as small as possible \cite{deng_coherence_2017-1}.

To mitigate the lack of spatial coherence of light sources with significant emission areas, e.g., light emitting diodes (\ac{LED}s), the illuminating source is commonly spatially filtered by a pinhole, as also seen in \cref{fig:dhm_configurations}c. For holographic microscopy, pinholes with diameters in the order of the illuminating wavelength ($\leq$ 0.5-2 µm) are often used to ensure spatial coherence, though the exact size depends on the specific application \cite{garcia-sucerquia_digital_2006, lu_ultra-compact_2015, walcutt_assessment_2020, greenbaum_imaging_2012}.
\newpage
\section{Machine Learning for Image Classification}
\label{sec:ml_classification}

For imaging-based \ac{dWBC} techniques, machine learning (\ac{ML}) image classifiers have become the de-facto standard, providing incredibly fast classification\footnote{Once training is finished.}, and robustness to image variations. This section outlines the fundamental concepts of \ac{ML}, and its proficiency in image classification. Both a general overview is given, along with the working principle of the popular \ac{ML} model type for analyzing images, the convolutional neural network (\ac{CNN}). Certain parts and figures are sourced from the Master's Thesis of the Author \cite{madsen_algorithmic_2021-1}.

\subsection{Overview of Machine Learning}

The concept of machine learning — computers learning to perform advanced tasks without the explicit instruction of a programmer — has long been intriguing. In classical computing, a program is provided with previously determined rules and instructions from which it can derive an answer. In contrast, \ac{ML} allows for a mathematical model to learn the intricate relationships between inputs and outputs to perform tasks that would otherwise be too challenging for a human to write instructions for manually. A common application of \ac{ML} is classification, and especially image classification, which was also its main use in this PhD project. At its core, an image classification \ac{ML} model operates by identifying and learning patterns within data provided to it and using these learned patterns to generate predictions or classifications \cite{goodfellow_deep_2016}.

The training data, the dataset on which the model will learn, is comprised of both input images and corresponding output labels, e.g., monocyte, lymphocyte, etc. Through continued exposure to inputs and expected outputs, the model learns the appropriate associations between them by minimizing discrepancies between its predictions and the expected labels. This type of learning is typically denoted as supervised, as the model has access to both inputs and expected outcomes \cite{heaton_introduction_2008, chollet_deep_2018}.

A vital component in this learning process is the loss function, which quantifies the error of the model's prediction when compared to the target class label. The objective of training is then to adjust the parameters of the model such that the loss function is minimized, corresponding to a model which can accurately predict the correct label of a given image.

For the differentiation of \ac{WBC}s in holographic reconstructions, image classification \ac{ML} models may be well-suited for a variety of reasons. First, the \ac{WBC}s are difficult to differentiate, even by trained personnel. The similarity between the cell types is high, with natural variation in features, e.g., size and shape of the cell nucleus. With no staining applied, the subtle differences between the cells may appear even weaker. Secondly, with the addition of phase information of the cells, a whole new dimension is added to the classification task. By utilizing deep learning (\ac{DL}) model types such as \ac{CNN}s, which excel at classification on high-dimensional data, the additional phase information may aid in classification. Lastly, once a model is trained, it is capable of rapidly processing an immense number of images, owing in part to their capability to be run on \ac{GPU}s.

\subsection{Deep Learning for Image Analysis}

Deep learning is the subfield of machine learning which specifically concerns neural networks (\ac{NN}s); a network structure consisting of multiple individual layers arranged sequentially, such that the output of one layer becomes the input to the next. The deep in deep learning thus refers to the depth, or the number of layers, in a model \cite{goodfellow_deep_2016}.

Each layer in an \ac{NN} represents some transformation applied to the data presented to it. The successive application of transformations allows a \ac{DL} model to transform the input data into increasingly abstract representations, distilling the most important features of the inputs into a correct prediction. Specifically for computational tasks involving images, one \ac{DL} model type has long been the most popular choice: The convolutional neural network \cite{krizhevsky_imagenet_2012,simonyan_very_2015,szegedy_going_2015,he_deep_2015-1,tan_efficientnet_2019,tan_efficientnetv2_2021}.
\begin{figure}[t]
\centering
\captionsetup{format = plain}
\includegraphics[width=.65\linewidth]{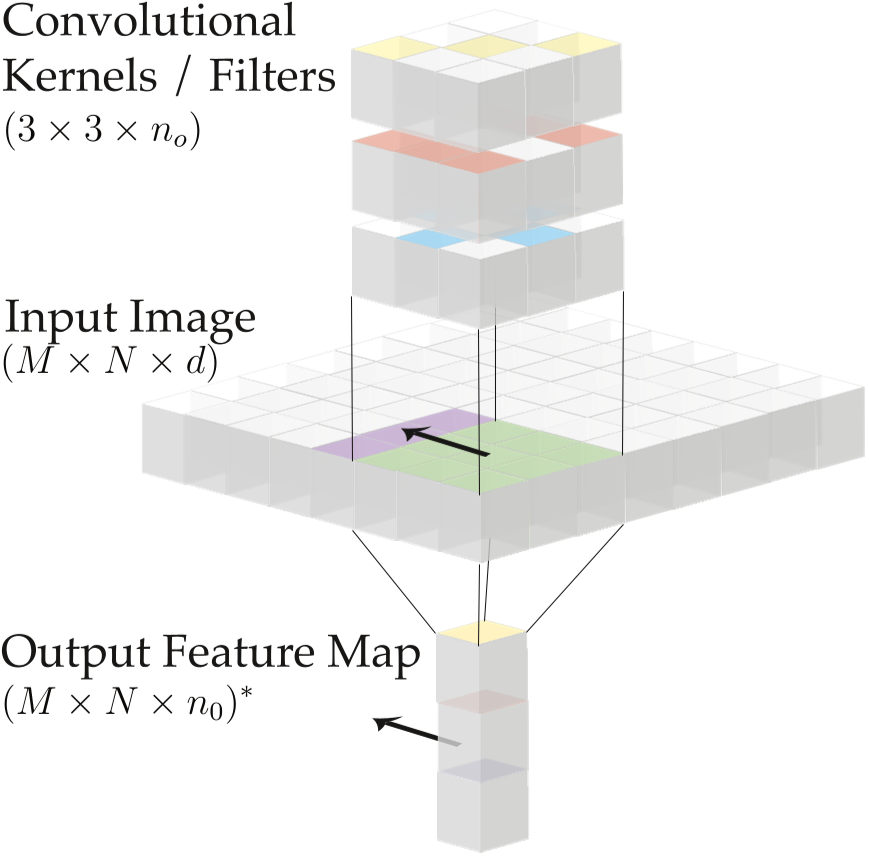}
\caption{Illustration of the convolution between an input image and three convolutional kernels. The convolution between the input and each of the kernels results in separate feature maps \cite{madsen_algorithmic_2021-1}.}
\label{fig:cnn_convolution}
\end{figure}

\subsubsection*{Convolutional Neural Networks}

In a \ac{CNN}, each layer consists of multiple filters, commonly called kernels. These are \ac{2D} weight matrices, usually $3 \times 3$ or $5 \times 5$ pixels, that emphasize specific local attributes of their input images. When an input image is presented to such a convolutional layer, the image is spatially convolved with each kernel. Each kernel can be thought of as shifting across the input image, pixel by pixel, and for each shifted location, the sum of the element-wise product between the kernel and the overlapping area in the image is computed and saved in a new \ac{2D} image, commonly referred to as a feature map. This operation is illustrated in \cref{fig:cnn_convolution}, in which a stack of $n_0$ convolution kernels shift across an input image, resulting in a feature map with $n_0$ channels.

\begin{figure}[t]
\centering
\captionsetup{format = plain}
\includegraphics[width=.8\linewidth]{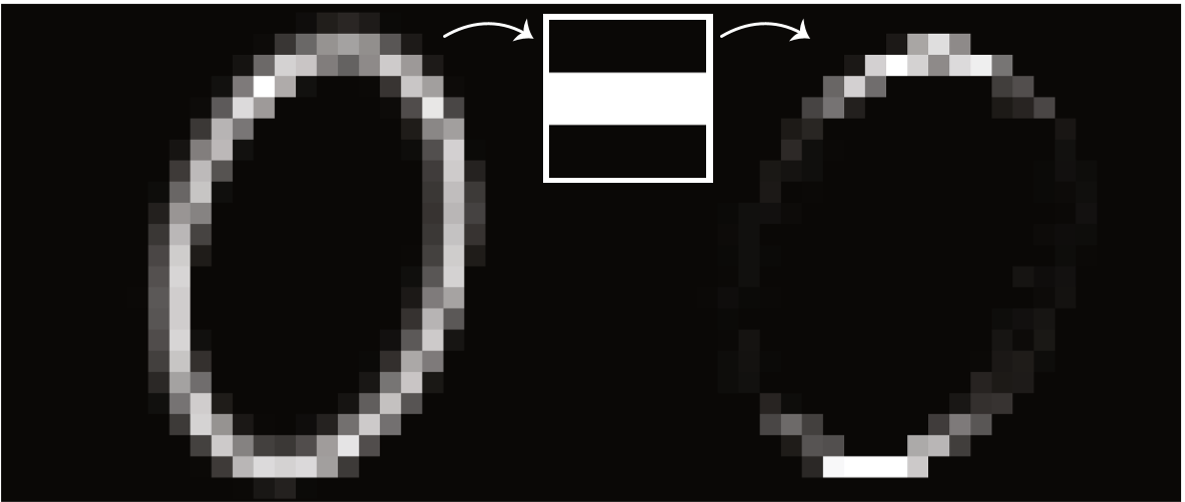}
\caption{Illustration of the effect of the \ac{2D} convolution. A $3 \times 3$ kernel, is convolved with the pixelated digit ``zero''. The convolution results in a feature map that highlights the horizontal edges of the digit and suppresses the vertical \cite{madsen_algorithmic_2021-1}.}
\label{fig:edge_detection}
\end{figure}

Depending on the numeric values within the kernels, numerous transformations of the input data can be performed. As an example, the convolution kernel depicted in \cref{fig:edge_detection} functions as a horizontal-edge detector. When convolved with the input image (the pixelized ``zero''), the weighted sum yields the greatest result when a horizontal line of pixels align with the center three values of the kernel.

Importantly, the actual values of the numerous convolution kernels within a \ac{CNN} are learned through the training of the network. As such, it becomes possible for the network to be able to extract and emphasize features of the input images which relate best to given class label. While the first layer may only learn to extract main features such as edges or brightness values, the successive application of new convolution layers and dimensionality reduction layers allow for increased abstraction. As each layer in a \ac{CNN} processes its inputs from the previous layer's feature maps, the information density increases. Deep layers may learn to encode abstract notions such as ``roundness'', ``transparency'', or even ``monocyte-ness'' in the increasingly high-dimensional latent space.

A significant advantage of \ac{CNN}s for the classification of images is that they are spatially invariant. Since the kernels are applied via convolution, it does not matter where in the input image a certain feature is located, only the degree to which it is present. This is an extremely powerful feature, as the classifier becomes robust to translation of the subject in the images.

\subsection{Classification using Convolutional Neural Networks}

The previous section on \ac{CNN}s dealt mainly with the internal workings of the convolution operations, and not with the classification itself. The final feature map in the sequence of convolution layers in a \ac{CNN} is not directly interpretable by humans as a classification. Instead, it is a high-dimensional representation of the input, encapsulating the important classifiable features.

To actually produce a useful classification, the final feature map is often passed through a fully-connected network (\ac{FCN}), consisting of one or two fully-connected (\ac{FC}) layers of artificial neurons. This \ac{FCN} is responsible for learning an interpretation of the high-dimensional feature map, such that the correct classification is achieved. As a simplified example, it may learn that if two specific values in the final feature map are high, the input image was more likely to have contained a monocyte.

This \ac{FCN} takes as input a version of the final feature map which has been flattened to a one-dimensional vector. The high-dimensional features are converted to the one-dimensional vector via a pooling layer [62–64], i.e., reducing the dimensionality of the data while retaining key extracted features. The final layer within the \ac{FCN} consists of a number of neurons equal to the number of classes into which the input data should be classified. For a three-part diff, the final number of neurons would be three, and five for a five-part diff. The prediction of the total classifier when presented with an input is then the index of the neuron with the highest numeric output value in the final layer.

\subsection{Transfer Learning}

The training of a \ac{CNN} capable of image classification is an expensive endeavor in terms of time and computational resources. Instead of gathering possibly millions of training examples, designing, and training a large network from the ground up, a common practice in classification is the concept of transfer learning. Transfer learning involves the use of a pre-existing network — one trained for a different classification task — and adapting it for a new, related task. As such, certain parts of the pre-trained network are reused, and new layers specific to the given classification are added. In practice, it is common to retain the trained convolutional block of the network, while the classifying \ac{FCN} is replaced and trained specifically for the task at hand.

This allows the new classifier to benefit from the knowledge already encoded in the pre-trained network, hence minimizing the need for learning fundamental image features, e.g., edges, corners, contrast, etc., such that training can focus on learning the specific abstract features of the new dataset. Most classification tasks share these low-level features, and transfer learning allows for a generalization to new categories \cite{goodfellow_deep_2016}. This is especially useful in tasks where training data is scarce, since the low-level features have already been learned on a much larger and diverse dataset.

\subsection{Assessing Performance}

To make sure training progresses as intended, and the network is actually learning, it is wise to monitor certain performance metrics. These metrics are usually different from the loss function used in training, as this is not always easily interpretable. Common metrics to determine the performance of any classifier include accuracy, precision, recall, and the F1-score.

Accuracy refers to the proportion of correctly classified samples to the total number of samples:

\begin{equation}
\text{Accuracy} = \frac{\sum_{i=1}^{N} (\hat{y}_i = y_i)}{N}
\end{equation}

where $N$ is the total number of class labels, and $\hat{y}_i$ and $y_i$ are the predicted and true labels, respectively, for the $i$th sample. The accuracy is particularly useful, as it presents an easily understood value for the performance of a classifier. However, the accuracy alone does not represent the full story of the performance of a classifier. For instance, a classifier able to detect a rare disease (affecting one in a million people) can easily be made to have an accuracy of 99.9999\% by simply always predicting that a patient is healthy \cite{goodfellow_deep_2016}. While an extreme example, it does reflect the problem of the relative frequencies of \ac{WBC}s in a blood sample, as noted in Section 2.1.4.

To properly measure performance in these cases where certain classes are rarer than others, precision, recall, and the F1-score are useful tools. Precision is a measure of true positives (TP) to false positives (FP):

\begin{equation}
\text{Precision} = \frac{\text{TP}}{\text{TP} + \text{FP}}
\end{equation}

In the context of \ac{dWBC}, a classifier with high precision for monocytes has a low rate of misclassifying other cells as monocytes. Recall on the other hand, is a measure of true positives to false negatives (FN):

\begin{equation}
\text{Recall} = \frac{\text{TP}}{\text{TP} + \text{FN}}
\end{equation}

Similarly, a classifier with high recall for monocytes exhibits a high sensitivity to detecting monocytes. It will rarely miss an actual monocyte, but may occasionally misclassify other cells as monocytes.

When training a classifier, it is often necessary to make a compromise between recall and precision. Comparing the performance of two classifiers is difficult if one exhibits low recall and high precision, and the other the inverse. The F1-score is often used to merge these metrics into one. Expressed as the mean of the two metrics, it provides a single quantifiable value, along with accuracy, on which to evaluate and compare the performance of classifiers.

\chapter{LENS-BASED HOLOGRAPHIC WHITE BLOOD CELL DIFFERENTIATION}
\label{ch:lens_based}

This chapter describes the development and results of an initial proof-of-principle digital holographic microscope and \ac{ML}-based image classifier for \ac{WBC} imaging and classification. 
As discussed in Chapter 2, \ac{DHM} exhibits several characteristics that makes it a promising approach for \ac{PoC} \ac{WBC} analysis, including \ac{3D} amplitude and phase imaging in a compact and simple mechanical setup.
Although inline holographic microscopy can be performed without magnification by an objective lens, this initial investigation employs objective lens magnification for several key reasons. 
By first establishing a foundation with a lens-based system, the experience gained can readily be translated to address the challenges of lensless \ac{DHM}. 
Specifically, this initial lens-based prototype allows the testing of \ac{DHM} for \ac{dWBC} in a best-case scenario, in which the spatial resolution is not expected to be the limiting factor. 
The experience gained from its implementation, including hologram reconstruction, determining cell focus, developing a \ac{WBC} classifier enables an easier transition to the more challenging lensless system.
The chapter begins with a description of the design and implementation of the lens-based system, equivalent to the configuration shown in \cref{fig:dhm_configurations}a, followed by the techniques used to acquire and reconstruct the cell images. 
Finally, the development and application of a \ac{CNN} capable of performing a 3-part diff is presented and tested, demonstrating the feasibility of \ac{DHM} for \ac{WBC} classification.

\section{Experimental Design and Hologram Capture}
\label{sec:lens_experimental}

The lens-based holographic analyzer consists of three basic assemblies;
\begin{wrapfigure}[27]{r}{5cm}
\centering
\captionsetup{format=plain}
\includegraphics[width=5cm]{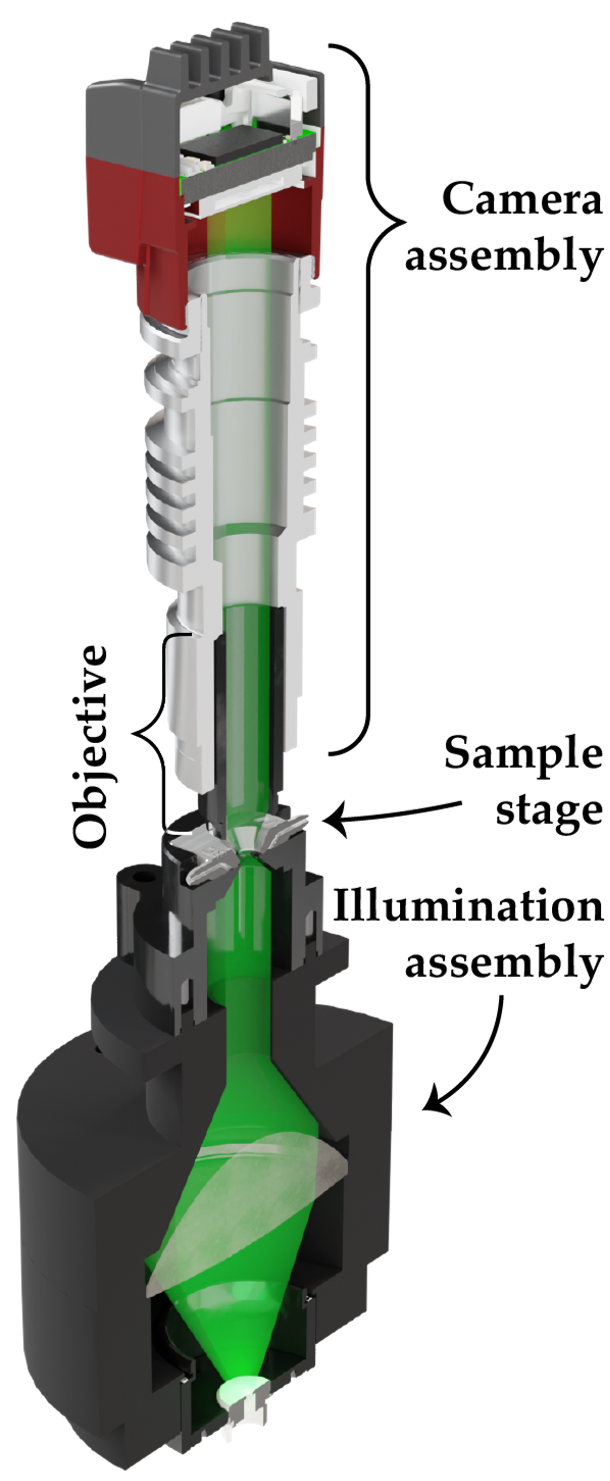}
\caption{Render of the experimental system for the lens-based holographic \ac{WBC} analyzer.}
\label{fig:lens_based_system}
\end{wrapfigure}

\paragraph{The illumination assembly} consists of a fiber coupled \cite{noauthor_p1-460b-fc-1_nodate} 520 nm laser \cite{noauthor_cps520_nodate}, collimated in an aspheric condenser lens \cite{noauthor_acl4532u--_nodate}. The expanded and collimated laser beam is shone on the sample through an aperture (\diameter 4 mm).
\paragraph{The sample stage} consists of a custom flow cuvette with internal dimensions 5 mm $\times$ 5 mm $\times$ 90 µm. The cuvette has an inlet and outlet on opposite sides, which allows controlled flow of sample fluid.
\paragraph{The camera assembly} consists of a monochrome camera \cite{noauthor_alvium_nodate} with 5496 horizontal pixels 3672 vertical pixels of size 2.49 µm. The defocused sample plane is magnified in a custom objective lens (seen in black in the metal tube) at 5$\times$ magnification, yielding an effective sample pixel size of 0.49 µm. The camera is mounted on a motorized stage, allowing absolute axial translation.

\newpage
\subsection{Hologram Capture Sequence}

A sample, e.g., \ac{WBC}s prepared according to Section 3.1.2, is injected into the cuvette via a syringe connected to the inlet, while the outlet discards the sample into a vial after examination. Following injection, an electro-mechanical clamp stops the fluid flow, rendering the sample essentially motionless for the duration of the capture\footnote{Aside from slow sedimentation of the cells. The holograms are captured while the cells are still sedimenting.}.

Before the capture sequence, the top of sample is brought into focus to establish a plane from which propagation distance can be measured. With the zero-plane defined, the hologram capture sequence proceeds as follows: new sample liquid is injected into the cuvette and a clamp stops its movement, the camera is translated axially to appropriate defocus plane(s), in which a hologram is captured by the camera. Specifically, the camera is translated axially to three axial distances, 700 µm, 450 µm, and 235 µm, wherein a hologram is captured in each. As such, a classifier trained on this data will be more robust to large changes in propagation distances compared to the cuvette path length (90 µm). For capturing the training dataset for the \ac{ML}-based classifier, these steps are repeated for each cell-type sample until a complete dataset is built.

\subsection{Sample preparation}

In order to generate a correctly labelled training dataset, i.e., samples with corresponding class labels such as monocyte, lymphocyte, etc., isolated cell samples were generated by way of cell separation kits (Stemcell Technologies, EasySepTM Direct Human Isolation Kits).

For each donor (n = 7), the cells are separated, each from its own whole blood sample, by introducing cell-specific antibodies linked to nano-scale paramagnetic particles. These antibodies recognize and bind to their corresponding cell type, thus linking the cells to the magnetic particles as well. Afterwards, the sample is placed in a magnetic field, which allows the separation of the antibody-affected cell type. In practice, this procedure is far more complex, and was carried out by experienced laboratory technicians.

Following separation of the three \ac{WBC} types for the 3-part diff; monocytes, lymphocytes, and neutrophils, the samples were diluted to a common concentration of 1.3 $\times$ 10\textsuperscript{9} cells per liter. Cell concentration was measured using a Beckman Coulter hematology analyzer. Standardizing the concentration for the training data is necessary to ensure that no classification bias can exist purely from observing the concentration of cells in a sample. However, the isolation procedure is not flawless. A nearly unavoidable small percentage of other cells will be included in each isolation.

\section{Hologram Processing and Cell Extraction}
\label{sec:lens_processing}

\subsection{Hologram Reconstruction}
\begin{figure}[t!]
\centering
\captionsetup{format = plain}
\includegraphics[width=.9\linewidth]{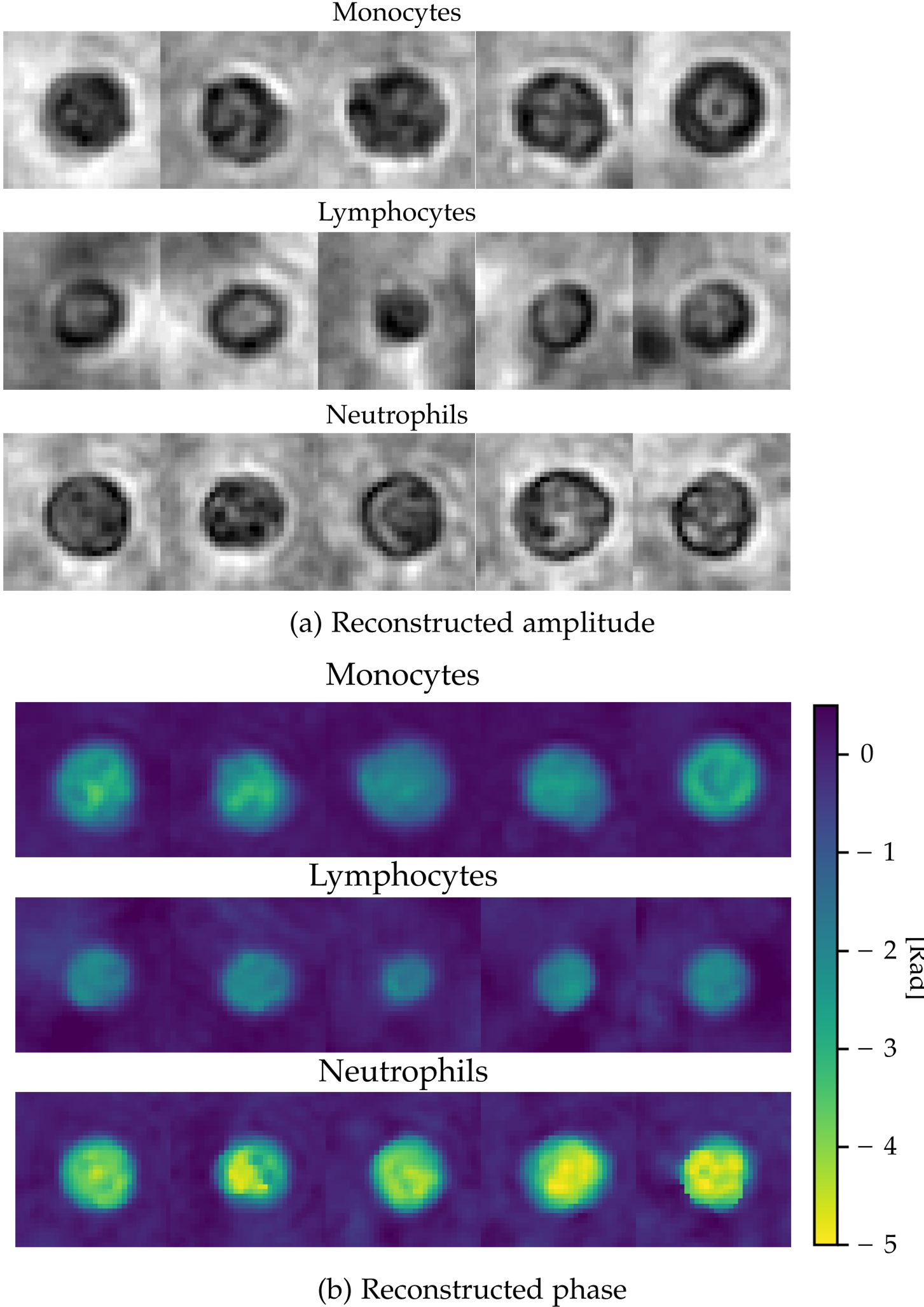}
\caption{Example reconstructions of the cell types in the 3-part diff in their focus planes. (a) The reconstructed amplitude profiles. (b) The reconstructed phase profiles.}
\label{fig:cell_reconstructions}
\end{figure}

With the numerous reconstruction methods yielding impressive results that have been developed in recent years \cite{madsen_-axis_2023, huang_quantitative_2024}, it is pertinent to consider which is most appropriate for the problem at hand. Given unlimited computing power and processing time, it would certainly be wise to choose an algorithm which spares no expense with regards to reconstruction quality, i.e., twin-image attenuation, noise reduction, etc., as more detail and precision of the sample wavefront may improve classification results. However, the practicalities of medical devices, hospital budgets, and the impatience of the health of a patient must also be considered. A \ac{PoC} analyzer must, first and foremost, be compact enough for ease-of-use in acute-care facilities, while also being able to quickly yield sufficiently accurate and clinically relevant results, at a reasonable purchasing cost.

With these constraints in mind, this initial proof-of-principle analyzer simply utilizes the raw numerical reconstruction of free-space propagation of the hologram via the angular spectrum method (\ac{ASM}) \cite{goodman_introduction_2017}, thus prioritizing on-device reconstruction speed at the cost of reconstruction quality and lack of twin-image attenuation.

In \cref{fig:cell_reconstructions}, amplitude and phase reconstructions of examples of the three cell types are shown. The smaller size of the lymphocytes stand out, as well as the greater recovered phase shift due to the neutrophils. The background noise seen especially in the amplitude reconstructions are a result of the artifacts involved in the reconstruction, including twin-image effects from the cell itself, but also from nearby cells.

\subsection{Reconstruction Autofocus and Cell Extraction}

To accurately classify the \ac{WBC}s, the correct reconstruction plane of each cell must be determined. The axial location of any particular cell depends on its depth within the cuvette, as well as the axial translation of the capturing camera. Therefore, an autofocus method, inferring the correct propagation distance, is required. Such an autofocus method can work in several different configurations.

First, given a single reconstruction plane, chosen at a random propagation distance within a specified interval, or the hologram itself, the autofocus determines the propagation distance immediately. This single-step approach is found in recent \ac{ML}-based autofocus algorithms \cite{cuenat_fast_2022, huang_holographic_2021,jaferzadeh_no-search_nodate}.

Second, given an entire stack of $N_z$ reconstruction planes on a propagation interval, e.g., the depth of the cuvette, the autofocus determines the correct focal index from a quantitative focus measure. This focal stack approach is expensive in terms of computer memory, as the entire stack must be pre-calculated and stored.

Third, given an initial plane reconstructed from the captured hologram, the autofocus algorithm iteratively optimizes the propagation distance using the quantitative focus measure, reconstructing to a new plane and re-calculating the focus measure in each iteration. This multi-step approach is effectively a hybrid of the first two, and is also a common approach in consumer cameras.

Despite the memory requirements of the focal stack method, it also presents several advantages. The cells are distributed throughout the cuvette and there is not one single plane of focus. Therefore, a multi-step approach would require optimization of cell-focus for each cell observed in the hologram. This is both infeasible computationally, as there can easily be thousands of cells in an image and each would require repeated propagation of a region of interest (\ac{ROI}) around the cell, and impractical as it would require precise lateral localization of the cells from the hologram alone. The latter becomes extremely challenging as the concentration of cells is increased even slightly.

As such, a customized focal stack algorithm is implemented, containing four separate steps as summarized here, and illustrated in \cref{fig:cell_extraction_process}:

\begin{enumerate}
\item \textit{Hologram partitioning:} To alleviate the memory concern of the focal stack method, and to increase parallelizability, the hologram is initially partitioned into several subholograms.

\item \textit{Subhologram propagation:} Each subhologram and propagated to $N_z = 50$ axially distributed reconstruction planes, spanning 150 µm — a range that accommodates slight axial translation errors and variations in cuvette depths exceeding the 90 µm internal thickness of the cuvette. This yields a stack of complex-valued reconstruction planes, $C_z(x, y, n\Delta z)$, where each reconstruction corresponds to a different propagation depth ($n\Delta z$). The corresponding axial sampling interval is then $\Delta z = 3$ µm.

\item \textit{Lateral cell-localization:} The lateral pixel coordinates of each cell center are extracted from a \ac{2D} map, $M(x, y)$, which emphasizes areas with strong spatial variation along the propagation axis. The specific implementation of the calculation of $M$ is found in Appendix B. The \ac{2D} map, $M$, takes into account both amplitude and phase contributions depth-wise across the focal stack, thus emphasizing cells while suppressing spurious noise.

\item \textit{Axial cell-localization:} For each cell center coordinate found in $M$, the reconstruction plane of best focus is identified by first defining a \ac{ROI} surrounding the cell. The \ac{ROI} is a small region, with a width and height of 64 pixels ($\approx$ 31 µm), thus completely containing any \ac{WBC}. A quantitative focus measure, $S_f$, is then calculated for each axial plane in $C_z$ within the \ac{ROI}. The focus measure evaluates the sharpness of the reconstruction at each plane in the stack. Once the best focus distance, $z_f$, is determined, a complex \ac{3D} volume of interest (\ac{VOI}), $C_{\text{voi}}$, of the cell is extracted from the reconstruction stack. This \ac{VOI} consists of the focus plane and 7 surrounding planes, forming a total axial span of $7 \times 3$ µm = 21 µm. This \ac{3D} \ac{VOI} is then used for further analysis and classification of the cell.
\end{enumerate}

Now, unlike conventional digital photography, where an autofocus algorithm must determine defocus in incoherent illumination, causing uniform blurring across the image plane, this algorithm must determine defocus under coherent illumination, where self-interference and twin-image distortions heavily influence the reconstructed field.

\begin{figure}[t]
\centering
\captionsetup{format = plain}
\includegraphics[width=\linewidth]{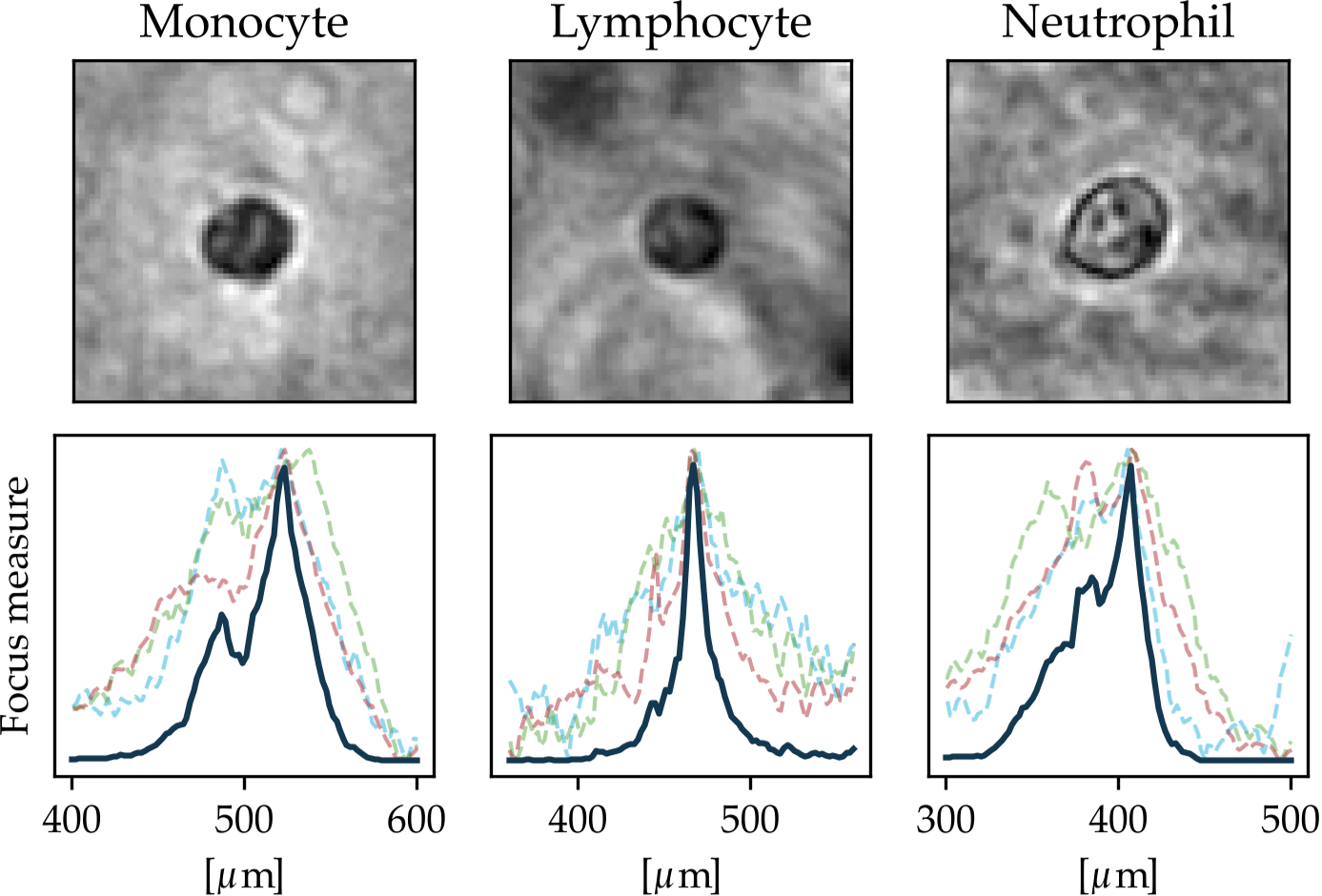}
\caption{Examples of the calculated focus measure $S_f$ (dark blue line), and its constituent measures (light blue, green, and red lines), for each cell type as it is calculated through the axial focal stack. The top row shows examples of the reconstruction absorption of a found focus plane for each cell type at the maximum value of the focus measure.}
\label{fig:focus_measure_examples}
\end{figure}

\begin{figure}[t]
\centering
\captionsetup{format = plain}
\includegraphics[width=0.8\linewidth]{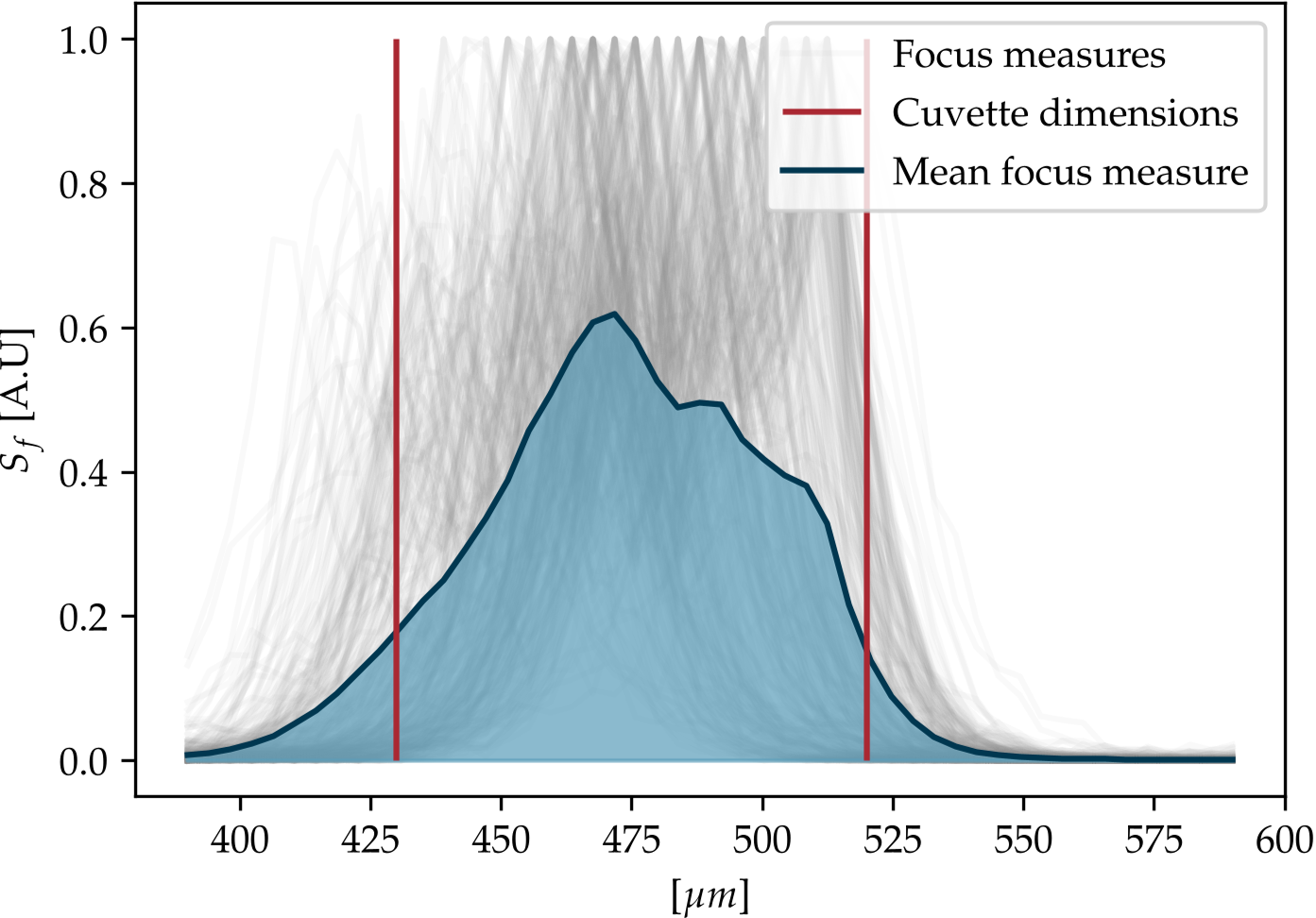}
\caption{Plots of the calculated focus measures, $S_f$, for 400 neutrophils in a sample (gray) and the mean focus measure (blue) across the dimensions of the cuvette.}
\label{fig:focus_measure_distribution}
\end{figure}

Therefore, there are numerous examples from the literature of development of holographic focus measures, and it is still an active research field \cite{cuenat_fast_2022,jaferzadeh_no-search_nodate,winnik_versatile_2021,li_terahertz_2019,liu_robust_2019,dharmawan_nonmechanical_2021,rathod_fast_2021,che_exploiting_2020,ilhan_digital_2014,brault_automatic_2022,ghosh_autofocusing_2021,tang_autofocusing_2020,zhang_robust_2018,malik_practical_2020}. Additionally, since the samples being reconstructed (\ac{WBC}s) can vary in both absorption strength and refractive index, a focus measure for this application must be robust enough to take this into account, i.e., one focus measure may work well on monocytes but not on lymphocytes. After extensive trial and error, the most effective focus measure for this application was found to be the product of three individual focus measures. These were selected for their ability to best indicate the focus of cells, regardless of their type, while minimizing the influence of the noisy background. The combined focus measure is defined as follows:
\begin{figure}[t]
\centering
\captionsetup{format = plain}
\includegraphics[width=\linewidth]{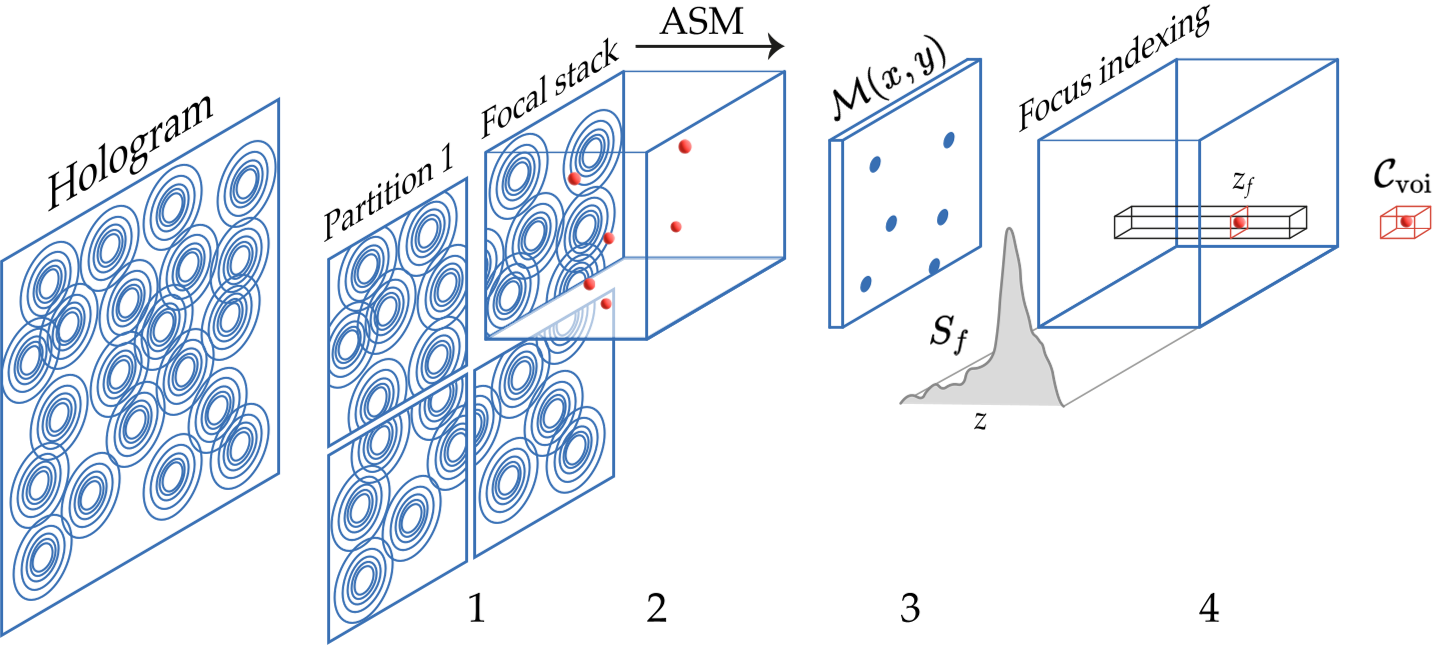}
\caption{Illustrated process of extracting cell \ac{VOI} for the lens-based holographic analyzer, according to the steps described in the main text. 1: Captured hologram is partitioned into sub-holograms. 2: Each sub-hologram is propagated through a focal stack of $N_z$ planes. 3: A lateral cell location map $M$ is calculated, emphasizing areas of strong variation along the propagation direction. Each cell location is determined by a peak finding algorithm. 4: A column defined by the cell location is extracted from the focal stack. The cell focus plane is determined from evaluation of each plane in the focus measure $S_f$. From the cell focus plane, the final cell \ac{VOI} extracted.}
\label{fig:cell_extraction_process}
\end{figure}

\begin{align}
S_f(C_z(x, y, n\Delta z)) = &S_{\text{GradientSum}}(\text{Im}(C_z(x, y, n\Delta z))) \nonumber \\
&\times S_{\text{GiniOfGradient}}(\text{Re}(C_z(x, y, n\Delta z)) \nonumber \\
&\times S_{\text{TamuraOfGradient}}(|C_z(x, y, n\Delta z)|)
\end{align}
where the definitions of the three focus measures can be found in Appendix C. The final focus measure utilizes the amplitude, real, and imaginary part of any given reconstruction, maintaining robustness despite inherent differences in cell properties. Before computing this final focus measure, each focus measure is normalized to values in [0; 1] over the propagation range. The cell focus is then selected as the propagation distance, $z_f$, corresponding to the maximum value of $S_f$. An example of the focus measure for three instances of the cell types in the axial column through the focal stack is shown in \cref{fig:focus_measure_examples}. The reconstructed planes correspond well to cell-focus. In \cref{fig:focus_measure_distribution}, the focus measure has been calculated and plotted for 400 neutrophils in a sample. The mean focus score is also plotted, showing the distribution of axial cell locations within the cuvette. Due to the suspected laminar flow of the sample liquid upon injection, the cells appear to be populated more densely in the center of the cuvette.

\section{White Blood Cell Classification using a Convolutional Neural Network}
\label{sec:lens_classification}

\subsection{Classification Datasets}

\ac{ML}-based image classification typically requires a labelled dataset. Following the cell-type isolation and cell extraction, as explained in Sections 3.1.2 and 3.2.2, a dataset of \ac{VOI}s surrounding the \ac{WBC}s has been generated. Each cell \ac{VOI} is labelled according to its specific isolation.

To facilitate training of a \ac{CNN} classifier, the dataset is divided into training and validation data. Training data, as the name suggests, is used by a classifier during training to identify and learn key features that best distinguish the cell classes. Validation data, in this case, consists of a mixture of data from the same dataset as the training data that has been withheld from use in training, and data from separate donors. The purpose of the different datasets is mainly to monitor and prevent overfitting. For instance, if the training loss decreases without an accompanying loss in validation loss, it is an indication that the classifier has overfitted to the data, i.e., learned to classify based on specific features and biases of the training set, instead of capturing generalizable features that are applicable to unseen data.

Overfitting is best prevented by acquiring a large and diverse training set with nuanced examples of samples from each of the classification classes. Using the isolated cell samples, each cell reconstruction is localized in x, y, and z within the cuvette, and a complex cell \ac{VOI} is extracted from the focal stack according to the method previously described. After processing the captured holograms from each of the isolated cell types, the acquired dataset consists of 31000 total cell \ac{VOI}s. The complete dataset breakdown is shown in \cref{tab:lens_dataset}.

\begin{table}[htbp]
\centering
\caption{Dataset overview for the lens-based classifier.}
\label{tab:lens_dataset}
\begin{tabular}{lcc}
\hline
Category & Training & Validation \\
\hline
Monocytes & 8819 & 1595 \\
Lymphocytes & 8770 & 1475 \\
Neutrophils & 8761 & 1580 \\
\hline
Total & 26350 & 4650 \\
\hline
\end{tabular}
\end{table}

Notably, as neural networks are typically designed to operate only on real-valued data, each complex \ac{3D} \ac{VOI} is separated into two stacks of amplitude and phase profiles, respectively. These stacks are concatenated channel-wise prior to training. Each \ac{VOI}, as it will be presented to the \ac{CNN} classifier is thus of size $64 \times 64 \times 14$ (height $\times$ width $\times$ channels).

\subsection{\ac{CNN} Design}

As described in Section 2.3.3, transfer learning is a popular method within the image classification community as a way of utilizing large pre-trained models for new, related tasks. Transfer learning allows the exploitation of the already learnt low-level features, drastically reducing training time and computational resources.

\begin{figure}[t]
\centering
\captionsetup{format = plain}
\includegraphics[width=\linewidth]{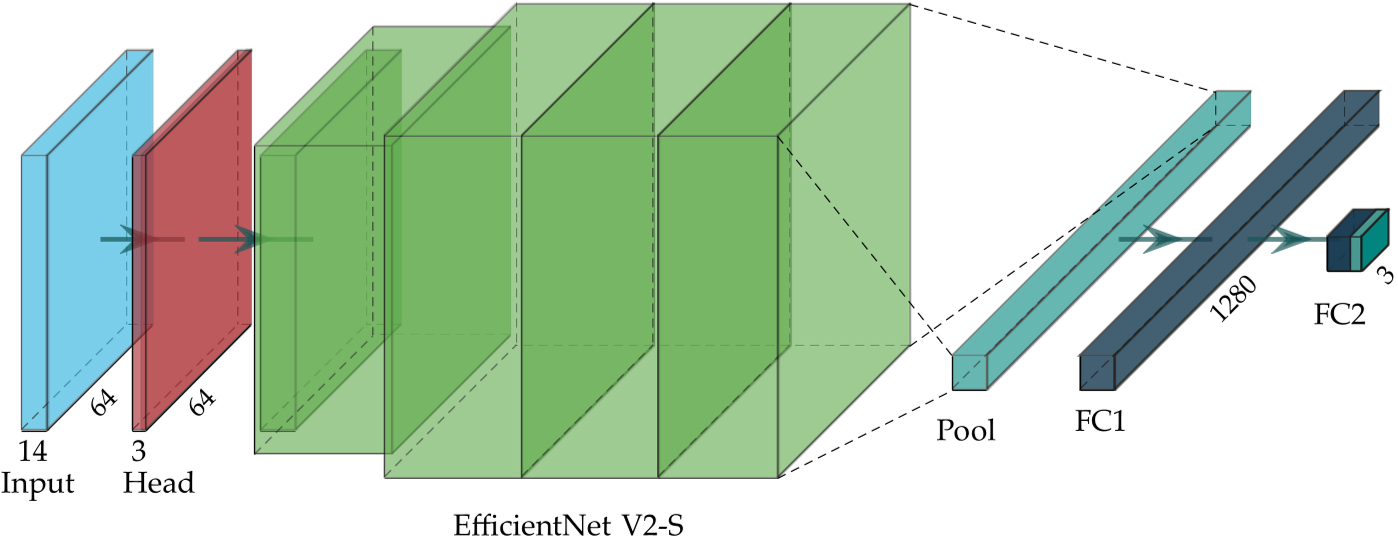}
\caption{Illustration of the classifier for the lens-based analyzer. The input \ac{VOI} is passed through the head and the pretrained network, after which it is finally flattened and evaluated in the two \ac{FC} layers to yield the classification.}
\label{fig:cnn_architecture}
\end{figure}

For the purpose of classifying \ac{WBC}s, the choice of pre-trained network fell on the so-called EfficientNet architecture \cite{tan_efficientnet_2019,tan_efficientnetv2_2021}. EfficientNet is a network architecture family specifically designed for image classification tasks. The network employs a so-called compound scaling to appropriately balance the network depth, width, and resolution to increase accuracy while maintaining a low computational cost. Specifically, the pre-trained EfficientNetV2-S model is selected. In transfer learning, it is common to add two custom blocks to the pre-trained network: a head and a tail (See \cref{fig:cnn_architecture}).

The head is responsible for adjusting the extracted cell \ac{VOI} into the format that the pre-trained network expects. Through a convolutional layer, the head learns to optimally condense the 14 reconstruction planes in each cell \ac{VOI} into a 3-channel \ac{RGB} image (the input requirement of the EfficientNet architecture), encapsulating the most important features.

The tail, which performs the actual classification, is an \ac{FCN}, as described in Section 2.3.3. The final layer within the tail then consists of three output neurons, one for each of monocytes, lymphocytes, and neutrophils.

\subsection{Network Training}

The proof-of-principle \ac{WBC} classifying network is implemented using the \ac{ML} library PyTorch \cite{paszke_pytorch_2019} for Python. As the loss function, a cross-entropy loss \cite{mao_cross-entropy_2023} is configured, which penalizes incorrect predictions by computing the negative log-likelihood of the true class, encouraging the network to output higher probabilities for the correct classes \cite{goodfellow_deep_2016}. Training of the network is performed over 10 epochs\footnote{An epoch corresponds to one training cycle in which the network has been shown the entire training dataset.} with the Adam optimizer \cite{goodfellow_deep_2016, chollet_deep_2018} on a MacBook Pro with an M1 Pro System-on-a-Chip. Intermittently during training, the validation data is classified by the network, such as to gauge its accuracy during training.

\section{Classification Results}
\label{sec:lens_results}

\begin{figure}[t]
\centering
\begin{subfigure}[t]{0.49\linewidth}
\centering
\includegraphics[height=5.5cm]{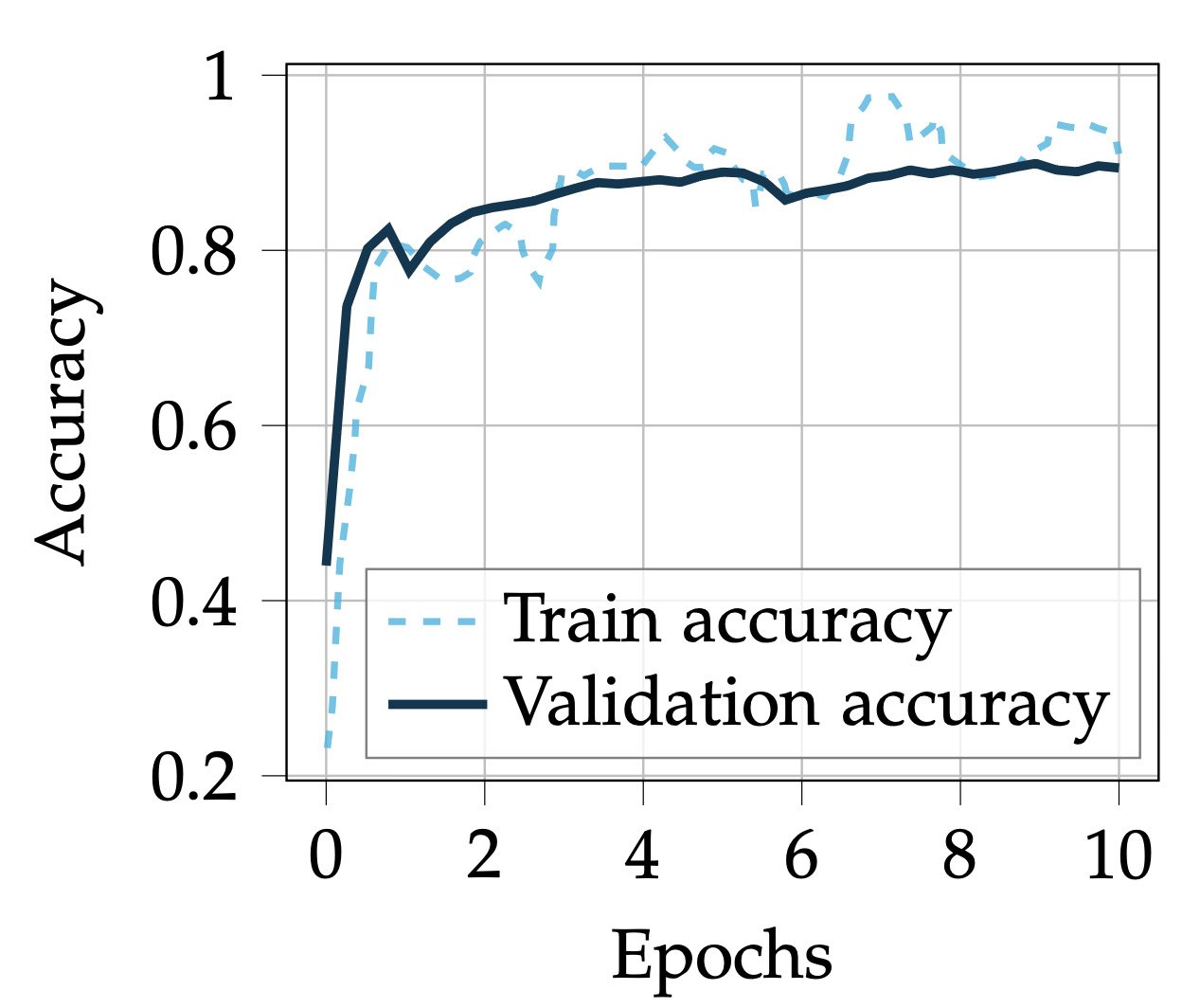}
\caption{Evolution of training and validation accuracy over the course of 10 training epochs of the lens-based classifier.}
\label{fig:training_progress}
\end{subfigure}
\hfill
\begin{subfigure}[t]{0.49\linewidth}
\centering
\includegraphics[height=5.5cm]{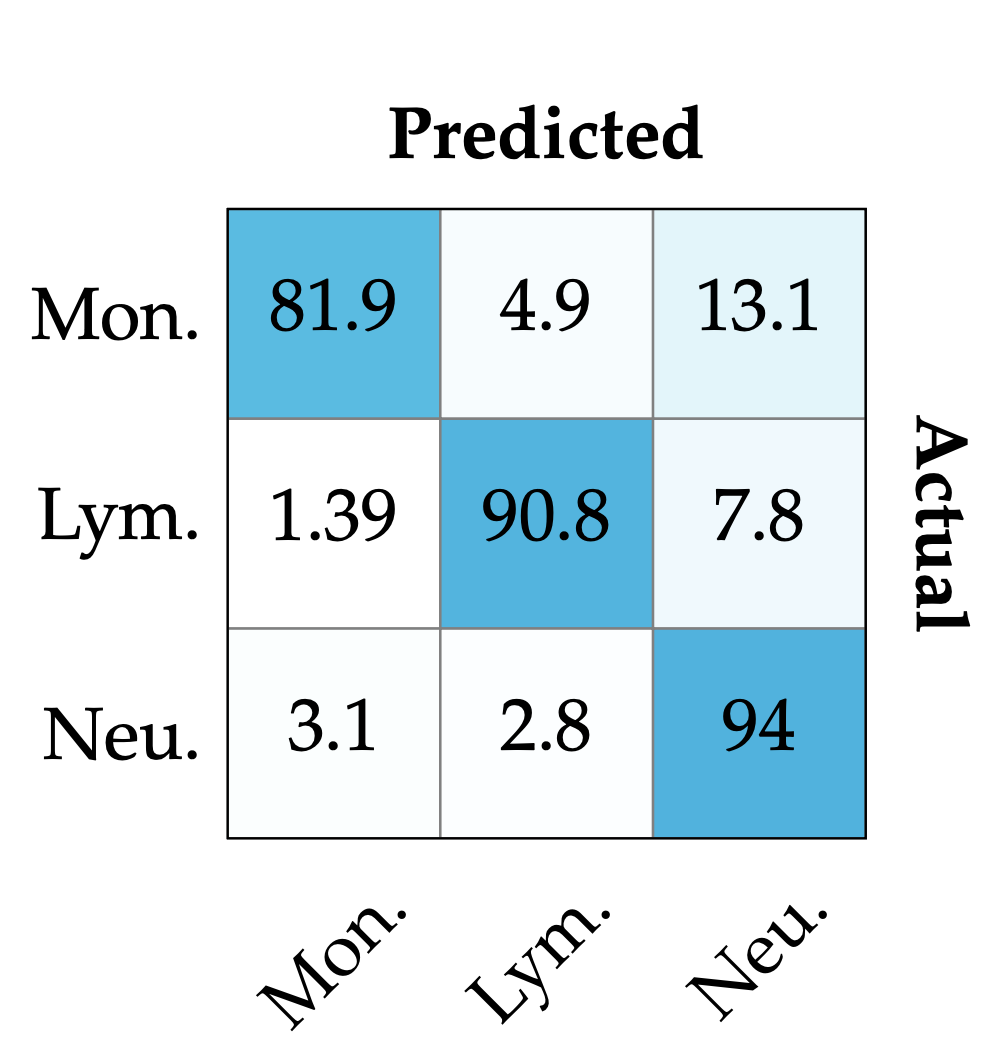}
\caption{Confusion matrix calculated from inference of the lens-based classifier on the unseen validation data.}
\label{fig:confusion_matrix_lens}
\end{subfigure}
\caption{Training results for the lens-based classifier.}
\label{fig:lens_training_results}
\end{figure}

The training progress of the lens-based \ac{WBC} classier is illustrated in \cref{fig:training_progress}. Over the course of the 10 epochs, the network learns to classify the three \ac{WBC} types with 89.6\% overall accuracy on the unseen validation data, with an F1 score of 0.89. It is apparent that the model is not significantly overfitting to the training data, as the validation curve approaches the training curve asymptotically, indicating stable generalization to unseen data. If overfitting were occurring, the curves would be expected to show strong divergence, with the validation curve plateauing or decreasing, while the training curve continued to improve. Following training, the portion of correctly classified samples for each \ac{WBC} type in the validation data is examined, and is illustrated in the confusion matrix in \cref{fig:confusion_matrix_lens}. The network shows strong ability to correctly classify neutrophils and lymphocytes, whereas monocytes are correctly predicted only 81.9\% of the time, with 13.1\% misclassifications as neutrophils. The misclassifications can perhaps be explained by the simple fact that lymphocytes are more different from neutrophils and monocytes than these two are from each other. Lymphocytes are, on average, approximately 8-12 µm in diameter, with nuclei occupying most of the cell volume. In contrast, monocytes and neutrophils are typically larger, with diameters ranging from 12-16 µm and 15-30 µm, respectively, and both contain nuclei with distinct shapes that may be mistaken for each other if spatial resolution is limited, and the cell is particularly ill-oriented, even if this misclassification should be rare, according to \cref{fig:wbc_misclassification_matrix}.

However, there may be more to the reason for the monocyte misclassification. First of all, the cell-isolation procedure described in Section 3.1.2 cannot guarantee perfect isolation. Therefore, there is a chance that the ground truth labels are incorrect for a small, but non-trivial, percentage of cells. These mislabeled cells will, even for a perfect classifier, cause a decrease in classification accuracy. Second, the isolation is not a harmless procedure for the cells. The sheer mechanical stress applied to the cells may cause damage to the cells; causing deformation, activation, or even cause the cell wall to rupture. Damaged cells may not have the same characteristics as their intact counterparts, and will, inevitably, lead to more difficult classification. Ruptured cells may be detected by the cell localization and autofocus algorithms, however, they bear little to no resemblance to their intact counterparts.

Finally, the diversity of samples contributing to the training data is limited, potentially introducing learned biases. For instance, donor-to-donor variations in cell morphology is large, and the monocytes in the validation set may be too different from the examples in the training set. Therefore, in order to improve on the classification accuracy, increasing the number of donors, and raw training samples, will be of significant importance.

\section{Chapter Conclusion}
\label{sec:lens_conclusion}

This chapter documented the development of a lens-based proof-of-principle holographic analyzer for 3-part \ac{dWBC}. The promising results, achieving 89.6\% overall accuracy on the unseen validation data, indicate that this \ac{DHM} implementation captures sufficient cell information for accurate categorization of the three \ac{WBC} types. The classifier shows a strong ability to distinguish neutrophils and lymphocytes, achieving high accuracy for both. However, monocytes presented a harder challenge, with some misclassifications as neutrophils.

As discussed in the results section, several factors may contribute to the monocyte misclassification, including morphological similarities, mislabeling due to sample preparation, and donor-to-donor variations. Especially the latter has significant impact, as the limited number of donors could introduce biases in the classifier, affecting its ability to generalize to unseen monocyte samples.

As mentioned, this system was optimized for reconstruction speed and memory footprint, prioritizing its viability for a \ac{PoC} device. For instance, it relies on simple hologram reconstruction via \ac{ASM}. Consequentially, the classifier has to be able to extract the valuable information from the cell reconstruction, with as little influence as possible from the twin-image and self-interference term. This choice, while beneficial for \ac{PoC} classification, also presents a challenge for the classifier.

Naturally, with further development, the more complex and superior reconstruction algorithms should be employed \cite{madsen_-axis_2023} and tested to determine if even better classification performance can be achieved. Capable \ac{GPU}s are continuously decreasing in price and size and could in time be incorporated into a \ac{PoC} device to perform both reconstruction, focusing, and classification in parallel. Alternatively, powerful cloud computing solutions could be employed such that only capture and data transmission is performed on-device.

\chapter{Lensless Holographic White Blood Cell Differentiation}
\label{ch:lensless-wbc}

This chapter describes the development and testing of a prototype lensless holographic \ac{dWBC} microscope, capable of performing 3-part and 5-part diff.

With the foundational experiences and knowledge from the construction of the lens-based prototype in Chapter 3, an attempt can now be made to simplify the mechanical setup, and take advantage of the increased \ac{FoV} that a lensless system provides. While the setup described in Chapter 3 is likely to outperform a lensless system in raw differential accuracy with more research, the aim of this project is, in part, to optimize the mechanical setup of an analyzer for use in existing devices, and for acute care testing. Such a device requires compactness and robustness, especially for stressful environments such as emergency departments, where the care for a device is a second priority over the care of a patient.

Requiring only a light source and a camera sensor, a lensless \ac{DHM}-enabled \ac{dWBC} analyzer can be constructed compactly \cite{mudanyali_compact_2010, wu_lensless_2018-1}, cost-effective, and robust, making it a suitable fit for the \ac{PoC} testing environment. In addition, the large \ac{FoV} allowed by the lack of a magnifying objective lens may prove useful to the differentiation of more cells concurrently. This advantage naturally comes at the cost of reduced spatial resolution, which the classifier will be required to cope with.

The chapter begins with a description of the optical setup of a lensless microscope equivalent to the \ac{DHM} configuration illustrated in \cref{fig:dhm_configurations}b, followed by an exploration of the lateral resolution achievable with this lensless system. Then, an overview of the hologram processing steps is included, leading to a description of the development of the \ac{CNN}-based \ac{WBC} classifier. Several configurations of the input data are tested by performing a 3-part diff, and the resulting network is trained and analyzed further by expanding to a 5-part diff using unseen blood cell data from separate donors. Finally, an example is given for how this lensless microscope can enable the direct observation of other diagnostic health markers, specifically the monocyte distribution width (\ac{MDW}).
\newpage
\section{Experimental Design and Hologram Capture}
\label{sec:lensless-experimental-design}

The lensless holographic analyzer consists of three assemblies;
\begin{wrapfigure}[22]{l}{5.9cm}
  \centering
  \captionsetup{format = plain}
  \includegraphics[width=5.9cm]{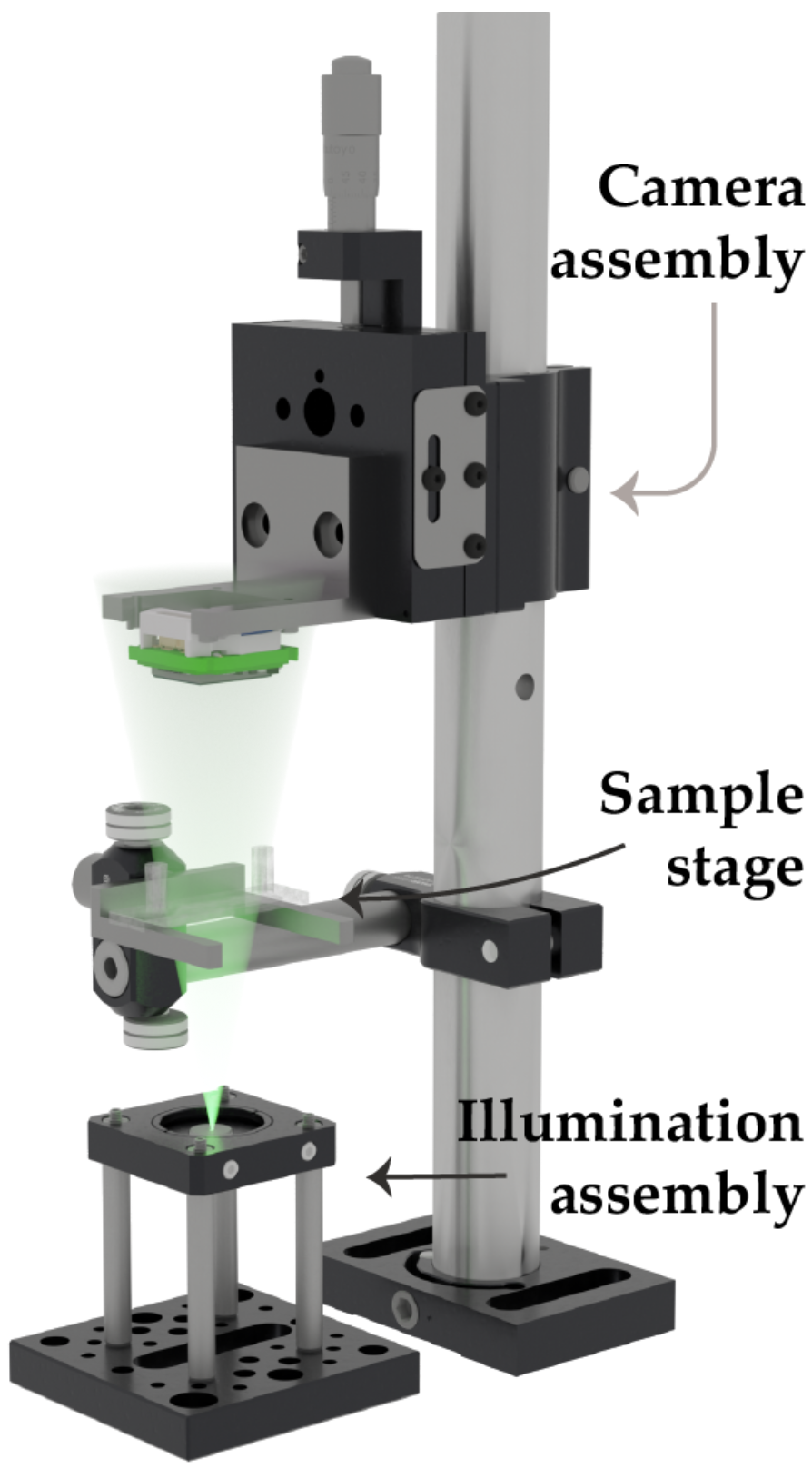}
  \caption{Render of the lensless holographic prototype microscope.}
  \label{fig:lensless-setup}
\end{wrapfigure}
\paragraph{the illumination assembly} consists of a fiber coupled 520nm laser \cite{noauthor_cps520_nodate}, terminated below the sample stage. The diverging laser beam is shone on the sample.

\paragraph{the sample stage} consists of a flow cuvette \cite{noauthor_flow_nodate} with internal dimensions 38mm $\times$ 9mm $\times$ 100µm. The cuvette has an inlet and outlet on opposite sides, and is mounted by double-sided adhesive tape on a custom machined fork. The fork-position on the post can be freely adjusted axially via the post-clamp.

\paragraph{the camera assembly} consists of a monochrome \ac{CMOS}-sensor-based camera \cite{noauthor_alvium_nodate} with resolution and pixel size of 5496 $\times$ 3672 and 2.49µm, respectively. The camera is mounted on a custom machined bracket, axially movable by adjustment of the translation stage.

\newpage
\subsection{Hologram Magnification and Lateral Resolution}

\begin{wrapfigure}{r}{6.2cm}
  \centering
  \captionsetup{format = plain}
  \includegraphics[width=6.2cm]{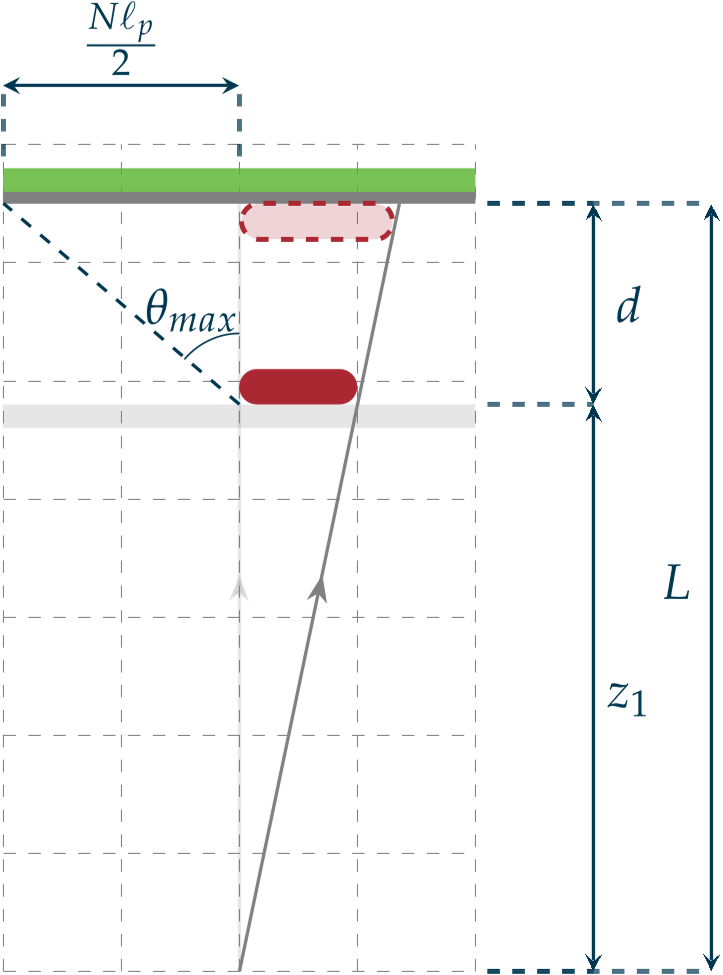}
  \caption{Illustration of the projected magnification principle. The object (red, solid) hologram is enlarged (red, dashed) on the detector due to the divergent illumination. The effective lateral resolution is dependent on the \ac{NA} of the sensor (via $\theta_{\max}$) and the distances d and L.}
  \label{fig:magnification-principle}
\end{wrapfigure}
With no objective lens providing explicit optical magnification present in the experimental setup, the lateral resolution in the holographic reconstructions is more limited than in the lens-based case. However, due to the uncollimated divergent nature of the laser beam from the fiber termination, it is still possible to achieve projected magnification of the hologram. Much like the shadow of a hand placed in the beam path of a flashlight appears larger when placed near the flashlight itself, so will the projection of the hologram on the camera sensor. The specific hologram magnification factor M, depends on the distances between the source and sample, and sample and camera sensor. As mentioned in Section 2.2.4, an object in the sample projects enlarged onto the camera sensor, as illustrated in \cref{fig:magnification-principle}. \\The proportionality constant between the original and projected object is the magnification, M, as expressed in \cref{eq:magnification-M}.

Lateral resolution of the hologram is then given by:
\begin{equation}
\delta_{\text{pixel}} = \frac{\ell_p}{M}
\label{eq:lateral-resolution-pixel}
\end{equation}
where $\ell_p$ is the pixel size of the camera sensor. From \cref{eq:magnification-M} it appears that the hologram can be magnified indefinitely, however, this is not the only factor.

Consider the setup in \cref{fig:magnification-principle} at d = 0, such that the sample is placed immediately in front of the camera sensor. In this case, no projected magnification occurs, and the lateral resolution is entirely limited by the pixel size of the camera sensor. As the sample is moved axially away from the camera, the magnification factor M becomes relevant. As this magnification increases, the hologram interference fringes gradually extend beyond the boundaries of the sensor area. This loss of high-frequency information leads to increasingly blurred reconstructions, effectively limiting the lateral resolution, despite the increased magnification. This can be expressed as the decrease in \ac{NA} that occurs as d is increased. From \cref{fig:magnification-principle}, the effective \ac{NA} of the system is given by the sine of the maximum angle $\theta_{\max}$ between the optical axial and the light reaching the edge of the sensor:
\begin{equation}
\text{NA} = \sin \theta_{\max} = \frac{\frac{N_{\min}\ell_p}{2}}{\sqrt{\left(\frac{N_{\min}\ell_p}{2}\right)^2 + d^2}}
\label{eq:numerical-aperture}
\end{equation}
where $N_{\min}$ represents the number of pixels along the smallest dimension of the camera sensor. This leads to an alternative lateral resolution expression, based on the Abbe diffraction limit \cite{abbe_relation_1882}:
\begin{equation}
\delta_{\text{NA}} = \frac{\lambda}{2\text{NA}}
\label{eq:lateral-resolution-na}
\end{equation}
With these two competing expressions for lateral resolution, a minimum effective pixel size \cite{zhang_resolution_2019} occurs at the intersection: $\delta_{\text{pixel}} = \delta_{\text{NA}}$.

\begin{figure}[t!]
  \centering
  \captionsetup{format = plain}
  \includegraphics[width=.95\linewidth]{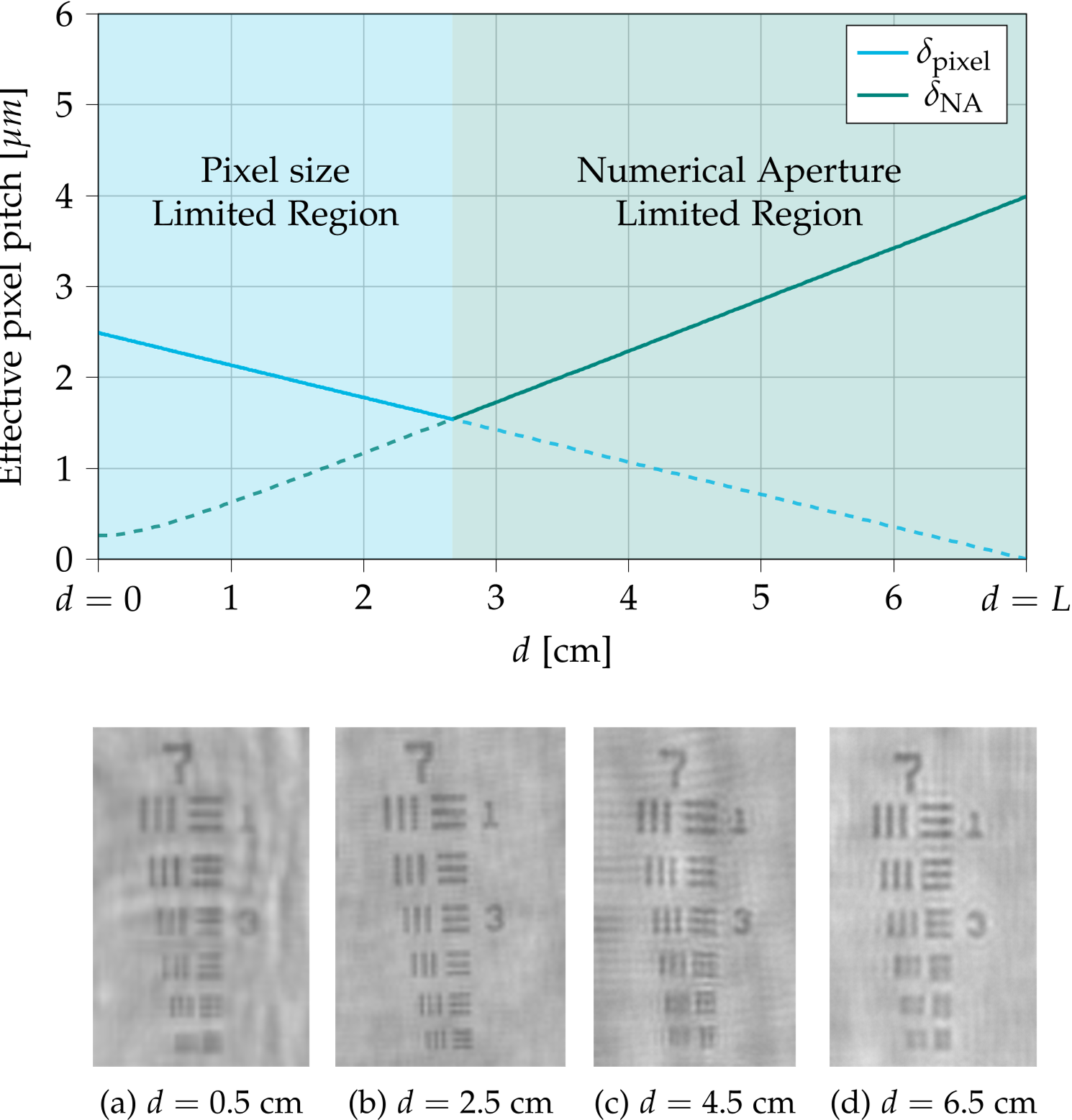}
  \caption{Analysis of effective resolution using the lensless holographic setup. (Top) plots of both \cref{eq:lateral-resolution-pixel,eq:lateral-resolution-na}, showing their overlap at d = 2.7 cm. (Bottom) Experimental reconstructions of a USAF resolution target at various values of d. From left to right: 0.5 cm, 2.5 cm, 4.5 cm, and 6.5 cm.}
  \label{fig:resolution-analysis}
\end{figure}

Now, an unknown in this system is the total distance L from light source to camera sensor. Due to the \ac{NA} of the optical fiber used ($\text{NA}_{\text{ill}} \approx 0.14$), L must be large enough for the divergent beam from the fiber to completely fill the area of the camera sensor. Therefore, the camera sensor is placed L = 7 cm behind the light source. Plotting the two lateral resolution expressions with this value fixed (see \cref{fig:resolution-analysis}) reveals a minimum effective lateral resolution of 1.54 µm (M = 1.62) at a sample-camera distance of d = 2.7 cm. To verify the existence of a minimum lateral resolution, a ThorLabs (R1L1S7P) 1951 USAF resolution test target is inserted in the sample holder of the lensless holographic assembly. Then, holograms of the target are captured at various sample-sensor distances, under constant illumination of the laser source. Reconstructions at d = 0.5 cm, 2.5 cm, 4.5 cm, and 6.5 cm are shown in a row below the analytical plot in \cref{fig:resolution-analysis}. Examining the reconstructions of group 7 in the test target, it does appear as though resolution is best preserved at d $\approx$ 2.5 cm, as expected. Here, the individual lines in the smallest element are discernible, where in the others significant blurring limits the fidelity.

\subsection{Sample Capture and Processing}

Holograms of each cell type; monocyte, lymphocyte, neutrophil, basophil, and eosinophil, from eight separate donors, are captured. Each sample is prepared using the cell isolation technique, as described in Section 3.1.2. Notably, due to the differentiation technique employed by the Beckman Coulter analyzer, which is used to determine the concentration of specific cells, it was not possible to determine the actual concentration of both eosinophils and basophils. The specific differentiation technique relies on relative metrics, based on the presence of the other more numerous cell types. Since these are not present in the isolation, the differentiation cannot proceed successfully. As such, a best attempt was made to match the concentration of the isolated eosinophil and basophil samples to the rest.

The captured holograms are reconstructed, and processed according to Section 3.2. 
Specifically, each hologram undergoes the same autofocus and cell extraction algorithm, in order to create a dataset of cell \ac{VOI}s. 
The \ac{VOI} are of dimensions 30 pixels $\times$ 30 pixels $\times$ 9 planes.

As described in Section 3.2, a critical feature of a \ac{PoC} analyzer is the speed of acquisition and processing. 
As such, the simplest reconstruction method is maintained, namely reconstruction via \ac{ASM}. 
The holograms are reconstructed using simulated plane wave illumination, thus maintaining the magnification of the cells.

\begin{table}[t!]
\centering
\captionsetup{format=plain}
\caption{Overview of the 3-diff dataset for the lensless classifier.}
\label{tab:lensless-3diff-dataset}
\begin{tabular}{lcc}
\hline
Category & Training & Validation \\
\hline
Monocytes & 42606 & 7586 \\
Lymphocytes & 42496 & 7572 \\
Neutrophils & 43117 & 7469 \\
\hline
Total & 128219 & 22627 \\
\hline
\end{tabular}
\end{table}

\section{Lensless White Blood Cell Classification}
\label{sec:lensless-classification}

\subsection{Classification Datasets}

Both a 3-diff dataset and a 5-diff dataset were created for this chapter, such as to easily compare to the lens-based proof-of-principle analyzer.

For the 3-diff dataset, a total of 128219 cell samples (85\%) are chosen for use in training, and 22627 (15\%) for validation. It is apparent that the increased \ac{FoV} of the lensless microscope makes it easier and faster to capture a greater number of cells in each capture. The 3-diff dataset is summarized in \cref{tab:lensless-3diff-dataset}.

The 5-diff dataset contains additional samples for each cell type, including \ac{VOI}s of basophils and eosinophils, thus forming a dataset of 416524 training samples (ca. 95\%) and 20000 (ca. 5\%) for validation. This training data consists of samples from seven donors, and the validation from a single separate donor whose samples are not present in the training data. This dataset is summarized in \cref{tab:lensless-5diff-dataset}.

\begin{table}[t]
\centering
\captionsetup{format=plain}
\caption{Overview of the 5-diff dataset for the lensless classifier.}
\label{tab:lensless-5diff-dataset}
\begin{tabular}{lcc}
\hline
Category & Training & Validation \\
\hline
Monocytes & 72670 & 4221 \\
Lymphocytes & 84437 & 4596 \\
Neutrophils & 89676 & 3014 \\
Eosinophils & 86252 & 3454 \\
Basophils & 83489 & 4715 \\
\hline
Total & 416524 & 20000 \\
\hline
\end{tabular}
\end{table}

\subsection{Network Input}

The same pretrained network, EfficientNet-V2s, acts as the backbone of the classifier. In Chapter 3, the input to the classifier was the concatenation of the reconstructed amplitude and phase profiles from the \ac{VOI}s. However, the complex \ac{VOI}s can be represented in several ways as input to the classifier, e.g., planes of amplitude, phase, absorption, etc. Therefore, an analysis to determine the combination of representations that yields the best performance is carried out. Five planar representations of the complex \ac{VOI}s are extracted, namely the amplitude, absorption, phase, and the real and imaginary part. These representations are of course not independent, being different representations of the same data. However, whatever the classifier deems as the most efficient representation of the data should be used. Although the same information is available in the amplitude and phase reconstruction as in the real and imaginary reconstruction, if one representation better provides the features that the classifier can base a decision on, there is no reason to waste computations in its layers trying to extract that same information. Each representation can be input to the classifier by itself or they can be input as combinations, concatenated channel-wise. With five features, there are 31 possible combinations that can be tested. To reduce the search space, a fractional factorial experiment design is employed. Specifically, a half-fraction factorial design with $2^{5-1} = 16$ combinations instead of the full 32 is constructed. The specific design construction allows the evaluation of the main effects of individual features on the classification accuracy. In \cref{tab:factorial-design}, the design scheme is defined, in which the presence or absence of each feature in the input is represented by a binary value (1 for inclusion, 0 for exclusion). For each input representation combination defined in the design matrix, the classifier was trained for two epochs using the complete training set. The performance of the classifier was then evaluated on the unseen validation set, and the resulting accuracy score was recorded.

\begin{table}[t]
\centering
\captionsetup{format=plain}
\caption{Half-fraction factorial ($2^{5-1}$) design matrix for selecting the best input representaions for lensless \ac{WBC} classification. Each run is defined by the included data representations, where 1 indicates inclusion in the input.}
\label{tab:factorial-design}
\begin{tabular}{lccccc}
\hline
Run & Amplitude & Absorption & Phase & Real & Imag \\
\hline
1 & 0 & 0 & 0 & 0 & 1 \\
2 & 0 & 0 & 0 & 1 & 0 \\
3 & 0 & 0 & 1 & 0 & 0 \\
4 & 0 & 1 & 0 & 0 & 0 \\
5 & 1 & 0 & 0 & 0 & 0 \\
6 & 0 & 1 & 1 & 0 & 1 \\
7 & 1 & 0 & 1 & 0 & 1 \\
8 & 1 & 1 & 0 & 0 & 1 \\
9 & 1 & 1 & 1 & 0 & 0 \\
10 & 1 & 1 & 0 & 1 & 0 \\
11 & 1 & 0 & 1 & 1 & 0 \\
12 & 0 & 1 & 1 & 1 & 0 \\
13 & 0 & 0 & 1 & 1 & 1 \\
14 & 0 & 1 & 0 & 1 & 1 \\
15 & 1 & 0 & 0 & 1 & 1 \\
16 & 1 & 1 & 1 & 1 & 1 \\
\hline
\end{tabular}
\end{table}

\begin{figure}[t!]
  \centering
  \captionsetup{format = plain}
  \includegraphics[width=\linewidth]{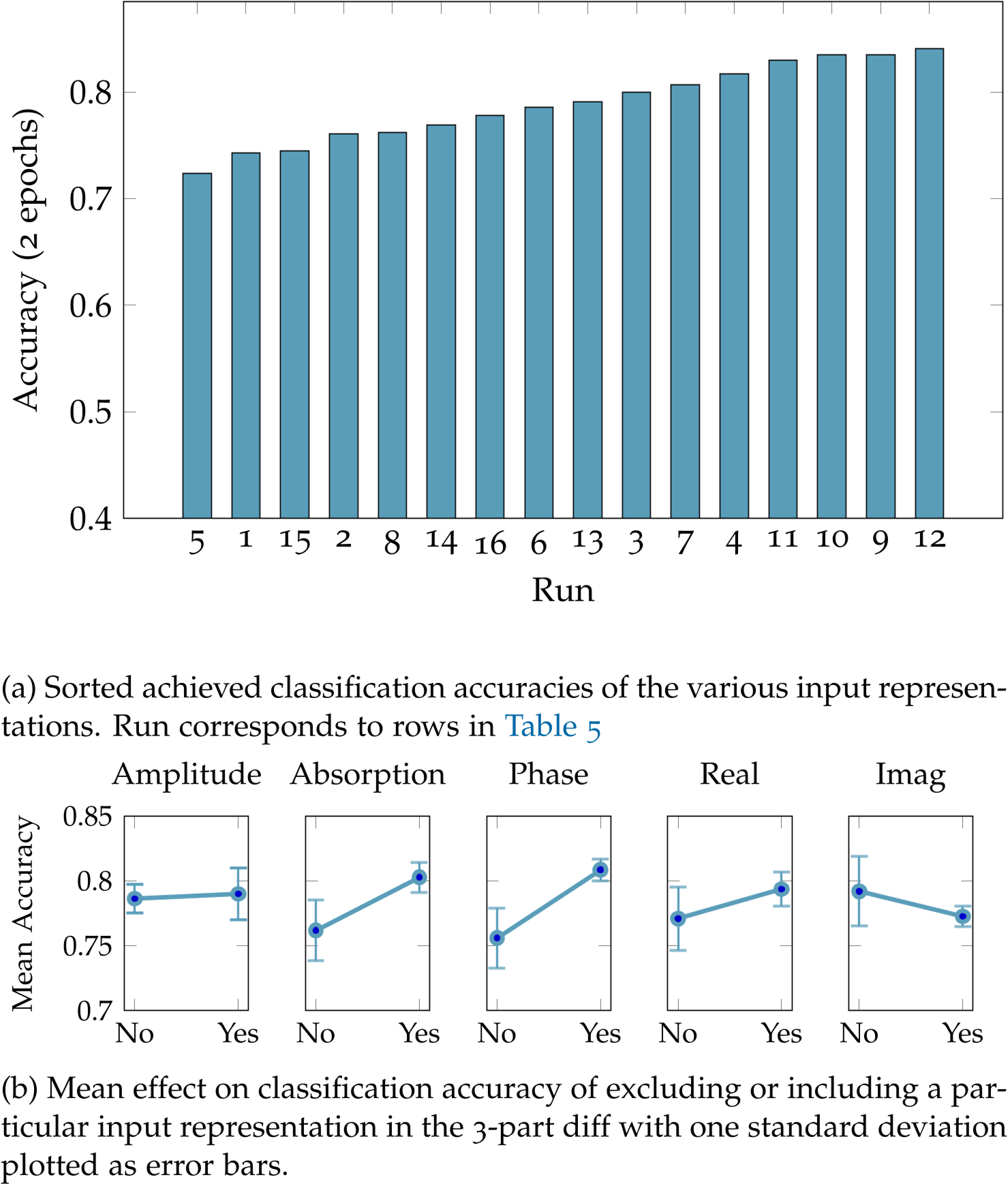}
  \caption{}
  \label{fig:input-representation-analysis}
\end{figure}

\cref{fig:input-representation-analysis}a shows the sorted accuracy scores obtained across all experimental combinations. Interestingly, the combination utilizing all available representations did not yield the highest performance. This can indicate several factors. For instance, the network may not be deep or complex enough to effectively process and learn from the increased dimensionality and information redundancy. Nonetheless, it is clear that certain combinations of input features perform better than others. In particular, the combinations utilizing planes of absorption, phase, and the real part of the reconstructions achieved the highest accuracy scores.

In \cref{fig:input-representation-analysis}b, the main effects of the presence of each feature is plotted. This demonstrates that, on average, including absorption, phase, or the real part results in a noticeable increase of classification accuracy. Conversely, the inclusion of amplitude or the imaginary part did not result in significant accuracy improvements. Based on these findings, the absorption-phase-real combination was selected as the optimal input feature representation for the lensless classifier, leveraging representations that demonstrably contribute positively to classification accuracy.

\begin{figure}[t]
  \centering
  \captionsetup{format = plain}
  \includegraphics[width=\linewidth]{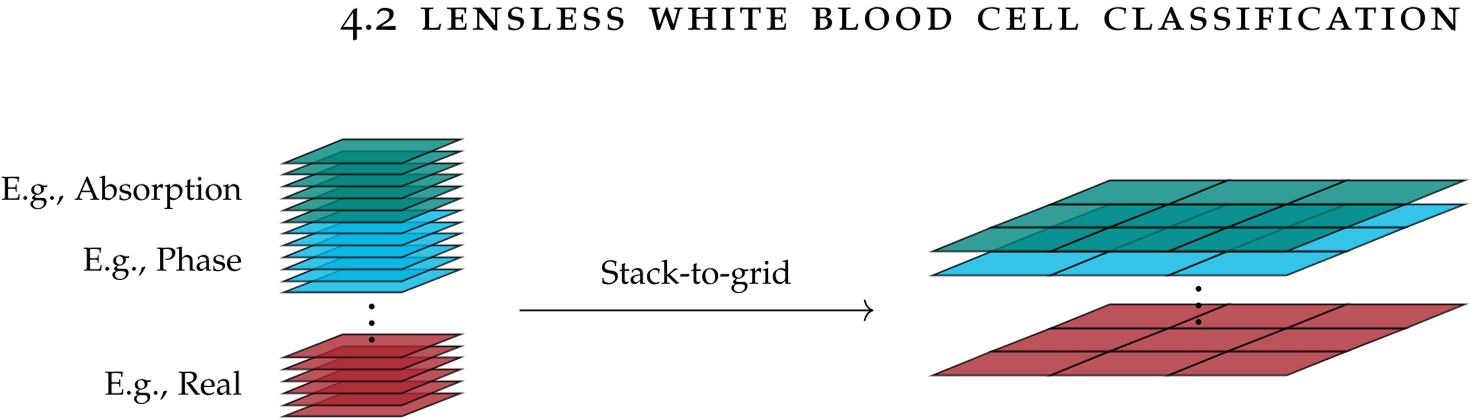}
  \caption{Change of input shape to the classifier. The various stacks of input representations from the \ac{VOI} are arranged in grids to better cope with features spanning multiple planes, focus location uncertainty, and better information transfer to the classifier.}
  \label{fig:stack-to-grid}
\end{figure}

\begin{figure}[t]
  \centering
  \captionsetup{format = plain}
  \includegraphics[width=\linewidth]{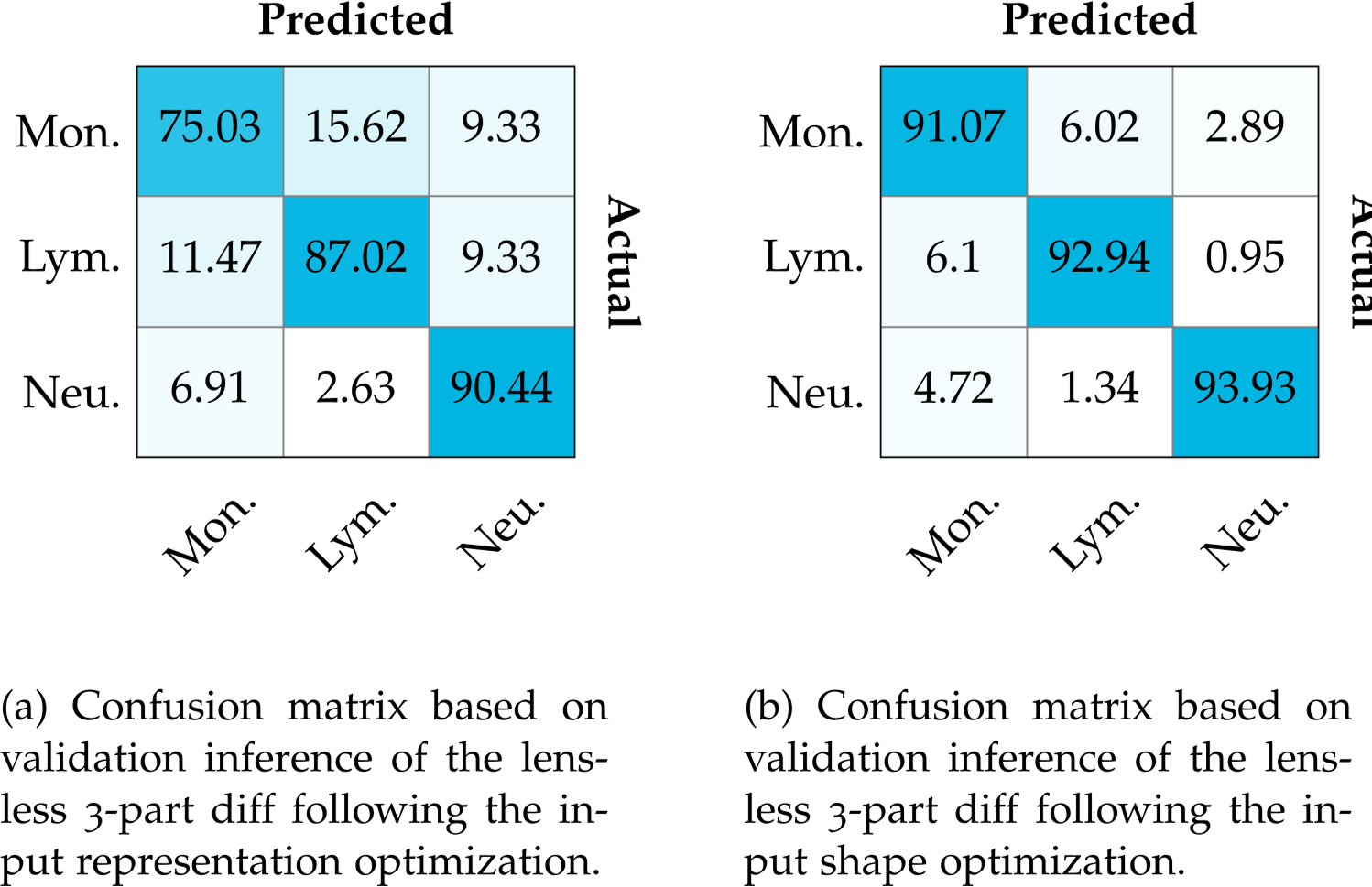}
  \caption{}
  \label{fig:confusion-3diff}
\end{figure}

\subsubsection*{3-Part Diff and Change of Network Input Shape}

With the input representation defined, the training of the lensless 3-diff classifier can be carried out with a similar approach as described in Section 3.3. For each sample in the training and validation dataset, the optimal data representations, namely the absorption, phase, real part, are extracted from the \ac{VOI}, and concatenated channel-wise. Specifically, each stack consists of nine planes, ideally centered around the focus of the cell. Thus, when concatenated, each input to the 3-diff classifier is of shape 30 $\times$ 30 $\times$ 27 pixels. The network is trained for 10 epochs, and finally evaluated using the unseen validation data.

The result of the evaluation is illustrated as a confusion matrix in \cref{fig:confusion-3diff}a. The 3-diff lensless classifier reaches an overall accuracy of 84.16\%, slightly lower than the previously described lens-based analyzer.

To improve the classification accuracy of the lensless classifier, an attempt at optimizing the utilization of the convolutional layers in the \ac{CNN} was carried out. Specifically, convolution kernels, such as the ones used in the head and pretrained blocks of the lensless classifier, are translation invariant, as described in Section 2.3. This means that it plays no role where, laterally, in an image a certain feature is located. Since the kernels are convolved with the input representations, each \ac{2D} coordinate in the input image is evaluated equally using a given kernel. They are not, however, depth invariant. The individual kernels are trained on singular input planes, and are not aware of the other planes. Therefore, if features span multiple planes, or if the focus plane is not found exactly, the kernels may not have been trained to compensate. In addition, as the task of the head block of the classifier is also to translate the representation stack to the expected input shape of the pre-trained network (3 input channels), there may be information loss associated with the compression of the larger stack of absorption, phase, and real planes. Therefore, it may prove advantageous to modify the shape of the input data slightly to both take fully advantage of the spatially invariant nature of the convolution kernels, as well as minimize information loss to the pre-trained network.

In \cref{fig:stack-to-grid}, this change of input shape is illustrated. Via a simple spatial redistribution of the focus planes spanning a cell sample, the focus planes can be arranged in a \ac{2D} grid. As the input stacks each consist of nine planes centered around the cell focus, these are readily arranged into 3 $\times$ 3 grids. Conveniently, as the best performing input type representation (Absorption-phase-real) consist of three stacks of planes in the \ac{VOI}, following the stack-to-grid redistribution, this corresponds to a 3-channel input image, which is exactly what the pre-trained network expects. Thus, the head block of the classifier is now not burdened by performing a similar information rearrangement.

The 3-diff classifier is once again trained with the same exact hyperparameters, but now with the 3-channel grid shape of the cell samples. In \cref{fig:confusion-3diff}b the result of the training is illustrated as a confusion matrix based on the inference on the validation dataset. The change of input shape results in a non-trivial 10.09\% improvement in classification accuracy on the unseen validation data, indicating better utilization of the data within the network. The 3-diff lensless classifier thus surpasses the result of the lens-based analyzer, with an overall accuracy of 92.65\%. These insights concerning both the input representation and shape are, naturally, applied to the 5-diff classification task as well.

\begin{figure}[t]
  \centering
  \captionsetup{format = plain}
  \includegraphics[width=\linewidth]{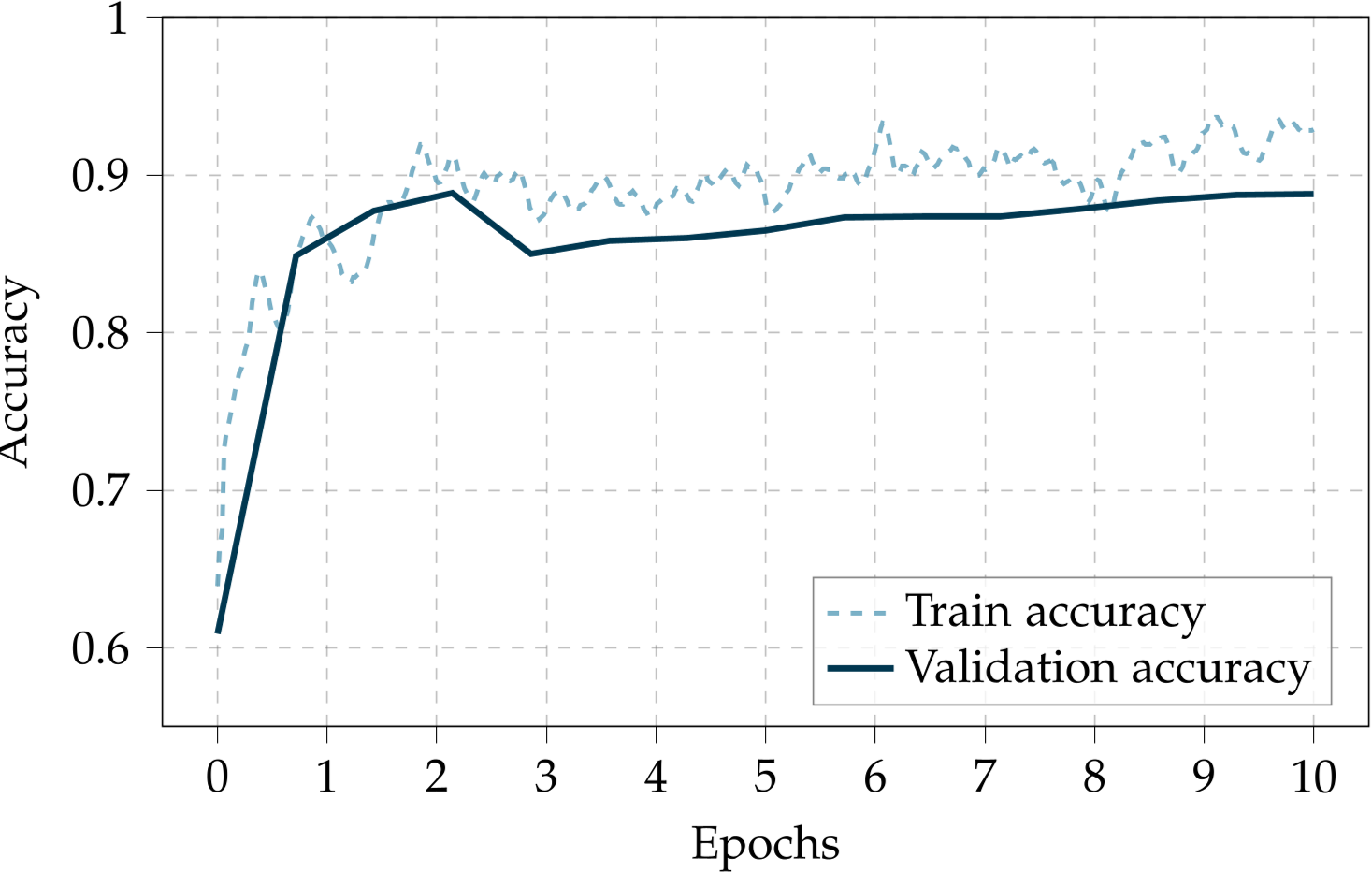}
  \caption{Training and validation accuracy of the 5-part classifier as a function of training epoch. Both accuracy measurements plateau after approximately 5 epochs.}
  \label{fig:training-progress}
\end{figure}

\begin{table}[t]
\centering
\captionsetup{format=plain}
\caption{Performance Metrics for the 5-diff lensless classifier.}
\label{tab:performance-metrics}
\begin{tabular}{lccc}
\hline
Cell type & Precision & Recall & F1-Score \\
\hline
Monocytes & 93.2\% & 86.1\% & 89.5\% \\
Lymphocytes & 89.9\% & 88.3\% & 89.1\% \\
Neutrophils & 94.9\% & 89.9\% & 92.3\% \\
Eosinophils & 95.7\% & 90.1\% & 92.8\% \\
Basophils & 77.0\% & 92.8\% & 84.2\% \\
\hline
Average & 90.1\% & 89.4\% & 89.6\% \\
\hline
\end{tabular}
\end{table}

\subsection{Expansion to 5-Diff Classification}

Increasing the possible outcomes of the classifier from three to five does not entail a significant change in architecture of the classifier. The basophil and eosinophil samples are included in the training and validation dataset summarized in \cref{tab:lensless-5diff-dataset}, and the samples are randomly selected and passed through the network during training. Practically, the most significant change consists of increasing the number of output neurons of the final \ac{FC} layer in the tail block of the classifier to match the new number of classification categories. More holistically, the increase in classification categories requires the network to learn more complex mappings to distinguish the separate categories. This also increases the requirement for a larger and more diverse training dataset in order to adequately describe the classes, as well as to prevent overfitting.

\section{Classification Results}
\label{sec:lensless-results}

The 5-diff lensless classifier is trained for 10 epochs on the training dataset summarized in \cref{tab:lensless-5diff-dataset} on a Lenovo Legion Pro laptop with an NVIDIA GeForce RTX 4080 \ac{GPU} for faster training and inference. In \cref{fig:training-progress}, the evolution of the classification accuracy in both training and the intermittent validation steps are shown. Both the training and validation accuracies increase steadily to a plateau, showing only minor signs of overfitting, with a difference in accuracy of 4.4\%.

\begin{figure}[t!]
  \centering
  \captionsetup{format = plain}
  \includegraphics[width=\linewidth]{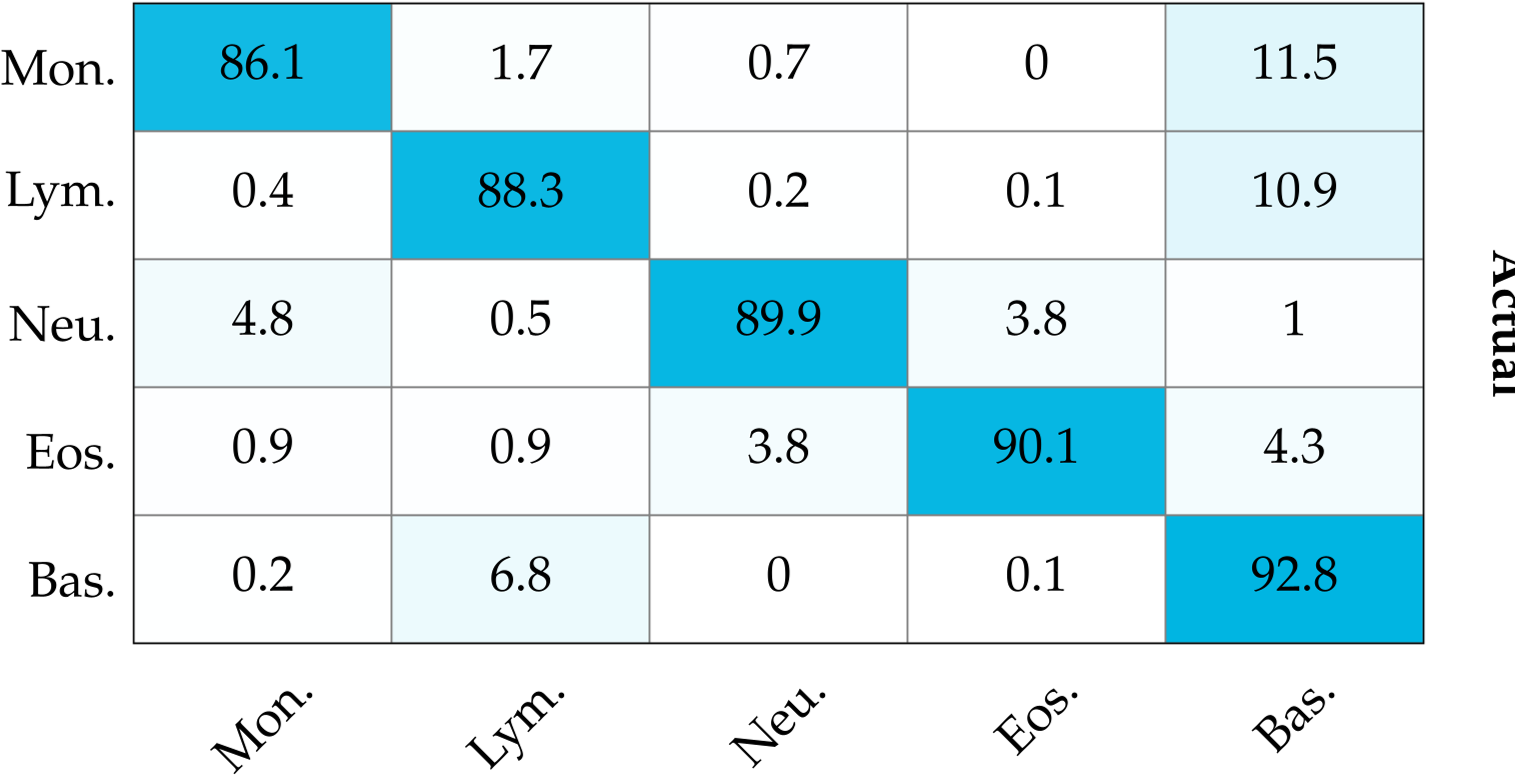}
  \caption{Confusion matrix based on validation inference of the lensless 5-part diff showing high accuracy for the majority of classification-pairs. Most misclassification occurs between basophils and the monocytes and lymphocytes.}
  \label{fig:confusion-5diff}
\end{figure}

In \cref{fig:confusion-5diff}, the results of the final validation epoch are shown as a confusion matrix. 
The classifier shows promising results in most combinations, yielding an overall accuracy of 89.44\%, with 16 of 25 fields showing misclassification rates of $\leq$ 5\%. 
Most notably of the misclassifications seem to be the basophils. 11.5\% of monocytes and 10.9\% of lymphocytes are misclassified as basophils. 
Unfortunately, these misclassifications fall within the medium-impact misclassification category (as described in \cref{fig:wbc_misclassification_matrix}), as high basophil counts can indicate health concerns such as an autoimmune disease allergic reactions, or even some types of blood cancers \cite{medicalnewstoday_everything_nodate}.

Inversely, there is also a tendency for the classifier to misclassify basophils as lymphocytes at a rate of 6.8\%; a misclassification with significantly lower impact, as the basophil count is typically much lower than the lymphocyte count.
To inspect the performance of the classifier further, additional performance metrics can be analyzed. In \cref{tab:performance-metrics} the calculated precision, recall, and F1-score are displayed for the validation epoch. As expected, the precision in classifying basophils, the measure of true positives to false positives, is significantly lower than the other cells (77.0\%), due to the higher rate of misclassifications of monocytes and lymphocytes as basophils. The classifier is, however, very good at detecting basophils when they are actually present in a sample, as indicated by the high recall score (92.8\%); the measure of true positives to false negatives. Overall, the classifier demonstrates strong initial performance, with a weighted average precision of 90.1\%, recall of 89.4\%, and F1-score of 89.6\%, yet with room for significant improvement.

One significant source of possible error is found in the sample preparation. As mentioned, the basophil and eosinophil counts were not possible to ascertain in the isolated cell samples. Therefore, it is a possibility that significant cross-contamination occurred in these samples. It is not expected, however, given the excellent performance of the isolation kits on the other cell types. Nonetheless, for future research, manual identification and labelling should be employed.

\subsection{Additional Classification Features}

In the current approach, all feature extraction is assigned to the classifier itself. While impressive, there may be additional features that are outside the scope of simple cell-images that carry significant information useful for classification. This is where human feature engineering could become an important factor. Aside from explicitly defined morphological features, e.g., size, granularity, cell membrane uniformity etc., one example could be exploiting the difference in densities of \ac{WBC}s. As soon as a cell sample is injected into the cuvette and the flow-clamp is engaged, the cells will begin to settle towards the base of the cuvette. As the sedimentation rate of objects in a medium, as determined by Stokes' Law \cite{norouzi_sorting_2017}, depends on the difference between the density of a given cell and the density of the medium, any differences in density between the cells themselves, can be exploited for use in classification \cite{larsen_apparatus_nodate}. The rate of sedimentation can be monitored by capturing a sequence of holograms, and determining the focus plane for each cell in each time step. Such a feature could be used for classification on its own, along with other engineered features, or injected into the \ac{CNN} classifier alongside the extracted cell images, and may lead to even more robust classification.

With the engineered features, and/or the embeddings from a feature extracting \ac{ML}-model, a statistical model could also aid in more robust classification and make the classification more interpretable by an operator. Examining a large dataset of cells and their features, populations according to the cell types could be formed in an n-dimensional space, either directly or through some dimensionality reduction algorithm, e.g., principal component analysis (PCA) or linear discriminant analysis (\ac{LDA}). The probability that a new sample belongs to a given population can be determined through statistical models such as gaussian mixture models (\ac{GMM}s) or Bayesian classifiers, given the location of the new sample in this n-dimensional space.

\section{Direct Diagnostic Capabilities}
\label{sec:direct-diagnostics}

In addition to the classification and count of the \ac{WBC}s, their other characteristics can play an important role as diagnostic tools. 
For instance, in recent years, the monocyte distribution width (\ac{MDW}) has become a promising biomarker for accurate and early detection of sepsis, a medical condition in which the immune system responds disproportionally to an infection, leading to inflammation, tissue damage, and organ dysfunction \cite{kralovcova_understanding_2024}. 
A delay in the detection of sepsis, and the resultant delay in treatment, can have mortal consequences. 
Early signs of sepsis can be vague, and many, equally serious, non-infectious diseases may mimic the symptoms of sepsis, which only emphasizes the importance of accurate, and early, detection. 
\ac{MDW}, as a parameter for the detection of sepsis, reflects a change of the distribution of the volume of circulating monocytes. 
During sepsis, monocytes are activated by pathogen-associated molecular patterns from microbes causing metabolic changes, altered protein synthesis, and general reorganization within the cell, leading to swelling. 
In combination with a \ac{WBC} count, \ac{MDW} has been reported to perform equivalently or better than commonly employed methods for the diagnosis of sepsis \cite{motawea_comparison_2023, meraj_monocyte_2023, woo_monocyte_2021}.

\begin{figure}[t!]
  \centering
  \captionsetup{format = plain}
  \includegraphics[width=\linewidth]{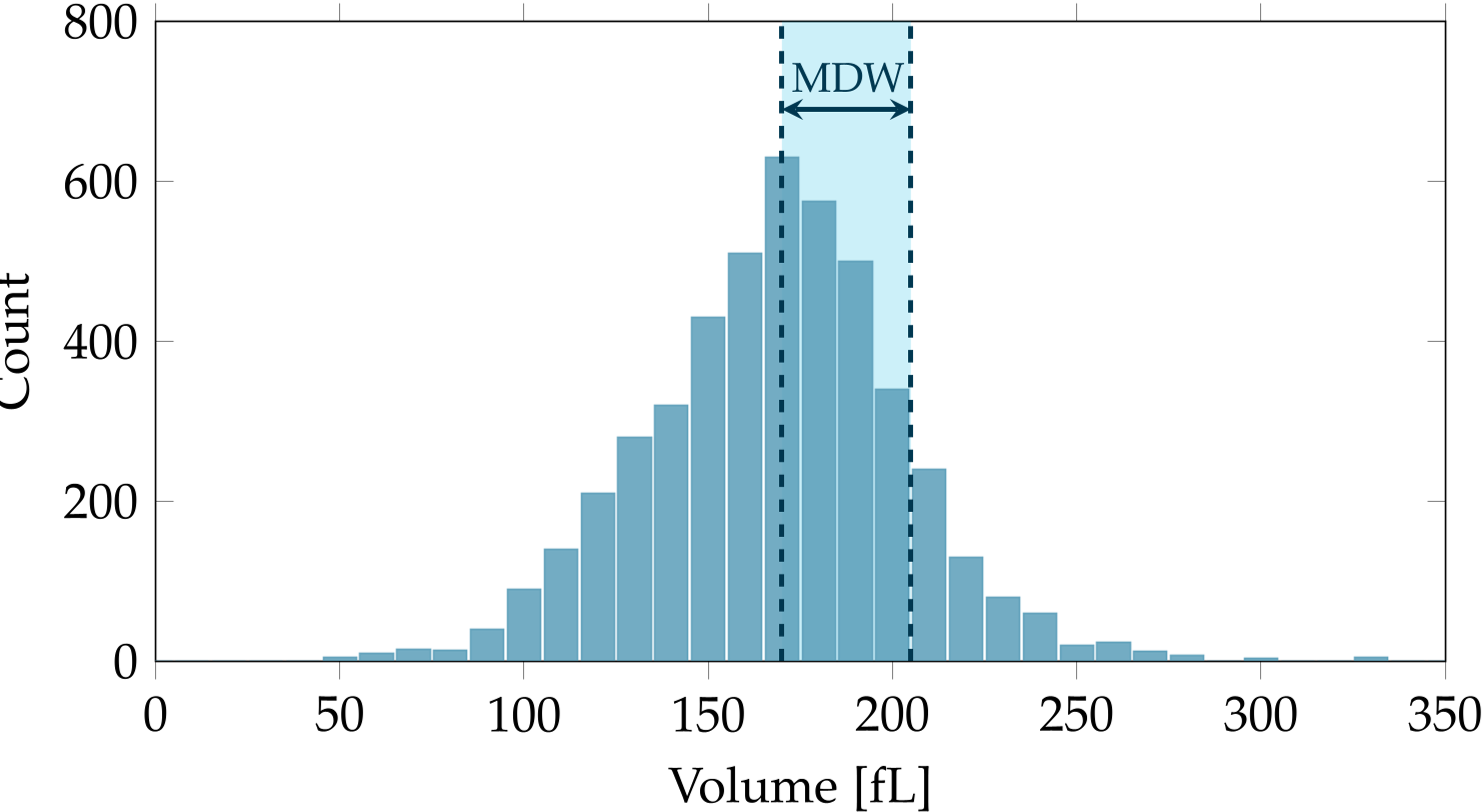}
  \caption{Histogram of the calculated monocyte volumes for one donor. The width of the light blue area denotes the \ac{MDW}. The \ac{MDW} is calculated to \ac{MDW} = 22.4\%.}
  \label{fig:mdw-histogram}
\end{figure}

To determine the \ac{MDW} using the lensless holographic microscope reconstructions is relatively straightforward. The cells labelled monocytes by the classifier from the validation dataset are extracted, and the cross-sectional areas of the cells are determined from the in-focus amplitude planes by conventional image processing techniques, e.g., thresholding, etc. Then, for each cell, the volume of a sphere of equivalent cross-sectional area is calculated and saved for analysis. \ac{MDW} is defined as the percentage ratio of the standard deviation of the calculated monocytes volumes to the mean of the same volumes:
\begin{equation}
\text{MDW} = \frac{\sigma_m}{\mu_m} \times 100
\label{eq:mdw}
\end{equation}

The results from the volume and \ac{MDW} calculation from a single donor is illustrated in \cref{fig:mdw-histogram}. The \ac{MDW} of the measured monocytes is calculated to be \ac{MDW} = 22.4\% which, according to the literature, is a somewhat elevated value for a healthy person \cite{kralovcova_understanding_2024, crouser_monocyte_2019}, indicating a possible small infection or, perhaps more likely, a slight activation of the monocytes due to stresses during sample preparation.

\section{Chapter Discussion}
\label{sec:lensless-discussion}

There are several areas which should be explored to further improve the performance of the lensless analyzer to bring it to the industry standard of a proper diagnostic tool. Focusing first on the classification process of the workflow, changes could be made to both the input, and the architecture of the feature extracting network, and how classification is performed on these extracted features.

A low-hanging fruit may involve adapting newer, and more complex, \ac{SOTA} \ac{ML} models. In recent years, the transformer architecture \cite{vaswani_attention_2023, turner_introduction_2024} has become the de-facto standard in Large Language Models, effectively revolutionizing the way we communicate with computers in the modern world. The transformer architecture has also made its way to image classification in the form of vision transformers (\ac{ViT}s) \cite{dosovitskiy_image_2021}, allowing for extraction of deeper connections between images than the local patterns (e.g., contours, textures) learned by \ac{CNN}s. Future research should be carried out to determine the capabilities of both state-of-the-art \ac{CNN}s, \ac{ViT}s, and possibly high-performing hybrid models.

A classifier is of course only as good and reliable as the data on which it has been trained. Therefore, gathering more cell data from diverse samples, e.g., from many more donors, captured in different environments, at different cell concentrations, with samples from both sick and healthy donors, etc., would help the generalizability and robustness of a classifier. Increasing both the number and diversity of a training dataset allows a classifier to learn complex underlying information about the cells and disregard sample specific biases. In addition, until this point, the classifier has only been trained on cell data from isolated samples, i.e., samples consisting exclusively of a given \ac{WBC} type.

To develop the analyzer from a laboratory prototype to a serious diagnostic tool, it must be able to handle blood samples that require minimum amount of mechanical, chemical, and biological manipulation, e.g., buffy coats\footnote{A fraction of centrifuged blood containing mostly leukocytes and thrombocytes.}, hemolyzed blood\footnote{A blood sample in which \ac{RBC}s have ruptured, releasing their contents into the surrounding plasma.}, or even raw blood, eventually. The additional complexity of these samples, containing unwanted noise and interference from particularly \ac{RBC}s, present a significant, but necessary, challenge to both the numerical reconstruction and localization, and the classifier.

To allow the classifier to gradually learn the added complexities of these samples, as the reliability of the classifier improves, cell data captured from these more complex samples, could slowly be introduced into the dataset. A gradual introduction of field-realistic training data, which should, eventually, replace the cell-isolation training data, would be immensely advantageous in terms of the functionality of a future \ac{PoC} device.

An unmentioned in this chapter, but important factor for a \ac{PoC} device is the \ac{PLT} count. Their diminutive size and transparent, nucleus-free cell structure, meant that it was not possible for the lensless microscope in its standard configuration to not only resolve them, but also localize them reliably, without mistaking spurious twin-image interference for \ac{PLT}s.

Lifting the limits on computational and mechanical complexity allows for the utilization of more sophisticated capture methods and reconstruction algorithms that could significantly improve both the possibility of resolving and localizing \ac{PLT}s, the accuracy of direct diagnostics, such as the \ac{MDW}, and \ac{WBC} classification and count. In Chapters 5 to 7, the use of lensless superresolution, multiwavelength reconstruction, and a physics-informed \ac{ML}-based reconstruction model will be examined.

\section{Chapter Conclusion}
\label{sec:lensless-conclusion}

This chapter detailed the development of a prototype lensless \ac{dWBC} analyzer. The setup simplifies the mechanical design by eliminating the need for an imaging objective. However, the simplification is accompanied by both advantages and disadvantages in terms of performance.

While the lensless system sacrifices spatial resolution, a significantly wider \ac{FoV} is gained, equating to approximately 31 mm$^2$. The increased \ac{FoV} allows for the capture of more cells concurrently, reducing the sample acquisition time, an important parameter for \ac{PoC} devices.

The effective magnification of the hologram was determined by a geometric derivation involving the location of the sample between the illumination source and the camera sensor. An optimal sample-sensor distance was found and verified experimentally to be d $\approx$ 2.7 cm, yielding an effective lateral resolution of 1.54 µm.

Both a 3-diff and a 5-diff dataset were generated from captured holograms of isolated cell samples, with the latter including all five major \ac{WBC} types (monocytes, lymphocytes, neutrophils, eosinophils, and basophils). The optimal configuration of input features was determined using a fractional factorial design experiment, ultimately resulting in an absorption-phase-real representation of the reconstructions. In addition, the classifier input shape was changed from a stack to a grid representation of the input planes, resulting in an accuracy score improvement of approximately 10\%.

The 5-diff classifier demonstrated promising results with an overall classification accuracy of 89.44\%, precision of 90.1\%, recall of 89.4\%, and F1-score of 89.6\%. However, notable misclassifications were observed for basophils, highlighting an area for future refinement.

Finally, this chapter demonstrated an additional potential use of the analyzer for direct observation of other diagnostic markers. Specifically, the \ac{MDW} — a promising biomarker for sepsis — was calculated from the monocytes detected by the classifier, showcasing the versatility of the system.

\chapter{Super-resolved holographic microscopy}
\label{ch:superresolution}

Increasing the resolving power of the lensless holographic microscope is immensely desirable, not only to improve possible classification accuracy by gathering more information about cells (e.g., nucleus structure), but also to possibly resolve and count \ac{PLT}s. To that end, a pixel super-resolution technique and algorithm is implemented. Pixel super-resolution \cite{zhang_resolution_2020-1,bishara_lensfree_2010-1, gao_pixel_2022}, denoting the process of utilizing multiple so called diversity measurements, captured at low-resolution, to facilitate the synthesis of a single, high-resolution measurement.

The high-frequency information in the modulated wavefront from a holographic sample, corresponding to spatial oscillations greater than the Nyquist sampling limit \cite{shannon_communication_1949-1} of the camera sensor, is lost during capture. However, by piecing together undersampled information from multiple measurements captured with differing sampling matrices, it has been demonstrated to effectively recover parts of the lost high-frequency information \cite{gao_generalized_2021}.

The mentioned diversity measurements can take several different forms. The forward model of the capture process of the lensless holographic system can be expressed as \cite{gao_pixel_2022}:
\begin{equation}
I_k = S_D A_k \mathcal{I}_h, \quad k = 1, 2, \ldots, K
\label{eq:forward_model}
\end{equation}
where $\mathcal{I}_h$ represents the unsampled hologram, $A_k$ is the sampling matrix of the $k$th diversity measurement out of $K$ total measurements, and $S_D$ is the downsampling operation of the discrete camera sensor with the downsampling factor $D$, which must be a positive integer.

\begin{wrapfigure}[26]{r}{5.8cm}
  \centering
  \captionsetup{format = plain}
  \includegraphics[width=5.8cm]{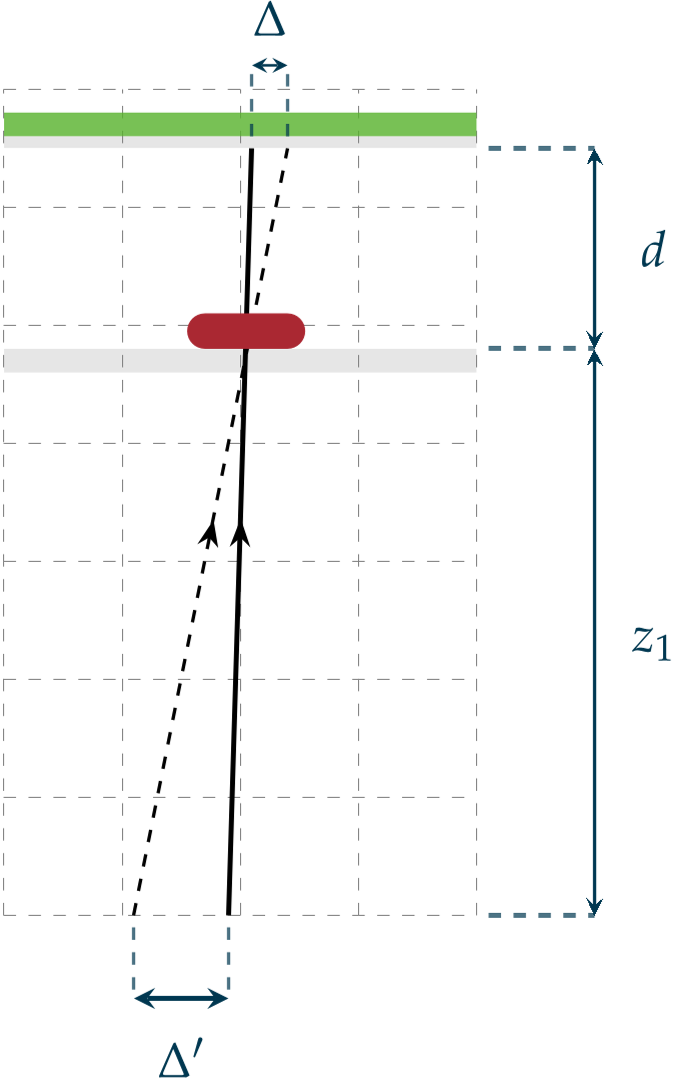}
  \caption{Principle behind the demagnification of the light source shift. A large translation of the light source $\Delta'$ corresponds to a smaller translation of the projected and captured hologram $\Delta$.}
  \label{fig:light_source_shift}
\end{wrapfigure}

Varying the sampling matrix $A_k$ for every measurement in the sequence by modulating physical parameters in the system allows for sampling different components of $\mathcal{I}_h$ that would not otherwise be present in their entirety in a single, low-resolution, measurement. 
Examples of variations of the physical parameters include varying the defocus distance \cite{shen_noise-robust_2018,guo_lensfree_2020}, illumination wavelength \cite{luo_pixel_2015,wu_wavelength-scanning_2024}, illumination pattern \cite{katkovnik_computational_2017,gao_structured_2013}, sensor position \cite{bishara_lensfree_2010-1,bishara_holographic_2011}, etc. \\

\noindent
The aim of a pixel super-resolution algorithm for lensless holographic imaging is then to recover a synthesized high-resolution hologram of size ($DN_1 \times DN_2$), from a stack of images of size ($N_1 \times N_2$), such that, when sampled digitally according to the known sampling matrix $A_k$ and downsampled according to $S_D$, the result is identical to the captured low-resolution hologram $I_k$, for all $K$ captures.

For the present lensless holographic microscope, the choice of diversity measurement fell on sensor position due to the simplicity of implementation, and reported excellent results from the literature. 
In this chapter, the sensor position pixel super-resolution process and algorithm is implemented and explained and its performance is evaluated on both test targets and blood samples.

\section{Pixel Super-Resolution Working Principle}
\label{sec:sr_working_principle}

\begin{figure}[t!]
  \centering
  \captionsetup{format = plain}
  \includegraphics[width=.98\linewidth]{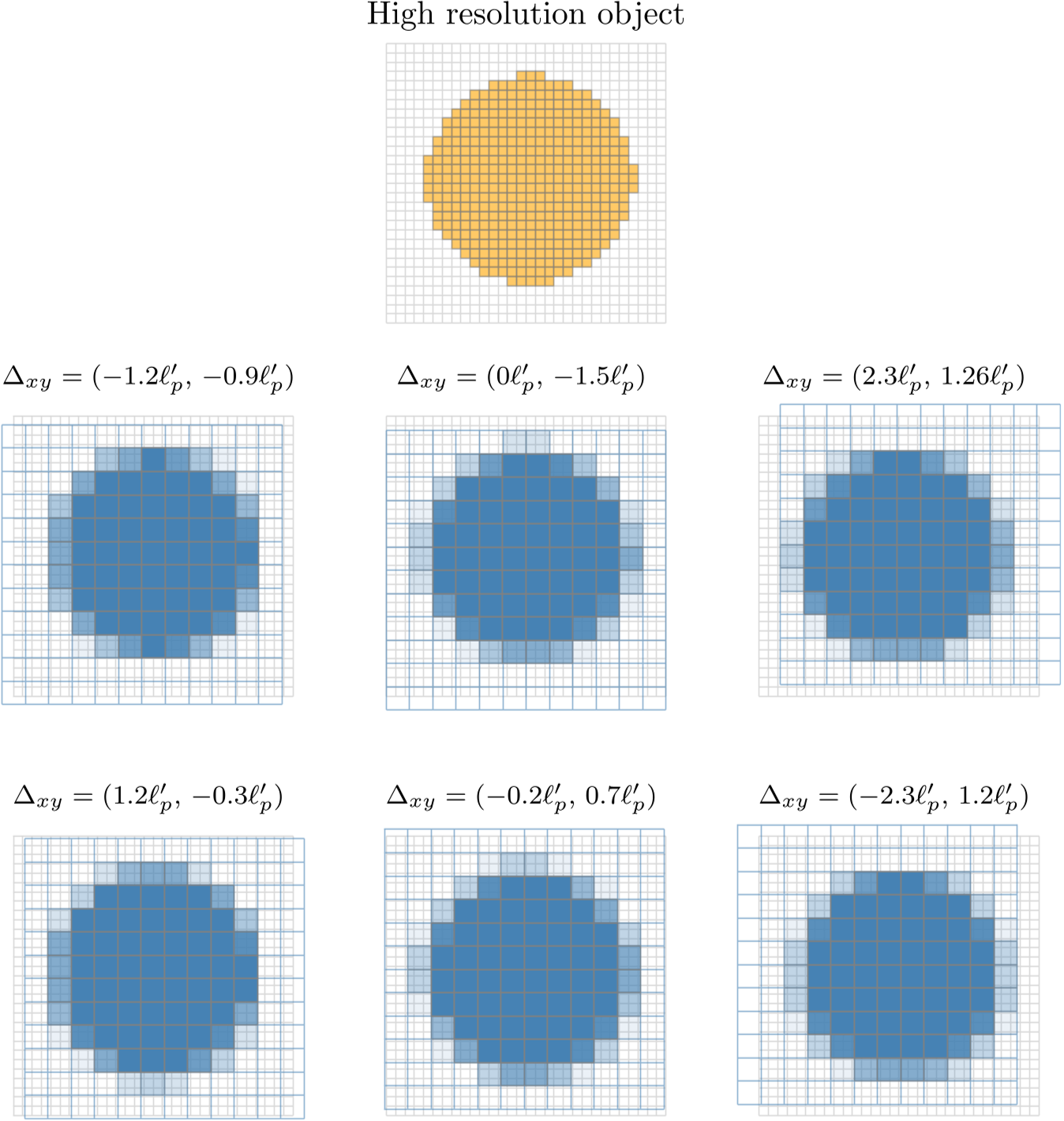}
  \caption{Illustration of the concept of translation diversity measurements. The top yellow circle represents an object sampled on a high-resolution grid. Below, a group of 6 measurements are made in a lower sampling grid. The blue circles represent the measurement of the top circle on the lower resolution grid. By changing the overlap between the high and low resolution grids, as indicated by the translation $\Delta_{xy}$ in low-resolution pixel steps $\ell'_p$, additional information about the top circle is gathered.}
  \label{fig:translation_diversity}
\end{figure}
Taking as a starting point the forward model in \cref{eq:forward_model}, the sampling matrices $A_k$ for the case of sensor position diversity, correspond to sub-pixel\footnote{Integer steps of the camera pixel pitch are equivalent and do not reveal additional information.} lateral translations of $\mathcal{I}_h$ by $\Delta_k^x$ and $\Delta_k^y$.

There exist multiple approaches for synthesizing a high-resolution hologram from a sequence of low-resolution sub-pixel translated captures. 
The simplest one being the shift-and-add algorithm, in which the low-resolution captures are upsampled to the desired resolution, inversely translated according to $A_k$, and superimposed to form a single high-resolution hologram with contributions from each low-resolution hologram. 
Alternatively, a high resolution discrete hologram $I_h$, approximating $\mathcal{I}_h$ can be synthesized by minimizing the following:
\begin{equation}
\begin{split}
\hat{I}_h &= \underset{I_h}{\text{argmin}} \, F(I_h) + \lambda R(I_h) \\
&= \underset{I_h}{\text{argmin}} \, \frac{1}{2K} \sum_k ||S_D A_k I_h - I_k||_2^2 + \frac{\alpha}{2}||I_{h\text{-fil}}||_F^2
\end{split}
\label{eq:sr_optimization}
\end{equation}

where $\hat{I}_h$ is the synthesized high-resolution hologram, $I_k$ is the $k$th captured hologram, and $F(I_h)$ is a data-fidelity term, ensuring that the generated hologram closely resembles the captured holograms after applying the sampling matrices and downsampling. 
Due to the ill-posedness of the data fidelity term, optimizing for this term alone can lead to significant high-frequency noise \cite{panagiotopoulou_super-resolution_2012}. 
Therefore, a regularization term, $R(I_h)$, is introduced that encourages more spatially smooth solutions. 
Specifically, regularization of the solution is carried out by calculating the Frobenius inner product (the sum of the element-wise squares) of $I_{h\text{-fil}}$, where $I_{h\text{-fil}}$ is a high-pass filtered version of of the recovered high-resolution $I_h$. 
Minimizing this regularizer corresponds to decreasing the influence of high frequency components of $I_h$. The regularization strength is controlled via $\alpha$, with greater values leading to more smoothing and smaller values to more high-detail information, in addition to high-frequency noise. 
\cref{eq:sr_optimization} can then be solved via an iterative optimization scheme, as will be outlined in \cref{sec:algorithm_design}.

\section{Experimental setup}
\label{sec:sr_experimental_setup}

For the implementation of the super-resolution technique, the lensless microscope is used, as shown in \cref{fig:lensless-setup}. The required sub-pixel translations of the hologram in relation to the camera sensor can be achieved both by moving the sample, the camera sensor, and even the illumination source. The latter is particularly interesting, as it does not require substantial mechanical accuracy. The principle of moving the illumination source to achieve hologram translation is illustrated in \cref{fig:light_source_shift}. First, the sample stage is moved to be immediately in front of the camera sensor ($d \leq 1$-2 mm), yielding effectively unit magnification. The difference in distance between the illumination source to the sample and the sample to the sensor plane demagnifies the effect of source movement on the recorded hologram. A large displacement of the illumination source translates to a significantly smaller shift of the interference pattern on the sensor. As such, relatively coarse adjustments made to the illumination source can easily achieve sub-pixel translation steps.

The fiber termination of the light source of the lensless microscope is mounted on an XY-stage with manual micrometer adjustment screws. The stage is translated on a grid of 16-25 positions, in each of which a low-resolution hologram is captured. The translations step size is not required to be particularly accurate due to the demagnification illustrated in \cref{fig:light_source_shift}, nor is it required to record the given step size as it will be determined numerically. Therefore, each translation step is chosen to be approximately 0.1 mm, such that the sub-pixel shift corresponds to $\Delta_{xy} \approx 0.1$ mm $(z_1/d) \approx 1.5$ $\mu$m.

As depicted in \cref{fig:translation_diversity}, each captured low-resolution hologram overlaps differently on the assumed high-resolution hologram grid. With enough captured information, it should be possible to reconstruct this underlying high-resolution hologram.

\begin{figure}[t]
  \centering
  \captionsetup{format = plain}
  \includegraphics[width=\linewidth]{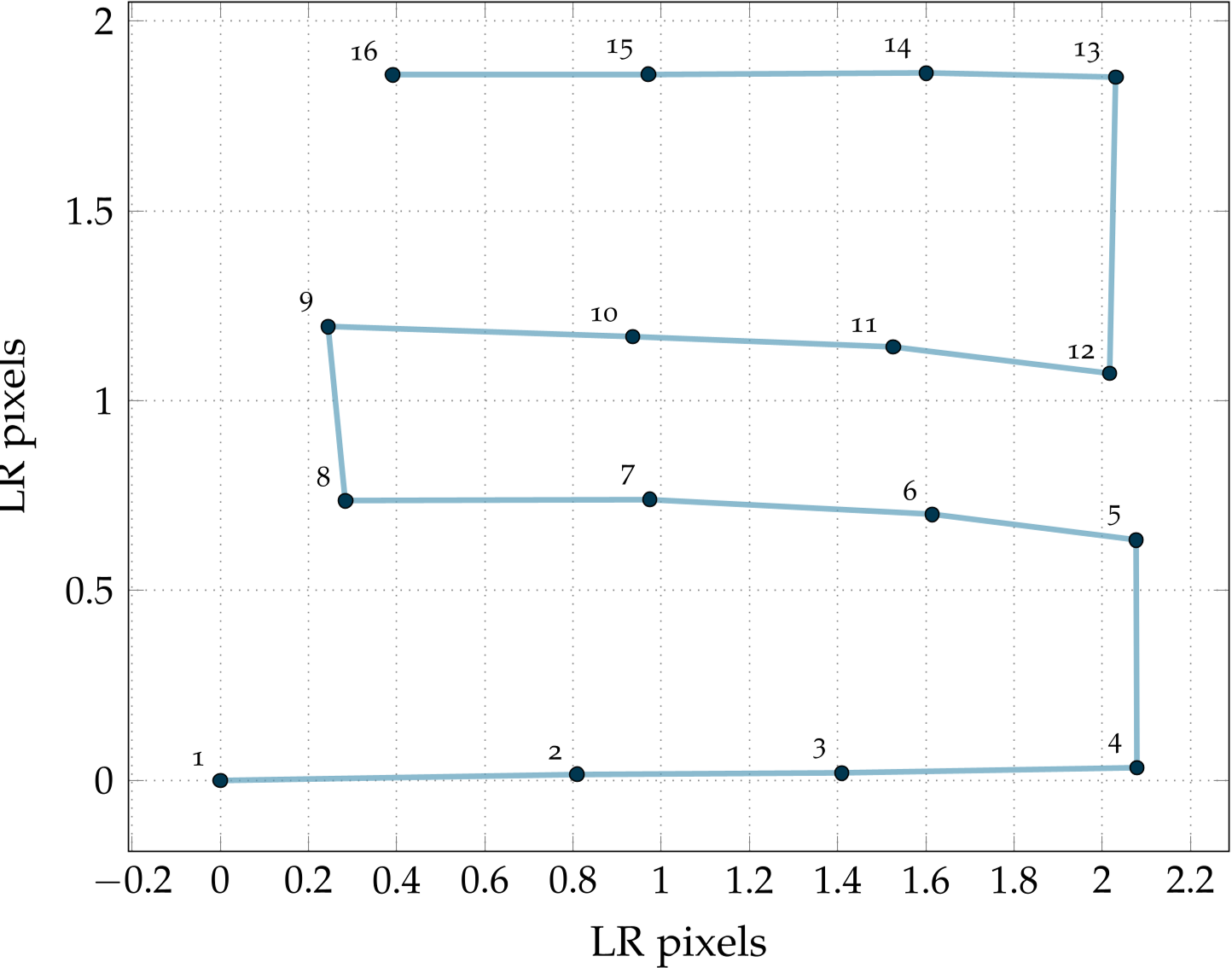}
  \caption{Numerically registered translation shifts in terms of the low-resolution pixel size. The translation is performed in a grid, with a hologram captured in each location.}
  \label{fig:registered_shifts}
\end{figure}

\section{Algorithm Design}
\label{sec:algorithm_design}
To facilitate the synthesis of the high-resolution hologram from multiple, laterally shifted, low-resolution holograms, the sub-pixel shifts must first be determined. 
To that end, an image registration algorithm, phase cross-correlation \cite{anuta_spatial_1970, kuglin_phase_1975, guizar-sicairos_efficient_2008-2}, is utilized. 
Given a reference image and an image whose lateral shift is to be determined, the phase cross-correlation algorithm calculates sub-pixel alignment errors between them in an upsampled spatial frequency domain. 
This step is repeated for all captured diversity measurements (lines 3--6 in \cref{alg:super_resolution}). In \cref{fig:registered_shifts}, the lateral shifts (given in terms of the low-resolution pixel size $\ell_p = 2.49$ $\mu$m) for a single captured sequence of $K = 16$ diversity measurements, as registered by the phase cross-correlation algorithm are shown.

\begin{algorithm}[t!]
\caption{Iterative pixel super-resolution optimization pseudocode}
\label{alg:super_resolution}

\KwIn{
Array of LR hologram captures $L = [I_1, I_2, \ldots, I_n]$\;
Regularization parameter $\alpha > 0$\;
Learning rate $\eta > 0$\;
Upscaling factor $D > 1$ (integer)\;
Maximum iterations $N_{\text{its}} \in \mathbb{N}$\;
Small constant $\varepsilon \ll 1$ to prevent division by zero\;
}
\KwOut{Synthesized HR hologram $\hat{I}_h$}

$\Delta \leftarrow [\,]$\tcp*[r]{Initialize empty array for sub-pixel registration}
$L_{\text{ref}} \leftarrow L[0]$\tcp*[r]{Initialize reference hologram}

\ForEach{$I_i \in L$}{
    $(\Delta_x,\Delta_y) \leftarrow \text{phaseCorrelation}(L_{\text{ref}}, I_i)$\tcp*[r]{Sub-pixel shift}
    append $(\Delta_x,\Delta_y)$ to $\Delta$\;
}

$\hat{I}_h \leftarrow \text{upscale}(L_{\text{ref}}, D)$\tcp*[r]{Initialize HR hologram}
$g^2 \leftarrow 0$\tcp*[r]{Accumulated squared gradient}

\For{$j \leftarrow 1$ \KwTo $N_{\text{its}}$}{
    $I_{h\text{-fil}} \leftarrow \text{laplace}(\hat{I}_h)$\tcp*[r]{Regularization}
    cost $\leftarrow 0$\;

    \ForEach{$(\Delta_x^i,\Delta_y^i) \in \Delta$}{
        $\tilde{I}_i \leftarrow \text{roll}(\hat{I}_h,\Delta_x^i,\Delta_y^i)$\tcp*[r]{Shift HR estimate}
        $\tilde{I}_i \leftarrow S_D(\tilde{I}_i, D)$\tcp*[r]{Downsample}
        cost $\leftarrow$ cost $+ \sum(\tilde{I}_i - I_i)^2$\;
    }

    cost $\leftarrow \frac{\text{cost}}{2} + \frac{\alpha}{2}\sum\lVert I_{h\text{-fil}}\rVert^2$\tcp*[r]{Add regularization}
    $\nabla \leftarrow \nabla(\text{cost})$\tcp*[r]{Gradient w.r.t.\ $\hat{I}_h$}
    $g^2 \leftarrow g^2 + \nabla^2$\tcp*[r]{Accumulate gradient}
    $\hat{I}_h \leftarrow \hat{I}_h - \frac{\eta}{\sqrt{g^2+\varepsilon}}\cdot\nabla$\tcp*[r]{Adagrad update}
}

\Return $\hat{I}_h$\;

\end{algorithm}

Following registration, the actual synthesis of the high-resolution hologram can commence. As an initial estimate, the high-resolution hologram $I_h$ is set to the reference hologram (the first hologram in the capture sequence), upscaled by a factor of $D$ by linear interpolation. For each iteration in the iterative optimization (lines 9--21 in \cref{alg:super_resolution}), the current estimate for the high-resolution hologram is high-pass filtered by way of a spatial convolution with a discrete Laplacian kernel, and saved as a separate \ac{2D} array. The Laplacian kernel is expressed as:
\begin{equation}
h_{\text{lap}} = \begin{pmatrix}
-1 & -1 & -1 \\
-1 &  8 & -1 \\
-1 & -1 & -1
\end{pmatrix}
\label{eq:laplacian_kernel}
\end{equation}

Then, for each numerically estimated shift resulting from the image registration, the current high-resolution estimate is sampled by the sampling matrix $A_k$. 
In practice, this corresponds to first shifting the high-resolution estimate by $\Delta_k^x$ and $\Delta_k^y$ pixels ($\Delta_k^{xy}$), followed by downsampling by a factor $D$ to match the low-resolution hologram. $\Delta_k^{xy}$ is measured in high-resolution pixels and denotes the closest multiple of high-resolution pixel sizes, $\ell_p/D$, that the hologram was originally shifted on the image sensor. 
The downsampling operation is carried out via a \ac{2D} averaging kernel with kernel size and stride of $D$ pixels. 
As such, contributions from all sub-pixels in each $D \times D$ area are averaged to form a resultant low-resolution pixel, mirroring the effect of the averaging of each pixel in the image sensor\footnote{The assumption that the photo-response of each image sensor pixel is perfectly square is an approximation, and modeling the actual photo-response could yield superior results \cite{bishara_lensfree_2010-1, isikman_giga-pixel_2012}.}. 
Following the digital sampling, the data fidelity cost is calculated and accumulated. Finally, the regularization cost is added, and the optimizer steps towards a minimum. 
Repeating $N_{\text{its}}$ iterations or until convergence is reached yields the final high-resolution hologram $\hat{I}_h$. 
The entire algorithm is implemented in Python using the PyTorch package in order to run the bulk of the algorithm on an NVIDIA GeForce RTX 4080 \ac{GPU}.

\section{Super-resolved lensless reconstructions}
\label{sec:sr_reconstructions}

\subsection{Resolution Target Reconstructions}

\begin{figure}[h!]
  \centering
  \captionsetup{format = plain}
  \includegraphics[width=.95\linewidth]{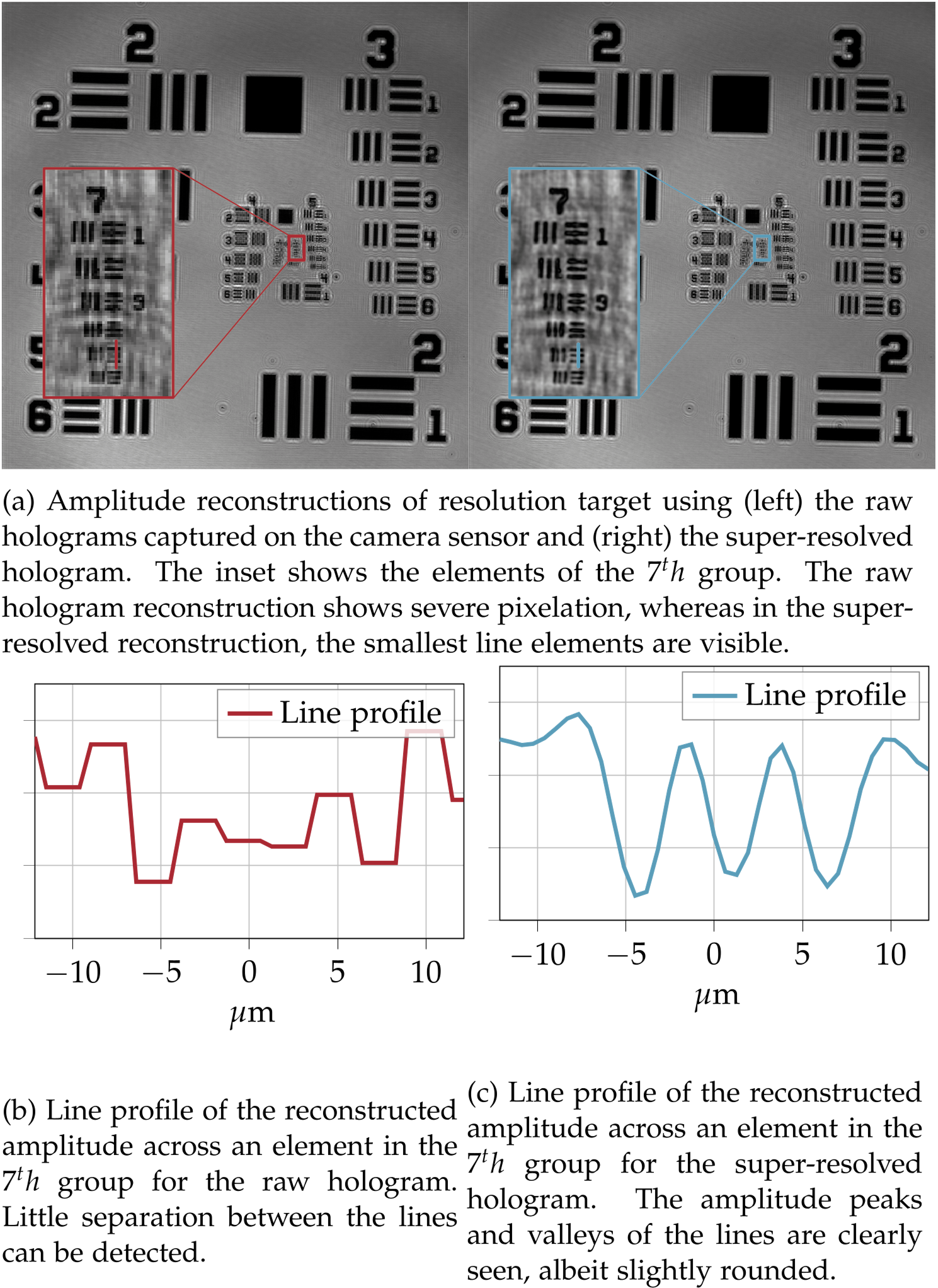}
  \caption{}
  \label{fig:resolution_target}
\end{figure}

Testing of the pixel super-resolution setup and algorithm begins with capturing holograms of a ThorLabs (R1L1S7P) 1951 USAF resolution target. 
The target offers a maximum resolution of 228 line pairs per millimeter, equating to single line widths of 2.19 $\mu$m in the smallest group. 
The test target is placed approximately 1.5 mm in front of the camera sensor, and 16 diversity measurements are captured, each shifted according to \cref{sec:sr_experimental_setup}. 
These captures are in fact the basis of the calculated numerical shifts in \cref{fig:registered_shifts}. In \cref{fig:resolution_target}a, the reconstructions of both the raw and a 4$\times$ super-resolved hologram of the resolution target are shown, along with zoomed \ac{ROI}s and line profiles measured across an element in the 7th group.

The raw reconstruction shows severe pixelation of the 7th group, with the digits 1 and 3 being nearly unrecognizable. 
The line profile shows virtually no distinguishable valleys from the absorbing lines in the element. 
This is in agreement with the Nyquist sampling theorem given the unmodified pixel size of the camera sensor $\ell_p = 2.49$ $\mu$m and the single line widths. 
The 4$\times$ super-resolved reconstruction, on the other hand, shows easily distinguishable lines in each element, and recognizable digits. 
The line profile shows three clear valleys as a result of the absorbing lines in the element.

\subsection{Buffy Coat Reconstructions}

With confirmation of resolution improvement given by the resolution target results, it is now relevant to examine if the same improvement is achievable on real, biological, samples. 
Therefore, a sample of buffy coat is prepared, containing both \ac{RBC}s, \ac{WBC}s, and \ac{PLT}s, and deposited on a microscope slide which is placed immediately in front of the camera sensor. 
The image capture is carried out according to \cref{sec:sr_experimental_setup}, with 16 holograms captured at each location in the grid scanning process. 
The pixel super-resolution algorithm is configured for an 8$\times$ upsampling, and is run for 20 iterations with regularization parameter $\alpha = 0.11$ and learning rate $\eta = 0.1$. 
For display purposes, a cropped view of the amplitude reconstruction of the resulting high-resolution hologram is displayed in \cref{fig:buffy_coat}a. 
The entire reconstruction spans 6.8 mm square, yielding a \ac{FoV} of 46.2 mm$^2$ with a synthetic pixel pitch of 0.46 $\mu$m\footnote{For these samples, only the green channel of a color camera sensor \cite{noauthor_alvium_nodate-1} was used, hence the larger physical pixel size of $\ell_p = 3.7$ $\mu$m.}.

Evidently, unwanted scratches or oils were present on the microscope slide, as indicated by the vertical streaks. 
Nonetheless, the reconstruction shows a uniform distribution of cells of various types across the entire \ac{FoV}. 
Indicated by the orange rectangles and shown in \cref{fig:buffy_coat}b--d are zoomed \ac{ROI}s which showcase the detail level of the high-resolution reconstruction. 
In \cref{fig:buffy_coat}b and \cref{fig:buffy_coat}d, regions containing both \ac{RBC}s and \ac{WBC}s are seen. 
Some of the \ac{RBC}s have arranged themselves in their characteristic rouleaux formation, resembling stacks of coins. 
In addition, it is also possible to distinguish the characteristic biconcave divot in the center of the \ac{RBC}s that are oriented flatly. 
This is particularly noticeable in \cref{fig:buffy_coat}c, which shows two \ac{RBC}s lying flat on the microscope slide, along with what is presumed to be a third \ac{RBC} leaning on another. 
As expected, by measuring the number of pixels across one of cells, the \ac{RBC} has a diameter of approximately 8.6 $\mu$m. In \cref{fig:buffy_coat_3d}, the phase profile of this subfigure has been extracted and plotted as a \ac{3D} surface, indicating the thickness of the cells.

An interesting side-effect of the super-resolution algorithm is also on display in these reconstructions. 
It appears as though a significant dark edge has been added to the reconstructed cells. 
One possible explanation for this lies in the regularization by the Laplacian filtering. 
As mentioned, this regularizer is included to limit high-frequency noise in the synthetic hologram. 
However, if it is not entirely successful, perhaps the high-frequency information in the measured holograms is amplified as well. 
Through testing, it was observed that lowering the weight of the regularization did increase this edge effect. 
However, increasing the weight too much leads to the loss of high-frequency information and therefore detail in the reconstructions. 
The value arrived at ($\alpha = 0.11$) was the result of these tests, and is, as such, the compromise between the sharpened edges and reconstruction fidelity.

\subsection{Platelet detection}

\begin{figure}[h!]
  \centering
  \captionsetup{format = plain}
  \includegraphics[width=0.8\linewidth]{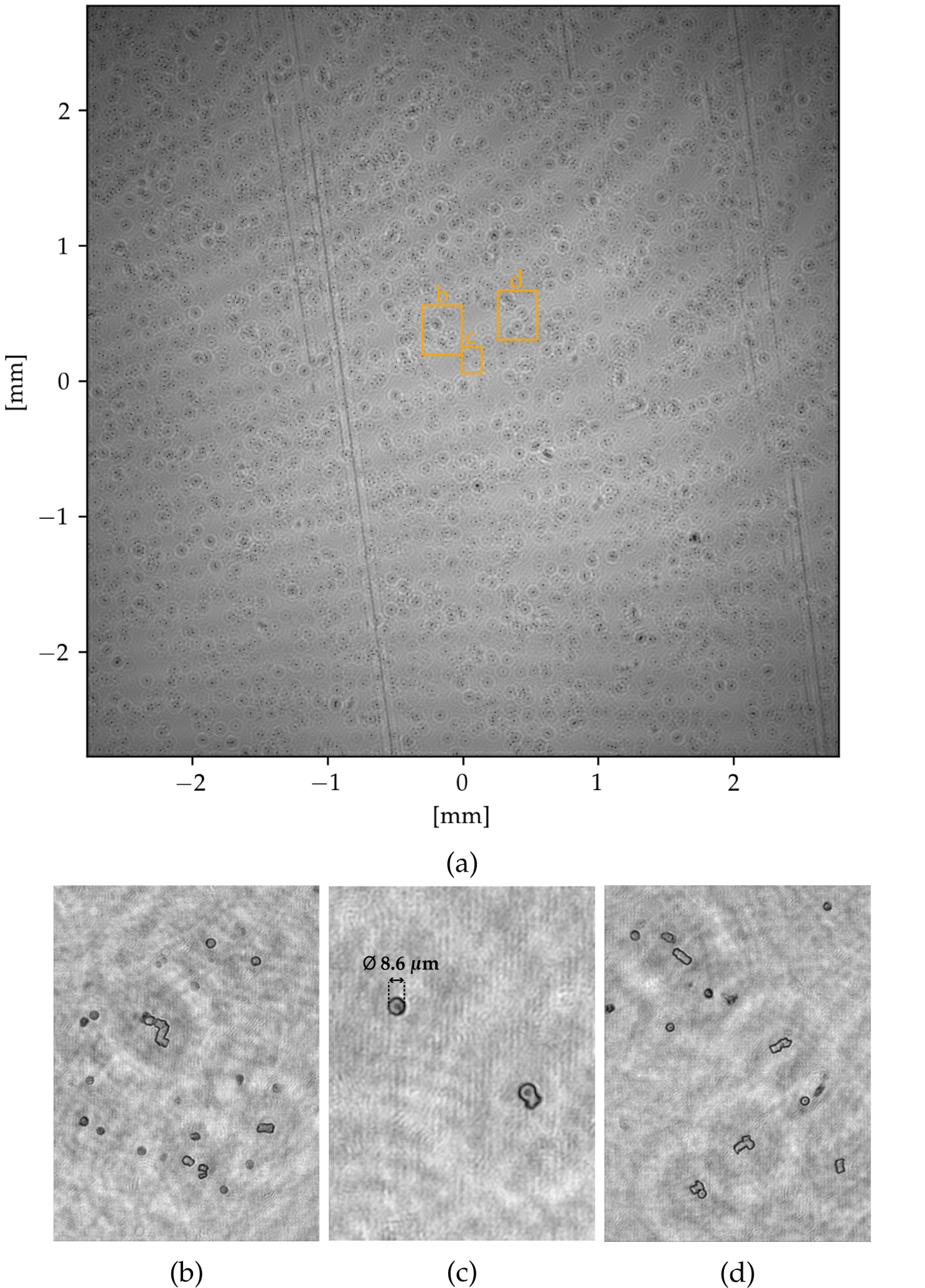}
  \caption{\ac{ASM} amplitude reconstruction of super-resolved hologram. (a) An \ac{ROI} with a \ac{FoV} of 46.2mm$^2$, highlighting the scale of the viewable area. (b-d) insets in the \ac{ROI} showing the amplitude reconstruction of blood cells.}
  \label{fig:buffy_coat}
\end{figure}

\begin{figure}[t]
  \centering
  \captionsetup{format = plain}
  \includegraphics[width=\linewidth]{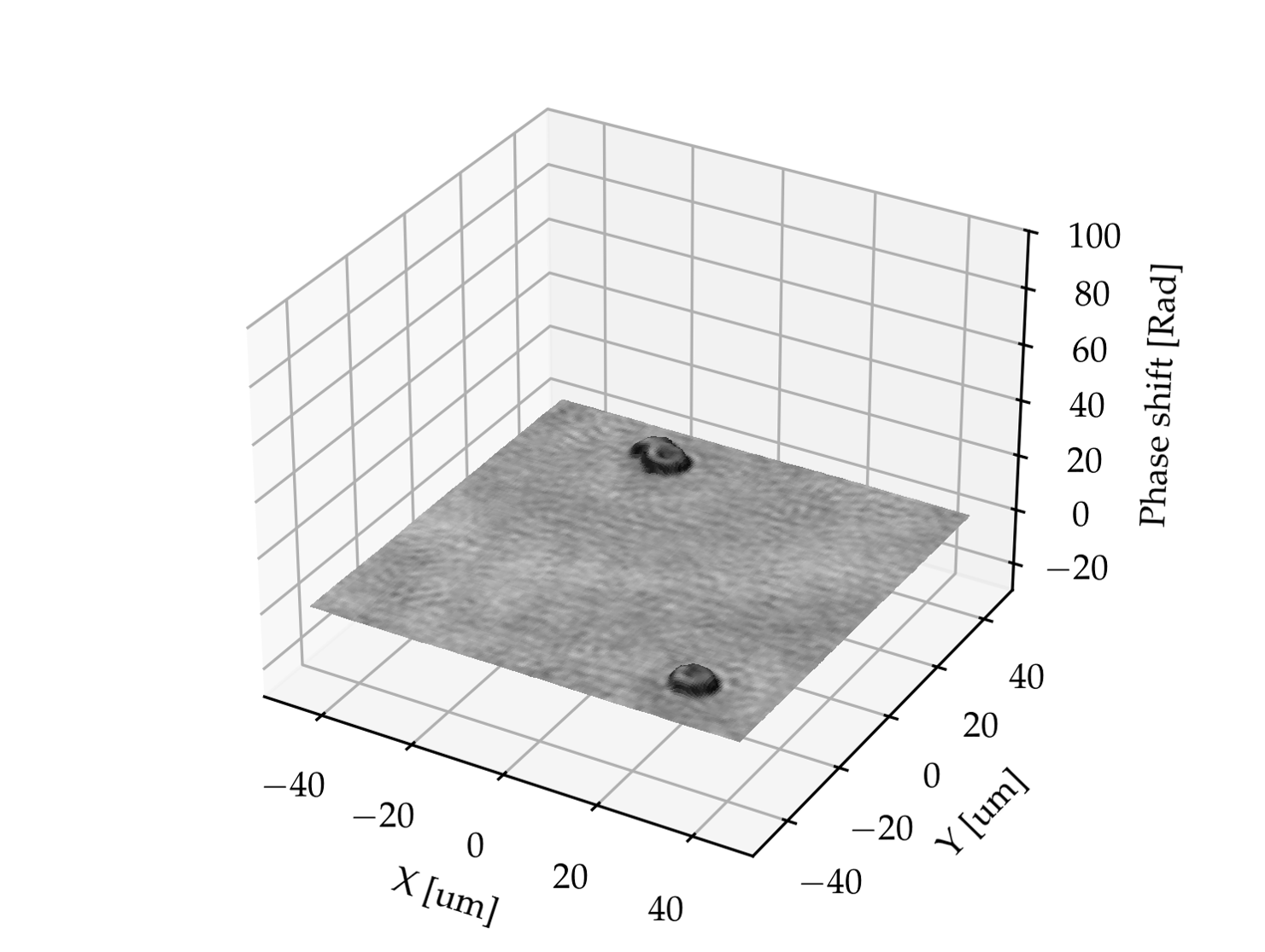}
  \caption{Zoomed \ac{ROI} of the cells in \cref{fig:buffy_coat}c. The reconstructed phase is plotted as the height value in \ac{3D} and the amplitude as the gray-values of the plane.}
  \label{fig:buffy_coat_3d}
\end{figure}

Perhaps surprisingly, there are no \ac{PLT}s visible in the high-resolution reconstructions shown in \cref{fig:buffy_coat}, even though the improved resolution certainly should be able to resolve them. 
One possible explanation as to why, is to once again consider their minute size. 
At diameters of approximately 2 $\mu$m, Brownian motion is not an insignificant factor. 
According to the Einstein-Smoluchowski equation \cite{einstein_uber_1905, von_smoluchowski_zur_1906}, the mean squared displacement of a particle in a medium is given by:
\begin{equation}
\langle \Delta x^2 \rangle = \langle \Delta y^2 \rangle = 2Dt
\label{eq:mean_squared_displacement}
\end{equation}

where $t$ is the elapsed time and $D$ is the diffusivity which for a sphere in translational diffusion is given by:
\begin{equation}
D = \frac{k_b T}{6\pi\eta r}
\label{eq:diffusivity}
\end{equation}
where $k_b$ is the Boltzmann constant, $T$ is the temperature, $\eta$ is the dynamic viscosity of the medium, and $r$ is the effective particle radius. 
The root mean square (\ac{RMS}) displacement represents a scale of displacement a particle will experience over time. 
For platelets at room temperature with effective diameters of 2 $\mu$m in blood plasma with dynamic viscosity 1.45 mPas\footnote{Assuming a solution of only \ac{PLT}s in plasma. 
The buffy coat may be more viscous.} \cite{mengana_torres_determination_2019}, this yields:
\begin{equation}
\sqrt{\langle \Delta x^2 \rangle} = \sqrt{2Dt}
\end{equation}
\begin{equation}
\simeq 2.96 \times 10^{-13} t \, \mu\text{m}
\end{equation}
\begin{equation}
= 0.54 \, \mu\text{m}
\end{equation}

Thus, after one second, the typical displacement of each of the thousands, if not millions, of \ac{PLT}s in the buffy coat due to Brownian motion will be approximately $\sqrt{\langle \Delta x^2 \rangle} \simeq 0.54$ $\mu$m in any random direction. 
This is certainly in sub-pixel territory, and given that the capture time of the multiple diversity measurements can take several minutes, the \ac{PLT}s cannot be assumed to be stationary for the entire duration. 
With the \ac{PLT}s in different locations in each captured low-resolution hologram, and not systematically as is the purpose of the translation stage, their contribution to the final high-resolution hologram will simply be averaged out.

It is possible to construct sensor-shift-based pixel super-resolution setups with capture times in the order of seconds, or even tens of milliseconds, that could make it possible to ``freeze'' the motion of the \ac{PLT}s. For instance, the setup demonstrated by Bishara et al. \cite{bishara_holographic_2011} in which individual illumination sources in a fiber-optic array are switched on and off in sequence, could, in theory, capture the diversity measurements as fast as the switching time of the sources and the frame rate of the camera sensor. This leads to an idea had for this project, in which the sub-pixel shifts due to the random motion of the \ac{PLT}s themselves are used to create the diversity measurements. Tuning the capture frequency, and capturing dozens of images without explicit sample translation, would lead to a stack of diversity measurements in which only the \ac{PLT}s are moving. With the random movements of the \ac{PLT}s, each image would capture new, sub-pixel shifted positions of all, from which a higher-resolution hologram could be synthesized. However, this was ultimately unfruitful for two key reasons:

\begin{enumerate}
\item The contribution to a captured hologram of each individual \ac{PLT} is virtually indistinguishable from the background.
\item Each \ac{PLT} moves independently from all others.
\end{enumerate}

The first point means that it is extremely difficult to localize individual \ac{PLT}s from just a captured hologram. The individual diffraction patterns overlap, and they are weak due to the physical properties of the \ac{PLT}s. The holograms could of course be reconstructed to make them appear clearer, but since the measurements are low-resolution holograms, they would still be difficult to robustly localize, which is the exact reason pixel super-resolution may be necessary in the first place. Even disregarding the problems of localizing \ac{PLT}s in the hologram, the second point nullifies the algorithm previously described in Algorithm \cref{alg:super_resolution}. As each \ac{PLT} moves independently, each must be tracked in \ac{2D} over the course of the capture sequence, and a pixel super-resolution algorithm will have to be run on a \ac{ROI} around it. This would have to be repeated for thousands or millions of \ac{PLT}s, making it computationally difficult to justify. Nonetheless, solving these problems would provide an elegant method of capturing super-resolved \ac{PLT}s.

\subsection{Effects of insufficient and uneven sampling}

The number and distribution of the diversity measurements is an important factor in synthesizing satisfactory high-resolution holograms. 
With insufficient measurements, there may not be enough information present in the overlapping low-resolution images, leading to local ambiguity and a blurring effect as the optimization will function more as a simple interpolation.
\begin{figure}[t!]
  \centering
  \captionsetup{format = plain}
  \includegraphics[width=\linewidth]{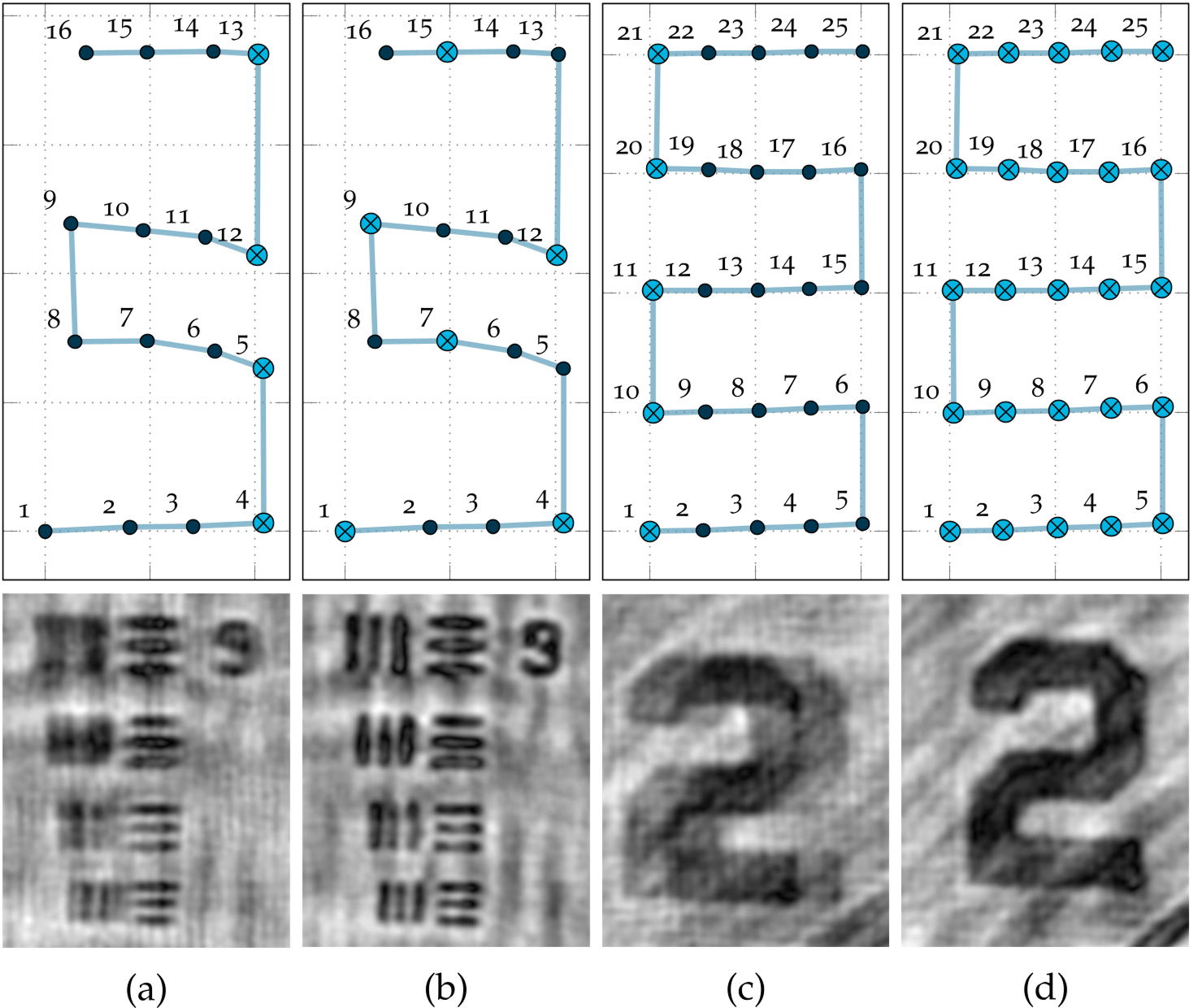}
  \caption{Effects of selecting different configurations of the sub-pixel shifts in terms of reconstruction quality. Light blue markers with a cross indicates inclusion in the algorithm, while the dark blue indicate exclusion. (a \& c) Using only translations in a single direction. (b) Using a diverse set of points. (D) using all points.}
  \label{fig:sampling_effects}
\end{figure}

This local ambiguity also presents itself if care is not taken to ensure that the diversity measurements are ``diverse'' enough. 
For instance, if the shifts only occur in one single direction, the camera sensor is not capturing any additional information for the perpendicular direction, and cannot therefore be expected to yield resolution above what is already captured in the low-resolution holograms. 
Examples of such cases are shown in \cref{fig:sampling_effects}a and \cref{fig:sampling_effects}c. 
Here, only a fraction of the diversity measurements are used in the synthesis of the holograms. 
Notably, only measurements captured along a single axis (y) are used, and thus, the algorithm lacks information along the perpendicular axis (x). 
The effect of the uneven distribution of points along the two axes is a blurring in the direction of the non-captured axis, as seen by the inseparable vertical lines in \cref{fig:sampling_effects}a, contrasted by the comparatively well defined horizontal lines. 
Similarly, the digit 2 in \cref{fig:sampling_effects}c is severely blurred along the direction of the non-captured axis. 
When all diversity measurements are used, as is shown in \cref{fig:sampling_effects}d, the digit appears of high-contrast in both directions.

\subsection{Limits of pixel super-resolution}

\begin{figure}[t!]
  \centering
  \captionsetup{format = plain}
  \includegraphics[width=\linewidth]{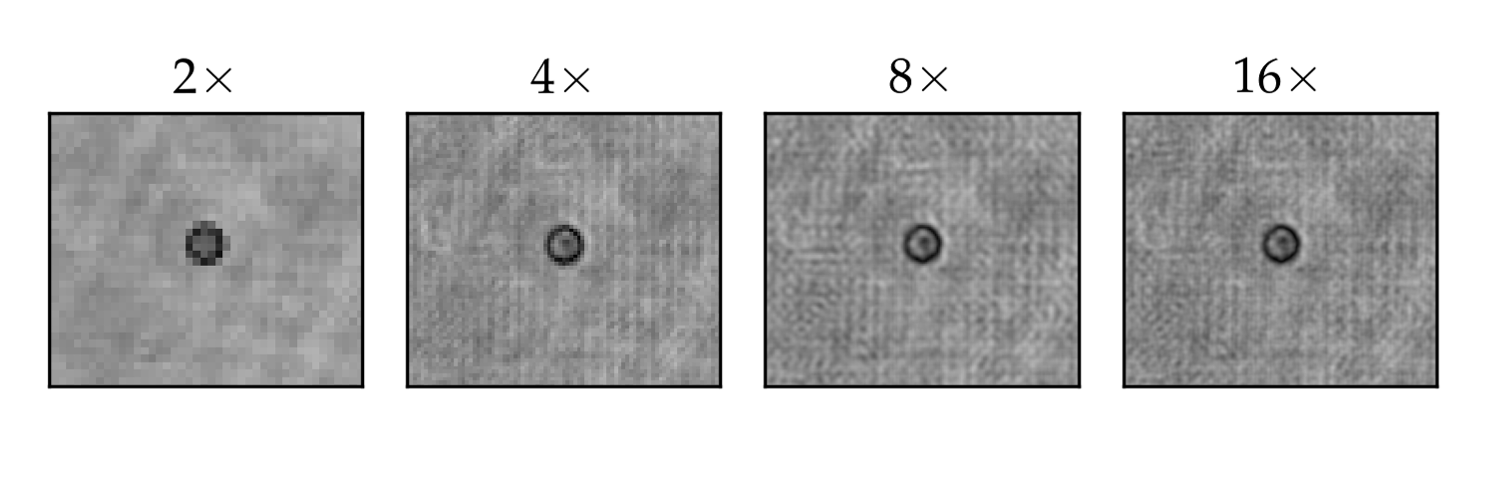}
  \caption{Illustration of the effect of choosing a higher super-resolution magnification factor. Resolution is increased effectively at $D \leq 8$, after which no noticable improvements are seen.}
  \label{fig:sr_limits}
\end{figure}

Naturally the question arises as to how far the super-resolution algorithm can be stretched, and if there is a point in choosing ever higher magnification factors $D$ at the cost of severely increased computational load. 
As such, taking as a foundation the 16 diversity measurements captured of the buffy coat sample, the hologram registration and super-resolution algorithm is tested at four different magnification levels; [2$\times$, 4$\times$, 8$\times$, 16$\times$]. 
The reconstructions of a single \ac{RBC} at the different magnification levels are shown in \cref{fig:sr_limits}. 
At 2$\times$ magnification, the \ac{RBC} is heavily pixelated, with limited visibility of the cell features. 
At 4$\times$, a significant improvement to the fidelity of the \ac{RBC} is seen. Only a slight improvement to the pixelation is seen as the magnification is increased to 8$\times$. 
Going beyond 8$\times$ yields insubstantial improvements, as is also seen in the 16$\times$ case. 
Going further than the 8$\times$ magnification may require additional diversity measurements.

\section{Chapter Conclusion}
\label{sec:ch5_conclusion}

This chapter detailed the principle and implementation of a pixel super-resolution algorithm, capable of achieving effective pixel sizes several times smaller than that of the image sensor. By utilizing multiple low-resolution holograms, captured in a sequence in which each hologram is shifted laterally by a sub-pixel distance, the pixel super-resolution algorithm synthesizes a high-resolution hologram. These sub-pixel shifts provide additional information about the underlying hologram, allowing the algorithm to determine a high-resolution hologram containing more pixel information than any single capture. The sub-pixel shifts were achieved by manually translating the source illumination -- a fiber coupled laser source. Because of the geometry of the lensless setup, large translations ($\approx$ 0.1 mm) of the illuminations source corresponds to much smaller translations of the captured interference patterns, on the range of the pixel size of the camera.

These sub-pixel translations were determined numerically using a phase cross-correlation algorithm. The pixel super-resolution algorithm was implemented as a regularized iterative optimization problem, tasked to synthesize the high-resolution hologram which could correspond to each of the sub-pixel shifted low-resolution holograms.

The algorithm was tested both on a USAF 1951 resolution target, and a sample of buffy coat. The high-resolution reconstructions from the synthesized holograms showed a substantial increase in resolution, with resolving power well above the capabilities of the image sensor itself, allowing better observations of \ac{RBC}s and \ac{WBC}s. With the diversity measurements captured, resolution improvements were seen with scaling up to approximately 8 times the original resolution.

With the additional detail granted by such a super-resolution algorithm, it may be possible to further increase the classification accuracy described in Chapter 4. The high-resolution reconstructions may allow a \ac{WBC}s classifier to take greater advantage of the differences in cell morphology. With more accurate cell dimensions and the potential to reconstruct the internal cell structure, intra-cellular differences may stand out clearer. However, the algorithm is heavy computationally. For a \ac{PoC} to make use of the algorithm while maintaining a reasonable rate of analysis, considerations of computing power or \ac{GPU} utilization should be made. Alternatively, the captured diversity measurements could, securely, be processed in a cloud computing service that then returns the synthesized high-resolution hologram.

\chapter{Multiwavelength holographic microscopy}
\label{ch:multiwavelength}

This chapter outlines the implementation of a lensless multi-wavelength holographic microscope, capable of both color reconstruction as well as super-resolution. Chapters 3 and 4 aimed at exploiting the potential of \ac{DHM} to reduce the complexity of hematological measurements by removing the need for a chemical staining for \ac{dWBC}. The results showed that a cell differential is possible, even without the need for staining. However, the system still has room for significant improvement in terms of classification accuracy.

In this chapter, the groundwork for evaluating the potential of combining the benefits of \ac{DHM}, e.g., reconstruction of phase information, large \ac{FoV}, and mechanical simplicity, and the classically used color information gained from cell staining. The additional color information would allow a classifier to not only take advantage of its capture of an objects reconstructed phase, but also its wavelength-dependent amplitude. Inter-cellular differences in absorption by the cell nucleus, cytoplasm, and membrane at different wavelengths can all contribute to separating the cell classes in the high-dimensional space of the classifier.

\section{Multi-Wavelength Experimental Setup}
\label{sec:mw_experimental_setup}

\begin{wrapfigure}{l}{6.2cm}
  \centering
  \captionsetup{format = plain}
  \includegraphics[width=6.2cm]{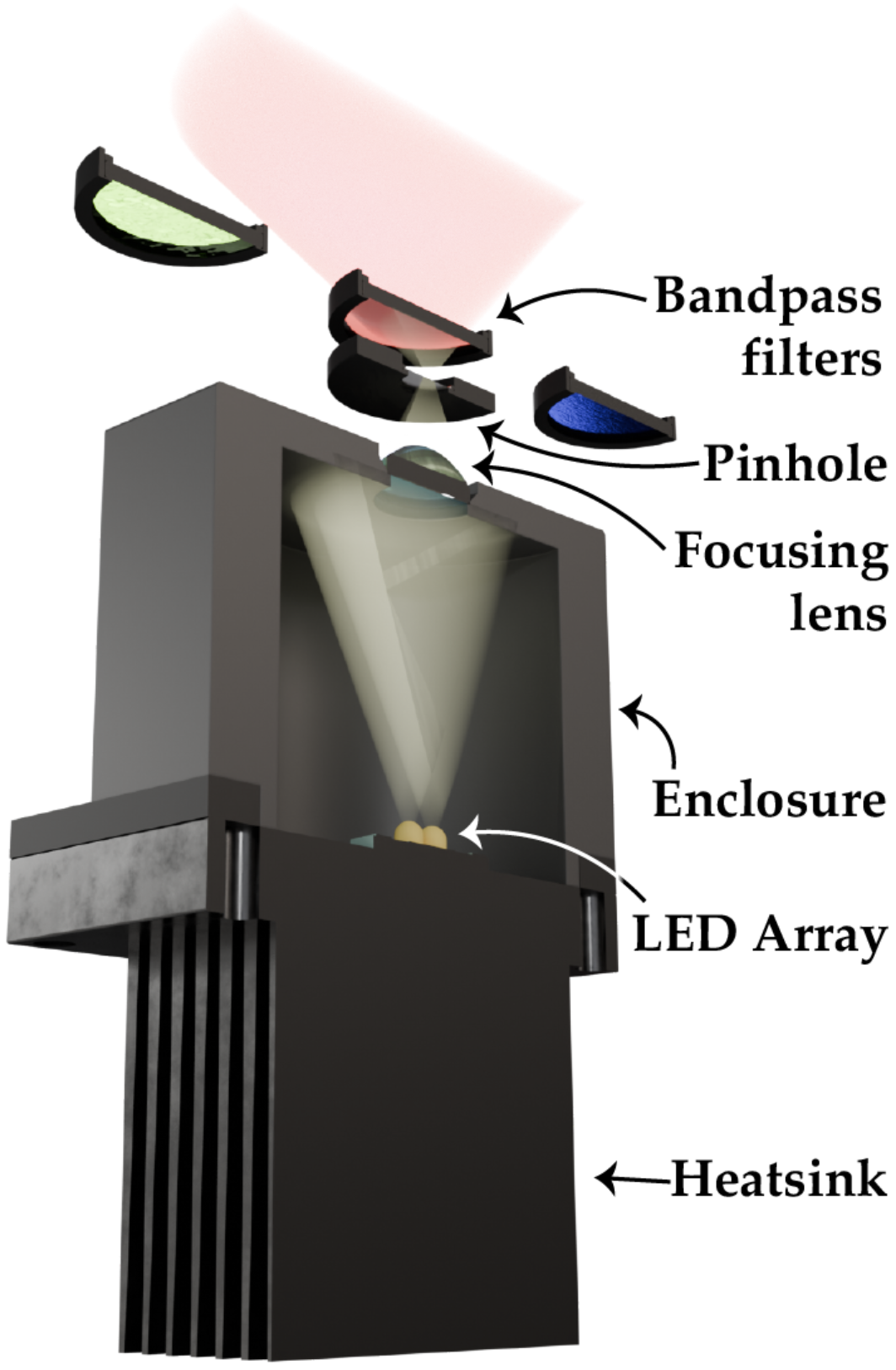}
  \caption{Render of the multiwavelength illumination unit. Light from a white \ac{LED} is focused by a lens and passed through a pinhole and interchangeable spectral bandpass filters.}
  \label{fig:mw_illumination}
\end{wrapfigure}

The lensless setup illustrated in \cref{fig:lensless-setup} is readily modified for multi-wavelength use, with modifications made to the illumination unit. The illumination unit adapted for multi-wavelength use is illustrated in \cref{fig:mw_illumination}, and is based around a neutral white (4000K) \ac{LED}-array \cite{noauthor_ilh-on04-nuwh-sc211-wir200_nodate} with four individual diodes. The \ac{LED}-array is mounted on an aluminium heatsink to cool the array during operation. To shield the image sensor from stray incoherent light, a \ac{3D}-printed enclosure covers the \ac{LED}s. Centered above one of the \ac{LED}s on the integrated circuit (\ac{IC}), a hole is made in the enclosure in which an $\diameter$18 mm plano-convex lens ($f = 15$ mm) can be placed. The lens focuses the light from the single \ac{LED} onto a $\diameter$100 $\mu$m pinhole in order to increase the illumination brightness. Immediately above the pinhole, spectral bandpass filters with center wavelengths of $\lambda_c = 450$ nm, 530 nm, and 630 nm and spectral bandwidth $\lambda_{\text{FWHM}} = 10$ nm \cite{noauthor_bandpass_nodate, noauthor_bandpass_nodate-1, noauthor_bandpass_nodate-2} can be placed manually to selectively extract narrow wavelength ranges from the wide-band \ac{LED} spectrum. The effect of the divergent beam exiting the pinhole on the efficacy of the bandpass filters are examined in Appendix D.

The camera assembly illustrated in \cref{fig:lensless-setup} is reused, apart from the substitution of a color \ac{CMOS} camera \cite{noauthor_alvium_nodate-1}. As was the case for the super-resolution configuration, any sample to be examined is placed immediately in front of the camera sensor with a propagation distance $\leq 1$-2 mm, yielding effectively unit optical magnification. The sample and camera are placed an axial distance of approximately 10 cm away from the pinhole aperture.

In a holographic setup utilizing \ac{LED}s, it is necessary to consider the coherence of the light source, or lack thereof, more stringently. As described in Section 2.2.5, the illuminating light source is required to exhibit a certain level of temporal and spatial coherence in order to facilitate the formation of interference patterns.

To make the illumination source temporally coherent in this multiwavelength holographic microscope (for each illuminating wavelength), the bandpass filters, illustrated in \cref{fig:mw_illumination}, limit the spectral bandwidth of the \ac{LED} source to a FWHM of $\Delta\lambda_{\text{fwhm}} = 10 \pm 2$ nm.
\newpage
\noindent
In addition, \ac{LED} sources are not spatially coherent due to the nature of their spontaneous photon emissions lacking a uniform phase relationship, and their extended emission area. The addition of a small pinhole (diameter of $\leq 0.5$-2 $\mu$m) can enforce spatial coherence. However, this combined spectral and spatial filtration of the illumination source drastically lowers the power of the illuminating light. To reduce the required exposure time of the camera for hologram capture, it is beneficial to select the largest possible pinhole while still maintaining sufficient spatial coherence. Using a correct axial configuration of the source, pinhole, sample, and sensor, it is possible to use pinholes with diameters more than 100 times larger than previously mentioned. \\ \\ \\ \\

By selecting a relatively large pinhole diameter (e.g., 100-200 $\mu$m), and placing the sample immediately in front of the camera sensor, the light can be considered approximately spatially coherent due to the apparent demagnification of the pinhole aperture \cite{mudanyali_compact_2010}. Reportedly, as long as the spatial extent of an illuminated object is smaller than the coherence diameter at the sensor plane, where the coherence diameter is defined as:
\begin{equation}
D_{\text{coh}} = \lambda z_1 / D
\label{eq:coherence_diameter}
\end{equation}
and $z_1$ is the source-sample distance and $D$ is the pinhole diameter, the source can be considered spatially coherent \cite{mudanyali_compact_2010, kim_lens-free_2012}. The experimental configuration outlined above yields a coherence diameter of $D_{\text{coh}} \geq 400$ $\mu$m for illumination wavelengths longer than 400 nm.

A side-effect of this approach is the unit magnification of the hologram on the camera sensor. However, as was shown in Chapter 5, this configuration, enables the option to perform the pixel super-resolution algorithm, and thus achieve super-resolved, multicolor, reconstructions at extremely large \ac{FoV}s, ultimately turning the magnification limitation into an advantage.

\subsection{Illumination Wavelength Selection}

Appropriate wavelength bands for multi-wavelength holographic reconstruction, for color sample renditions, can span from the typical red-green-blue (\ac{RGB}) channels commonly used in digital cameras and displays \cite{garcia-sucerquia_color_2016, martins_costa_rgb_2013, granero_single-exposure_2016}, and all the way to a near-continuum of wavelengths \cite{luo_pixel_2015,wu_wavelength-scanning_2024,sanz_improved_2015,zhang_object_2018,kemper_multi-spectral_2018,lariviere-loiselle_polychromatic_2020,brault_multispectral_2023}.

\begin{figure}[t]
  \centering
  \captionsetup{format = plain}
  \includegraphics[width=\linewidth]{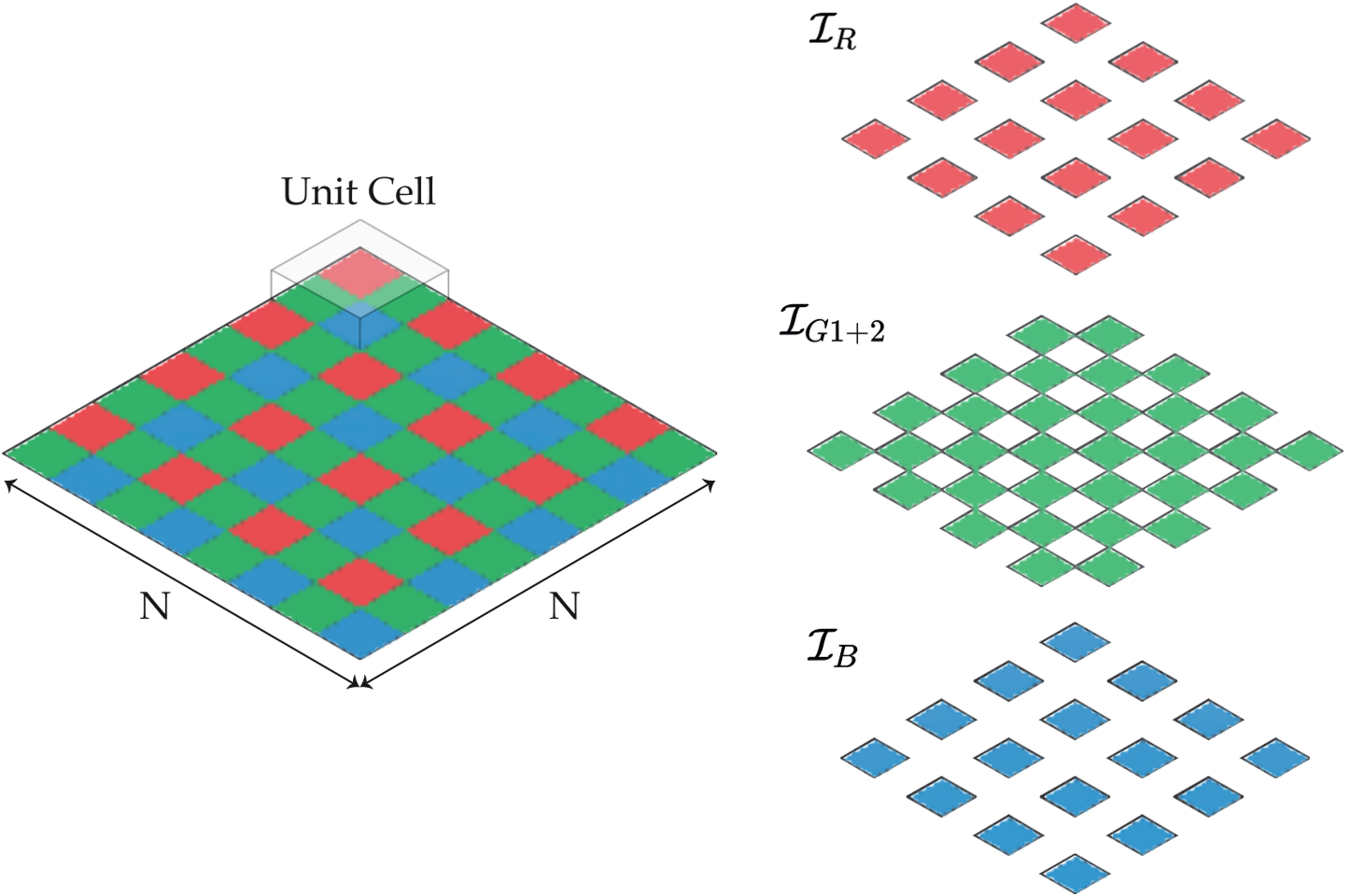}
  \caption{The structure of the Bayer pattern. Each unit cell consists of four sub-pixels, corresponding to sensitivity to either red, green, or blue light. To mirror the photo-response of human eyes, the Bayer pattern contains two green sub-pixels within the unit cell \cite{bayer_color_1976}.}
  \label{fig:bayer_pattern}
\end{figure}

Color \ac{CMOS} image sensors are readily available for commercial and industrial use. 
The pixels on these sensors are typically arranged in a Bayer pattern, as illustrated in \cref{fig:bayer_pattern}, in which four individual sub-pixels in each unit cell, are responsible for capturing color information.

\begin{figure}[t!]
  \centering
  \captionsetup{format = plain}
  \includegraphics[width=\linewidth]{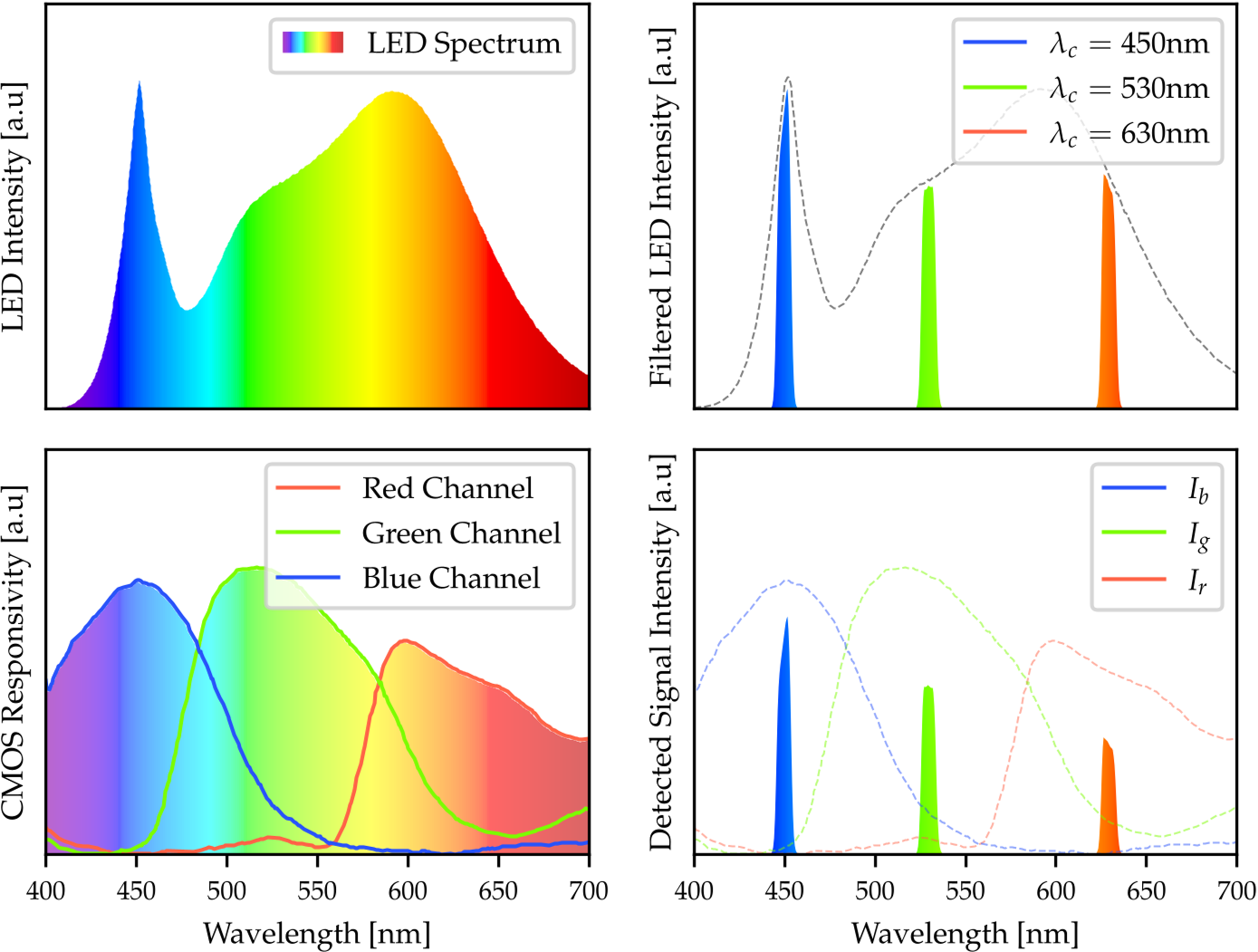}
  \caption{(Top left) The emission spectrum of the white \ac{LED} used for multi-wavelength measurements. (Top right) The combined emission spectrum of the \ac{LED} following spectral filtering by the three bandpass filters. (Bottom left) The responsivity curves of the sub-pixels in the color \ac{CMOS} sensor. (Bottom right) The relative spectrum of responsivity for each of the sub-pixels under illumination by the filtered \ac{LED}.}
  \label{fig:spectral_response}
\end{figure}

The Bayer pattern consists of one each of red and blue sub-pixels and two green on the diagonal. 
In conventional photography, a captured Bayer image undergoes debayering - weighting of the three colors - to produce a final \ac{RGB} image. 
However, the raw Bayer image, with its distinct unmixed color channels, provides an opportunity to capture and reconstruct \ac{RGB} color holograms in a single shot, without the need for wavelength scanning. In \cref{fig:spectral_response}, the bottom left hand plot shows the responsivity of the three color channels of the \ac{CMOS} sensor used for this experiment. 
Limiting the illumination source to only three colors; red, green, and blue, using spectral band pass filters whose central wavelength falls within the active spectral region of each sub-pixel, allows the capture of all three holograms at once. In the top right hand plot of the same figure, the filtered illumination wavelength bands are shown, given the transmission spectrum of the three selected bandpass filters. 
In the bottom right hand plot, the relative detected intensities of the three illumination wavelengths are shown, given both the spectral filtering of the \ac{LED} and the responsivity curves of the sensor. Since each sub-pixel in the Bayer pattern has a distinct spectral responsivity and the illumination bands are narrow ($\lambda_{\text{FWHM}} = 10$ nm), very limited spectral cross-talk occurs between the sub-pixels. 
Therefore, each sub-pixel effectively only measures the modulated light of its corresponding wavelength, and as a result, three distinct holograms can be captured at once.

In this instantiation of multicolor holographic reconstruction, however, the parallel capture of multi-wavelength holograms is simulated. 
The bandpass filter approach necessitates switching between the captured wavelengths. 
Nonetheless, the holograms are captured using the raw Bayer pattern, and an identical image processing is required.

\section{Simple Multi-Wavelength Reconstruction}
\label{sec:simple_mw_reconstruction}

There are many ways to utilize the additional sample data recorded in a stack of multi-wavelength holograms. 
The simplest approach involves treating the color channels entirely separate, i.e., each color channel hologram is processed and reconstructed independently, with or without the application of a phase retrieval algorithm, and the results are combined to form the final \ac{RGB} reconstruction. 
Naturally, the advantage of this approach is that any method of reconstruction can be easily repurposed for multiwavelength use, as it is simply a matter of applying the method in sequence to each hologram in the multi-wavelength stack. 
However, the same simplicity also limits the efficacy of this approach. 
As each wavelength acquisition is processed separately, the reconstruction m ethods are working ``blind'', figuratively speaking. 
No data is shared between the multi-spectral measurements, and thus the additional information is not given a chance to bring the reconstructions to a better minimum in the optimization. 
A further discussion of this is given in \cref{sec:additional_mw_data}.

As an initial test of the color rendition of the multi-wavelength hologram capture, however, the simplest reconstruction technique, propagation of the recorded hologram intensity by \ac{ASM}, is sufficient. On a microscope slide, three ink droplets (between 0.1 mm-0.5 mm in diameter) are deposited on a microscope slide. The ink droplets are red, green, and blue, respectively. The microscope slide is then placed in the sample holder immediately in front of the camera sensor, corresponding to a sample-sensor distance of $d \approx 1$-2 mm. Prior to hologram capture, the white balance of the camera sensor is set such that the mean sensor response is approximately equal for all three inserted band pass filters. To record the multi-wavelength holograms, each bandpass filter is inserted and a raw bayer image is captured, in sequence.

\subsection{Bayer Pattern Hologram Extraction}

The Bayer pattern as illustrated in \cref{fig:bayer_pattern} consist of four subpixels, arranged in an RGGB pattern (left-to-right, top-to-bottom), with responsivities shown in \cref{fig:spectral_response}. According to the inventor, Bryce Bayer, he designed the pattern with twice as many greensensitive elements to mimic the photo-response of the human eye \cite{bayer_color_1976}.

In conventional image-processing firmware, debayering (demosaicing) is the process of reconstructing a full-color image from the separate color filters incomplete output. To do so, some form of pixel interpolation is necessary to extrapolate the missing information from each color sample. In incoherent photography, it is common to include interpolation using gradient-based or frequencydomain techniques that exploit the implicit correlation between color channels to improve the full-color reconstruction.

However, applying these conventional techniques to a Bayerpatterned hologram captured under illumination of multiple wavelengths is not viable in practice, as it may introduce artifacts and non-existent features due to the color-aware interpolation \cite{wu_demosaiced_2016}.

Given a Bayer-patterned raw hologram of size $N \times N$, captured under multi-wavelength illumination, each sub-pixel image can be directly extracted into images $I_R$, $I_{G1}$, $I_{G2}$, and $I_B$, each of size $N/2 \times N/2$.

\begin{figure}[t]
  \centering
  \captionsetup{format = plain}
  \includegraphics[width=\linewidth]{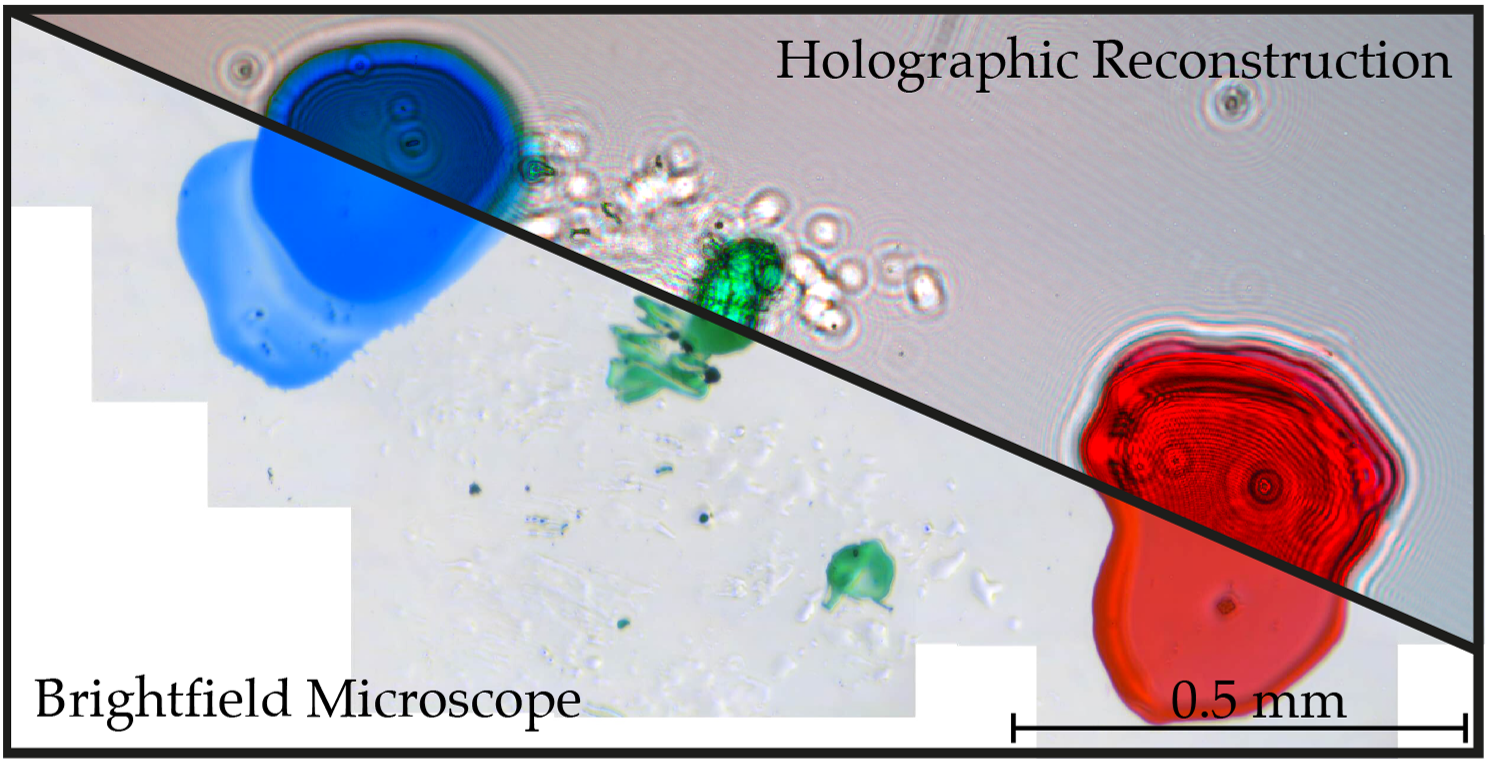}
  \caption{Ink droplets both imaged under a bright-field microscope and reconstructed by the multi-wavelength lensless microscope using \ac{ASM} reconstruction of the raw holograms extracted from the Bayer pattern.}
  \label{fig:ink_droplets}
\end{figure}

With this in mind, the individual color channels in the captured raw Bayer images of the ink drop holograms are extracted (the two green channel images are averaged) accordingly, and reconstructed to the focus plane using simple propagation by \ac{ASM}. The reconstruction and a comparison to the same markers captured in a multi-color bright-field microscope are displayed in \cref{fig:ink_droplets}.

It is evident that the holograms captured under multi-wavelength illumination are able to reconstruct the colors of the ink droplets. The particularly dominant oscillatory artifacts seen in the reconstruction are believed to originate from the sheer size of the ink droplets. In the shown reconstruction, a single reconstruction plane has been chosen. However, this reconstruction plane also includes amplitude and phase contributions from the field above and below the sample. Therefore, the reconstruction of the bottom of the droplets also contains the defocused reconstruction from their remaining volumes, leading to the internal brightness oscillations seen strongly under the coherent illumination. In addition, their size combined with the short propagation distance between the droplets and the camera sensor also affects the appearance of the twin-image. Figuratively speaking, the twin-image has not had a chance to significantly defocus in the reconstruction plane and therefore contains strong oscillations. The impact of the twinimage can be correlated to the Fresnel number, which itself is a function of both the droplet size and the propagation distance \cite{gluckstad_gabor-type_2024}. A much smaller object (on the size scale of the \ac{WBC}s) reconstructed with the same propagation distance is equivalent to a lower Fresnel number, and their twin-images will appear more defocused

It is noteworthy that the color renditions of \cref{fig:ink_droplets} are a result of the directly reconstructed amplitude of the samples at the illumination wavelengths. To properly display the reconstructions with accurate colors on display devices, additional processing is required, including gamma correction and conversion to an appropriate color space.

\section{Multi-Wavelength Super-resolution}
\label{sec:mw_super_resolution}

Before attempting to resolve objects in the approximate size scale as \ac{WBC}s using the multi-wavelength approach, the effective pixel size must be considered. For any of the extracted hologram images, the effective pixel size is twice that of the actual pixel size of the sensor due to the layout of the sub-pixels within the unit cell of the Bayer pattern ($2\ell_p = 3.7$ $\mu$m). With only half the resolution of the image sensor left in each color channel, the super-resolution algorithm presented in \cref{sec:algorithm_design} becomes even more relevant.

To implement this super-resolution for multi-wavelength holograms captured on a Bayer sensor, modifications must be made to both the experimental setup and the super-resolution algorithm.

\subsubsection*{Mechanical Adjustments}
The aperture translation mechanism introduced in \cref{sec:sr_experimental_setup} is easily adapted for the multi-wavelength experimental setup. 
Instead of translating the light source itself as was the case for the fiber-coupled laser, the pinhole in \cref{fig:mw_illumination} is mounted in the manual translation stage, and moved with the micrometer screws, yielding the same sub-pixel translation effect. 
For each sub-pixel shift introduced via the manual translation, an \ac{RGB} hologram can be captured.

\subsubsection*{Algorithmic Adjustments}
\begin{figure}[h!]
  \centering
  \captionsetup{format = plain}
  \includegraphics[width=.75\linewidth]{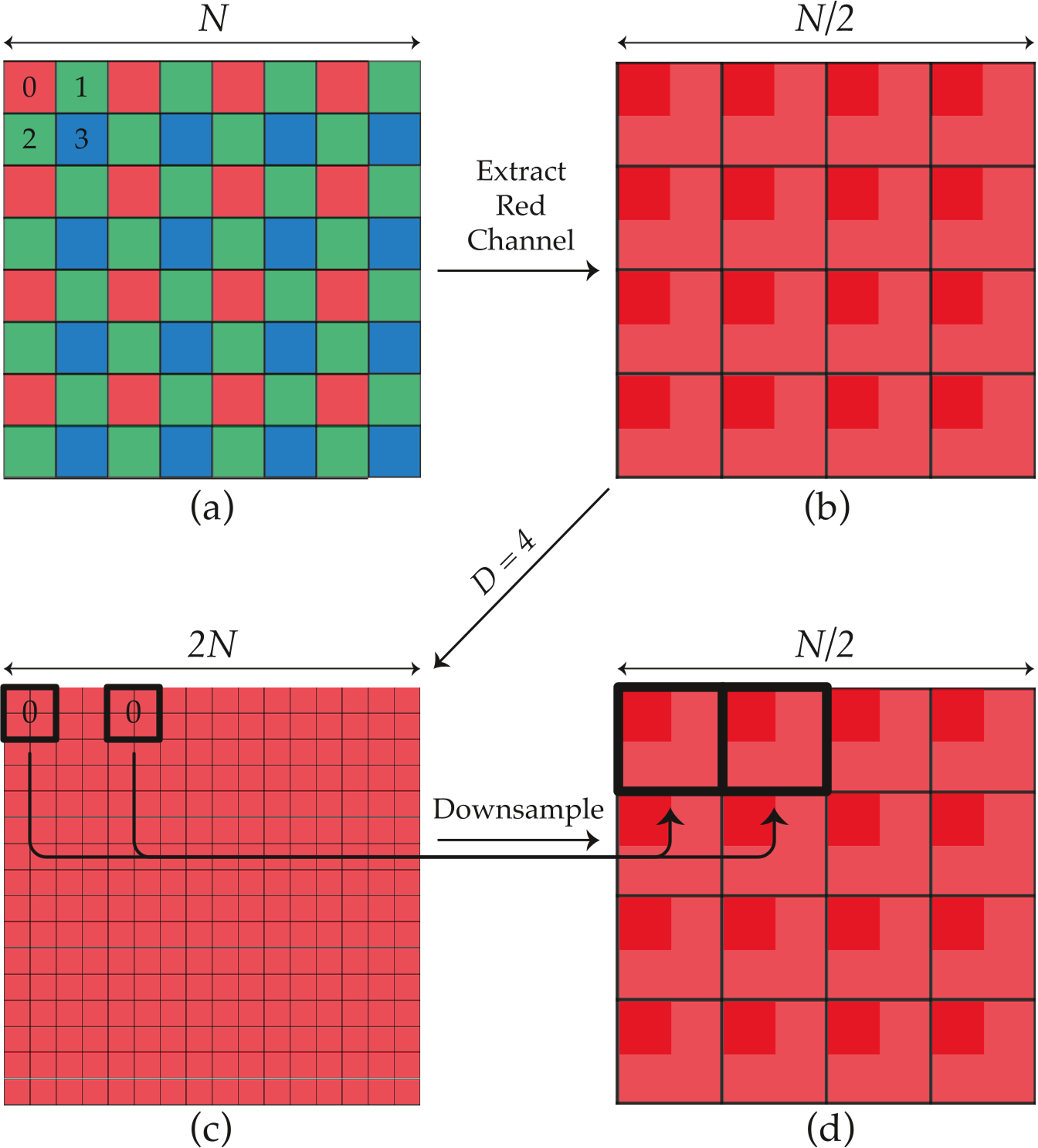}
  \caption{Illustration of the downsampling operation of the superresolution algorithm adapted for multi-wavelength holograms. (a) The raw captured Bayer pattern image with indexed sub-pixels according to their location (0-3). (b) The extracted hologram image using only the red sub-pixels. The extracted image has the same physical dimensions despite the doubling of the effective pixel size. The original and effective pixel sizes are indicated in dark and light red, respectively. (c) The super-resolved hologram estimate updated in each iteration. (d) The downsampled high-resolution estimate. Only the high-resolution pixels that overlap with the original red sub-pixels contribute to the sampling by the sensor. The low-resolution pixels are assigned the mean of overlapping high-resolution pixels.}
  \label{fig:bayer_downsampling}
\end{figure}

Usage of the super-resolution algorithm explained in \cref{sec:algorithm_design} is, in theory, as simple as applying it to each stack of color-separated hologram images. 
However, one small adjustment can be made to facilitate the synthesis of more accurate super-resolved holograms. 
Where in the monochrome case, the downsampling of the current estimate of the high-resolution hologram was implemented via a convolution with a \ac{2D} averaging kernel\footnote{This is valid since the monochrome pixel spans the entire unit cell, and the synthesized high-resolution pixels each contribute equally to the captured low-resolution pixel.}, the multi-wavelength downsampling operation must take into account the physical location of the sub-pixel within the unit cell of the Bayer pattern since not all of the synthesized high-resolution pixels actually contribute to the captured low-resolution pixel.
To explain this more clearly, a visual example can be made of one of the four sub-pixel images in the Bayer pattern, for instance $I_R$, which is shown in \cref{fig:bayer_downsampling}. 
In each iteration of the superresolution algorithm, the current estimate for the super-resolved hologram must be compared to the captured low-resolution hologram. 

To effectively compare them, the sampling by the camera sensor with its finite pixel size is mimicked by way of an averaging downsampling operation. 
In the example in \cref{fig:bayer_downsampling}, the super-resolution magnification factor is set to $D = 4$, and thus, the high-resolution hologram estimate has size $2N \times 2N$. 
Note that although the pixel count is different between the captured and synthesized hologram, their spatial extents are identical. 
Thus, it must be true that if the high-resolution hologram estimate is superimposed on the captured hologram, only the high-resolution pixels that spatially overlap with the physical sub-pixels sensitive to red light (marked by an offset, 0, in the unit cell in the top-left figure) can contribute to the low-resolution pixel. 
In this example, these are the top-left $2 \times 2$ high-resolution pixels, and each $2 \times 2$ block thereafter, with a spacing of 2 pixels in both directions. 
In \cref{alg:bayer_downsampling} in Appendix E, the pseudo-code for the downsampling operation is shown. 
The algorithm includes the calculation necessary for determining the correct location of the contributing high-resolution pixels, given a particular color channel.

Now, although only contributing high-resolution pixels are included in the downsampled hologram, the unknown space between the specific color-sensitive sub-pixels is effectively extrapolated by means of the sub-pixel shifts that are at the core of the super-resolution algorithm.

\subsection{Super-Resolved Multi-Wavelength Reconstruction}

To examine the capabilities of the super-resolved multi-wavelength approach, a test is carried out on a sample of blue-dyed polystyrene (\ac{PS}) beads of mean diameters 10 $\mu$m. As in the simple reconstruction case, the white balance of the camera sensor is adjusted such that each illumination wavelength yields the same approximate response.

\begin{figure}[t]
  \centering
  \captionsetup{format = plain}
  \includegraphics[width=\linewidth]{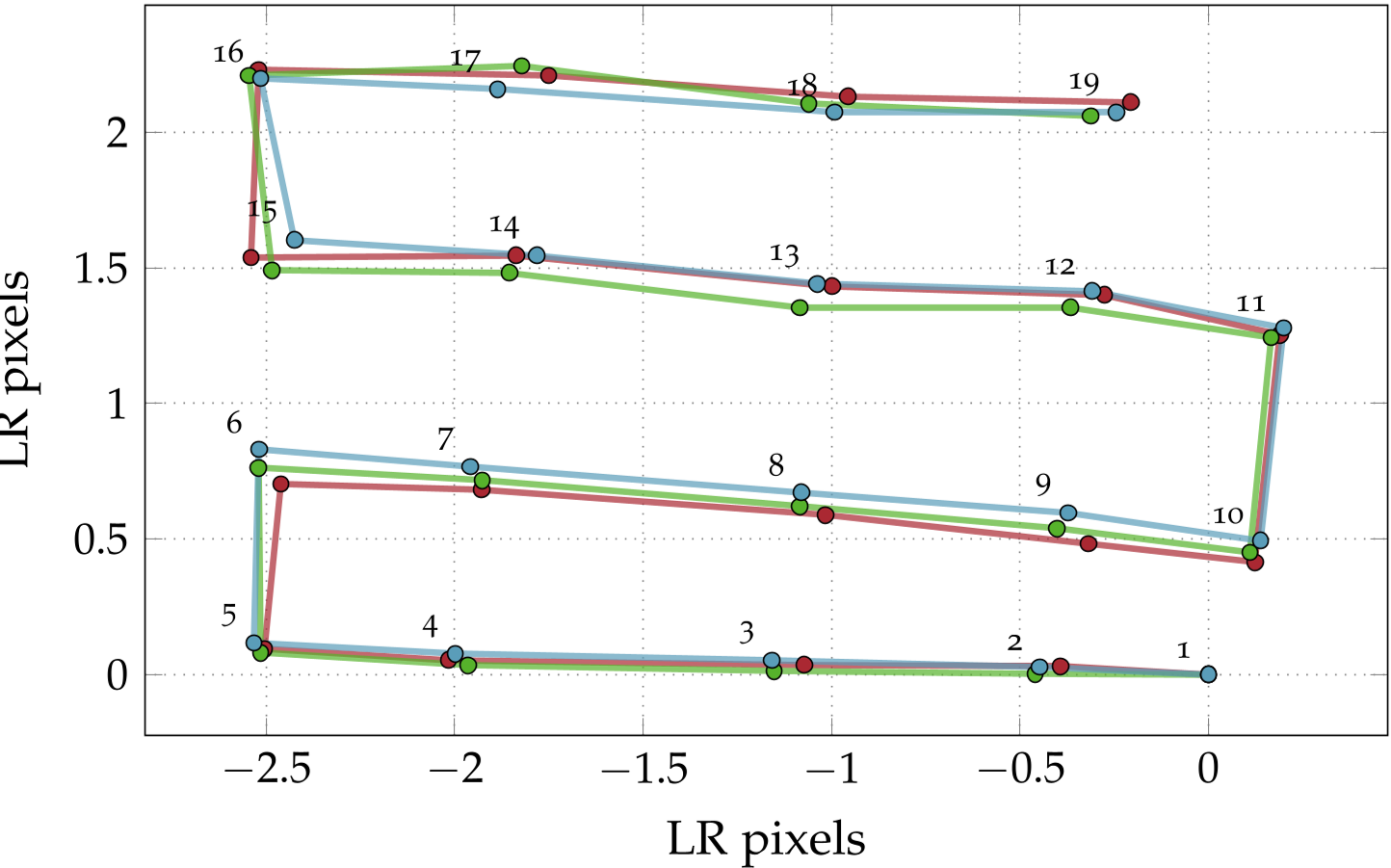}
  \caption{Registered sample translations for each of the illuminating wavelengths in terms of low-resolution pixels for use in the super-resolution algorithm.}
  \label{fig:mw_shifts}
\end{figure}

A droplet of the bead solution is applied to a microscope slide and a cover glass is laid on top. 
The sample is then examined using the multi-wavelength super-resolved lensless microscope at a super-resolution magnification of $D = 4$. 
In particular, for each translation step required by the super-resolution algorithm, an image is captured for each of the inserted bandpass filters. 
In \cref{fig:mw_shifts}, the numerically determined sub-pixel shifts used in the super-resolution algorithm are shown for each captured wavelength on the \ac{PS} bead sample. 
Although some inter-spectral variation in the shifts can be seen, which can be attributed to both numerical errors in the correlation algorithm and slight experimental variations due to the interchanging of bandpass filters, there is good correspondence between the paths. 
Additionally, even though the plot in \cref{fig:mw_shifts} shows the shifts in terms of low-resolution pixel sizes, within the actual super-resolution algorithm, only integer steps of high-resolution pixels are used. 
The small variations are thus reduced as the sub-pixel shifts are aligned to the high-resolution pixel grid.

The super-resolved stack of holograms, each corresponding to a different illumination wavelength, is processed one-by-one. 
Specifically, an amplitude and sparsity constrained phase retrieval (\ac{ASC-PR}) iterative reconstruction algorithm \cite{madsen_-axis_2023,li_sparsity-based_2019} is applied to the holograms in sequence, differing only in the associated numerical propagation wavelength. 
The algorithm is configured to run for 20 iterations per wavelength.

\begin{figure}[t!]
  \centering
  \captionsetup{format = plain}
  \includegraphics[width=.9\linewidth]{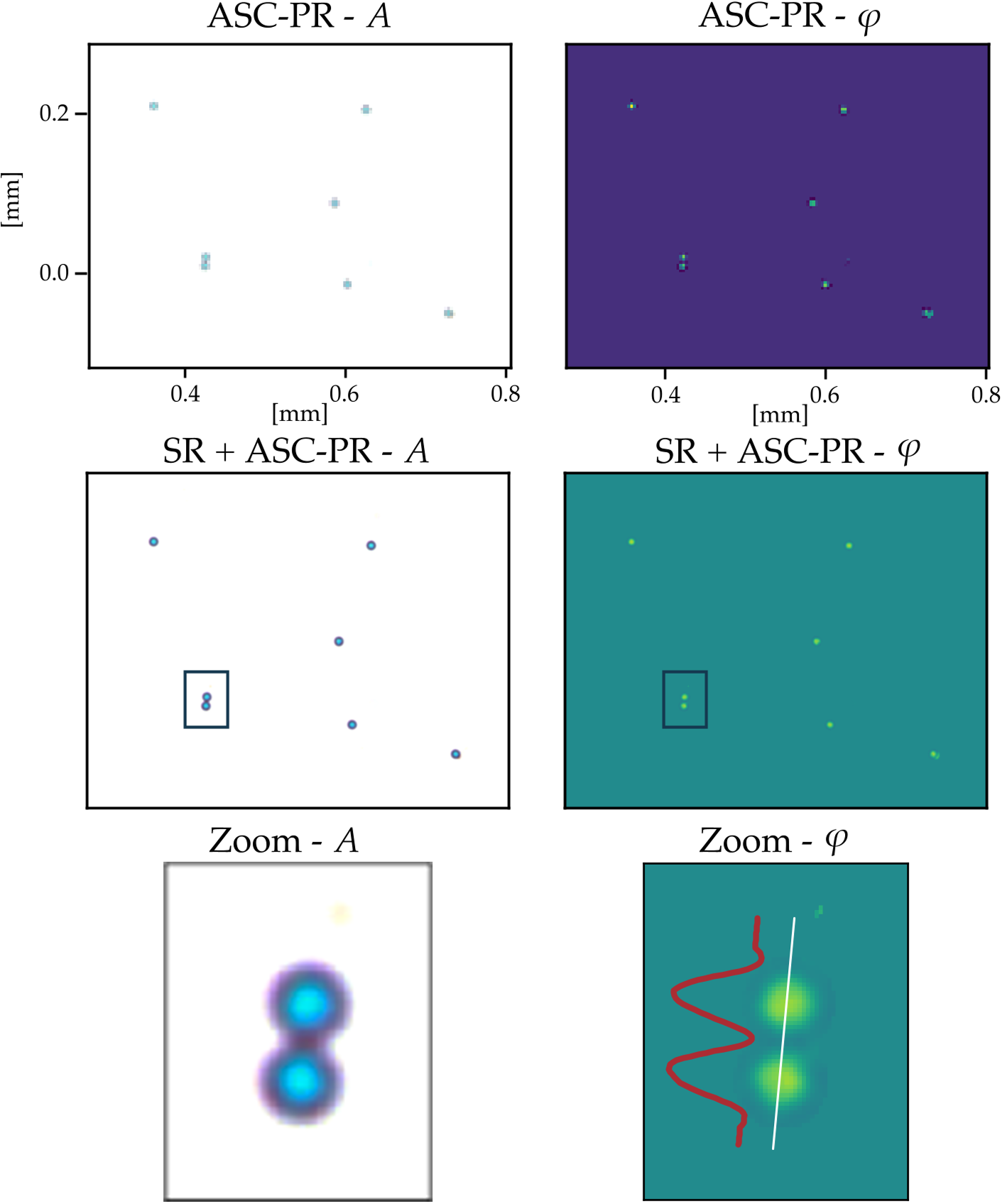}
  \caption{Amplitude and phase (from the blue color channel) profiles of dyed beads reconstructed using \ac{ASC-PR} (Top row) from raw Bayer pattern holograms and (Middle row) from the super-resolved holograms. (Bottom row) Zoomed on two adjacent beads from the super-resolved reconstruction. A line profile of the phase shift of the blue channel of the beads is superimposed.}
  \label{fig:ps_beads}
\end{figure}

In \cref{fig:ps_beads}, a cropped \ac{ROI} of the resulting recovered beadsample amplitude is displayed following the finished phase retrieval and reconstruction for all wavelengths in the hologram stack. 
Both reconstructions for the super-resolved holograms are shown (SC + \ac{ASC-PR}), as well as the holograms directly extracted from the Bayer pattern (\ac{ASC-PR}). 
The phase profiles correspond to the reconstructed blue channel, the channel in which the beads are most transparent.

As evident by the uniform background in the reconstruction amplitudes and phases, and clearly defined reconstructed beads, applying the \ac{ASC-PR} reconstruction algorithm has satisfactorily attenuated the twin-images. The application of the super-resolution algorithm (center row) has made it possible to resolve the beads, maintaining their correct blue color information. The raw Bayer reconstructions show a blue tint as well, however, the blue area is approximately the size of a single pixel in the reconstruction. In the bottom row of \cref{fig:ps_beads}, an \ac{ROI} zoomed in even further on the super-resolved reconstruction shows two beads next to each other. The blue centers of the beads are visible in higher resolution. In the \ac{ROI} of the phases of the two beads, a line profile is drawn through both and plotted on the image. The line profile shows the increasing phase shift due to both beads.

Now, despite the clear blue color recovered from the beads, attention is immediately drawn to the dark edges of the reconstructed beads. There are several possible explanations for this artifact. First, the refractive index of the beads is quite high ($n_b \approx 1.59$ \cite{noauthor_polystyrene_nodate}). Combined with the smooth curved surface of the spherical beads, significant internal reflections are likely to occur. The reflections may exit the bead back towards the light source or may become significantly attenuated by the repeated propagation around the bead volume.

Second, the edge darkness could be a consequence of the superresolution algorithm. As was observed in Section 5.2, the algorithm has a tendency to amplify the sharpness of edges of reconstructed objects.

While it may be a combination of the two hypotheses given here, the internal reflections seem the more likely. In the reconstructed amplitude of the raw Bayer captured holograms, the dark edges do seem to be present as well. However, the reconstruction is of too low resolution to be certain.

While these reconstructions only show a \ac{ROI}s, the entire reconstruction spans an effective \ac{FoV} of 46.15 mm$^2$. With the combined capabilities of super-resolution over such a large \ac{FoV} and multicolor reconstructions, this approach has the potential to provide more useful information to a cell classifier than both a conventional bright-field imaging approach and the monochromatic \ac{DHM} approach, while still maintaining the distinct advantages of \ac{DHM}.

\section{Additional Use of the Multi-Wavelength Data}
\label{sec:additional_mw_data}

It is clear that the multicolor holography approach presented in this chapter can yield reconstructions with correct color values. 
However, as mentioned, the holographic reconstruction algorithm is applied to each hologram in the wavelength stack independently. 
The capture of several holograms of the same subject, but with slight changes in the hologram formation model, denoted as diversity measurements, was utilized in Chapter 5 and in this chapter to facilitate super-resolution.

Aside from only facilitating super-resolution, the use of additional diversity measurements has been shown to enable better convergence and accuracy in iterative reconstruction algorithms, especially for densely populated samples. 

Options for diversity measurements include multi-height hologram capture \cite{rivenson_sparsity-based_2016,greenbaum_maskless_2012}, structured illumination \cite{guo_robust_2022}, phase-coded illumination \cite{katkovnik_computational_2017}, and, indeed, multi-wavelength illumination \cite{wu_wavelength-scanning_2024,wang_multi-wavelength_2022-1}, among others.
With the addition of more diversity in the captures, the solution space of the problem is limited significantly. 
The iterative algorithms are encouraged to find a reconstruction that obeys all of the measured holograms, given the particular hologram formation model.

Of the multi-wavelength phase retrieval approaches, a majority assume that the sample, being a weak scatterer, obeys:
\begin{equation}
A_{\lambda_1} = A_{\lambda_2} = \ldots = A_{\lambda_n}
\label{eq:amplitude_independence}
\end{equation}
\begin{equation}
\phi_{\lambda_{n+1}} = (\lambda_n / \lambda_{n+1}) \phi_n
\label{eq:phase_scaling}
\end{equation}
where $A_{\lambda_n}$ and $\phi_{\lambda_n}$ are the amplitude and phase of the sample, respectively, under illumination of the $n$th wavelength. 
Thus, the sample amplitude is assumed to be independent of illumination wavelength. 
Consequently, any convergent result of an algorithm working under this assumption is a monochrome amplitude and phase. 
While the multi-wavelength information is used to recover a common sample for all wavelengths, in a sense the color information of the sample is lost during optimization.
Now, this assumption may certainly be appropriate for a multitude of samples and applications, but for the specific problem of classification, even minute inter-cellular differences in the amplitude or phase modulation under varying illumination wavelengths may be important.

It is possible to utilize the color information for phase retrieval while maintaining that same information in the final reconstruction. 
One approach, as developed by Herve et al., regularizes the phase-retrieval optimization problem by way of a multi-spectral colocalization factor \cite{brault_multispectral_2023, herve_multispectral_2018}, i.e., the degree to which an object in the sample overlaps itself across the wavelength acquisitions. 
Disregarding the absolute amplitude dependence on the illumination wavelength, it is reasonably assumed that the edges of a recorded sample object should be super-imposed, meaning that the object should be observed at the same position independent of illumination wavelength, effectively limiting the solution space. 
However, as described in \cite{brault_multispectral_2023}, the efficacy of this co-localization regularization depends heavily on the aberrations in the system.

The use of diversity measurements for super-resolution and phase retrieval are also combined in certain cases \cite{gao_generalized_2021, guo_lensfree_2020, wu_wavelength-scanning_2024, isikman_giga-pixel_2012, greenbaum_imaging_2012}. For this application, One might imagine that the additional data gathered by the multi-wavelength illumination could lower the requirement for the number translation steps required in the superresolution algorithm. Unfortunately, these methods were not attempted in this project due to time constraints.

\section{Chapter Conclusion}
\label{sec:ch6_conclusion}

This chapter detailed the construction and implementation of a lensless multi-wavelength holographic microscope. The multi-wavelength extension to the initial monochromatic approach was explored, highlighting the potential advantages of adding additional information for use in classification, especially in cases where a simple staining procedure is acceptable. Specifically, bringing to light the wavelength dependent inter-cellular differences in amplitude and phase modulation.

The experimental design involved replacing the single laser source with bandpass filtered \ac{LED}s, and the camera with a color \ac{CMOS} sensor. As the multi-wavelength holograms are captured on a raw Bayer pattern sensor, additional processing was implemented in order to extract the relevant pixel data for each individual illumination wavelength. The effective hologram resolution as extracted from the Bayer pattern is halved, due to the sub-pixel layout in the Bayer pattern. Therefore, the chapter also combined the multiwavelength capabilities with the super-resolution algorithm laid out in \cref{alg:super_resolution}.

The addition of multi-wavelength hologram capture could have a significant impact on a \ac{DHM}-enabled \ac{dWBC} analyzer, an area that deserves more research. While maintaining a compact, simple, and robust mechanical system, it would enable a classification algorithm to take advantage of wavelength-dependent features of the \ac{WBC}s, e.g., color differences in cell nuclei, cytoplasm, granules, etc., following a simple staining procedure. The use of phase retrieval algorithms taking advantage of the additional information of the multi-wavelength hologram capture should be explored further, both to improve reconstruction quality and thereby possibly classification accuracy, and also to allow the reconstruction of higherconcentration samples.

\chapter{Physics-informed neural network for holographic reconstruction}
\label{ch:pinn}

Reconstructing holograms using \ac{ML} techniques has several desirable advantages over conventional iterative methods. First and foremost, assuming the training of some hologram-reconstructing network is complete, whatever its design may be, reconstruction of previously unseen holograms can be performed using a single pass through the network. This presents a massive possible computational speed-up in comparison to the iterative methods, in which it is common to perform multiple fast fourier transforms (\ac{FFT}s) on large holograms in each iteration, for tens of iterations.

Naturally, this advantage comes at a price. \ac{ML} models typically require enormous datasets, and a computationally intensive and time consuming training phase, and the result is effectively a blackbox algorithm with limited human-interpretable reason. Most of the approaches employ supervised learning, which are known to struggle with generalization to new, unseen, sample types.

Despite this, the field of \ac{ML}-aided holographic reconstruction techniques has become large over the past decade, showing truly impressive results \cite{madsen_-axis_2023,huang_holographic_2021,castaneda_video-rate_2021,jaferzadeh_holophasenet_2022,niknam_holographic_2021,wang_eholonet_2018,wang_y-net_2019}. Within this field, numerous approaches to both network design, training data collection and simulation, and overall philosophy, etc., have been explored. One trend, however, which seems to gain traction both in all scientific fields is the concept of physics informed machine learning, and physics-informed neural networks (\ac{PINN}s) \cite{cuomo_scientific_2022, baty_hands-introduction_2024}. \ac{PINN}s offer

an alternative method of training and using \ac{ML} for predicting outcomes and modeling physical phenomena, by embedding physical laws directly into the learning process. By only learning results that obey the fundamental physical principles, the network is better capable of generalizing across different sample types and conditions, possibly reducing the number and diversity requirement of training data massively.

This chapter describes the design, implementation, and testing of a hologram-reconstructing \ac{PINN}. The network is trained on fully synthetic data and tested on holograms captured using the multiwavelength lensless holographic system, as described in \cref{ch:multiwavelength}.

\section{Working Principle}
\label{sec:pinn_working_principle}

\subsection{Physically-Based Loss}

\begin{figure}[h!]
  \centering
  \captionsetup{format = plain}
  \includegraphics[width=0.85\linewidth]{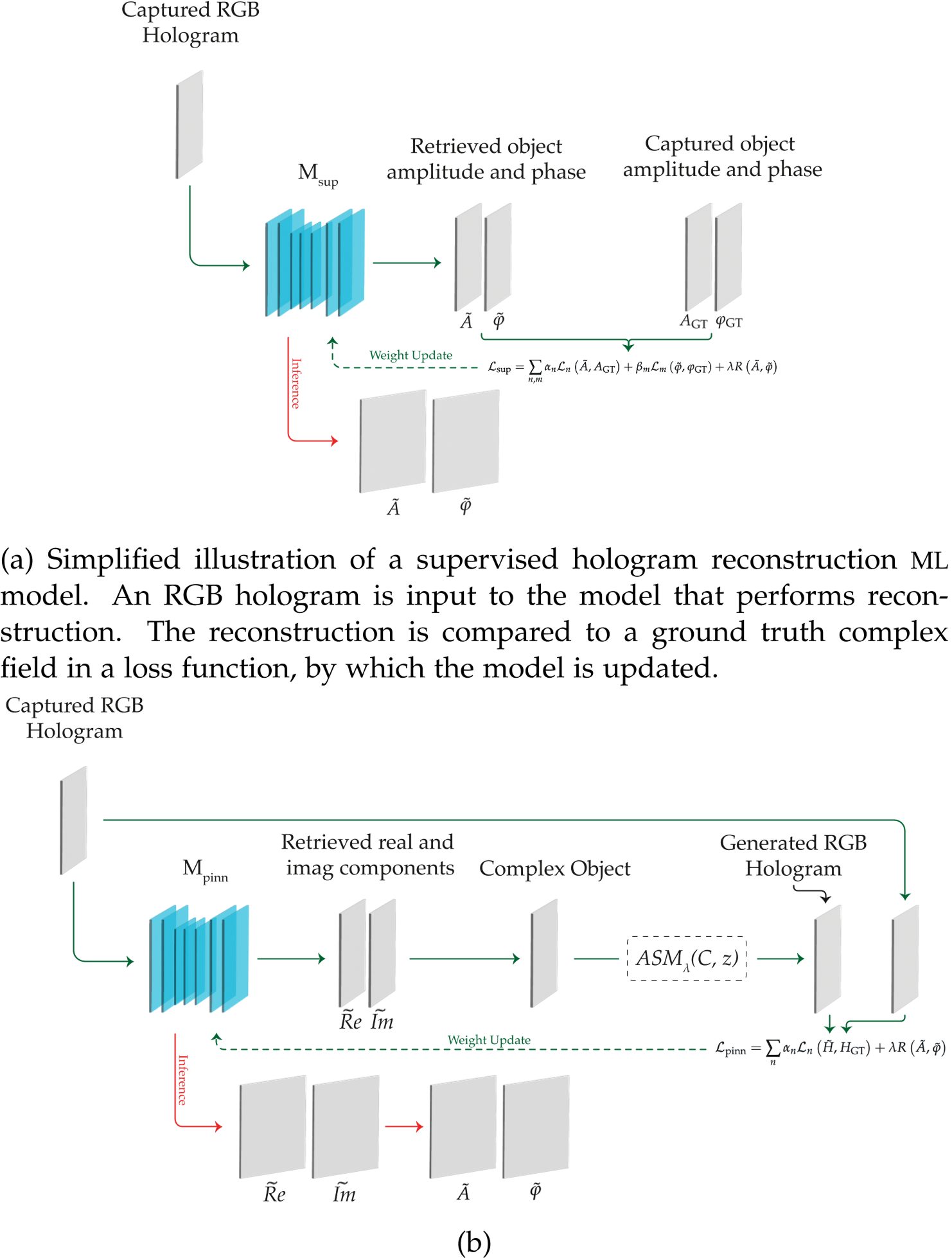}
  \caption{Simplified illustration of the \ac{PINN} approach for hologram reconstruction. An \ac{RGB} hologram is input to the model that performs reconstruction. The reconstruction is propagated back to the hologram plane where it is compared to the input hologram, by which the model weights are updated.}
  \label{fig:supervised_model}
\end{figure}

As mentioned, the majority of neural networks designed for holographic reconstruction are supervised in training. 
Supervised here meaning that the network has access to a ground truth reconstruction in its loss function, to which it can compare its output, as is illustrated in \cref{fig:supervised_model}a. 
As such, the training data consists of labelled data; complex fields of a sample, typically recovered via multi-height phase retrieval, off-axis holography, or similar, and their associated captured holograms. 
A loss function in such a supervised network may take on the form:

\begin{equation}
\mathcal{L}_{\text{sup}} = \sum_{n,m} \alpha_n \mathcal{L}_n (\tilde{A}, A_{\text{GT}}) + \beta_m \mathcal{L}_m (\tilde{\phi}, \phi_{\text{GT}}) + \lambda \mathcal{R}(\tilde{A}, \tilde{\phi})
\label{eq:supervised_loss}
\end{equation}
Where a number of error terms directly compare the generated amplitude $\tilde{A}$ and phase $\tilde{\phi}$ to the ground truth values $A_{\text{GT}}$ and $\phi_{\text{GT}}$. 

The generated amplitude and phase may be further constrained by regularization $\mathcal{R}$, e.g., non-negativity, sparsity, etc.
In contrast, a \ac{PINN} can be trained in a self-supervised manner, meaning that the ground truth is based on its input, as illustrated in \cref{fig:supervised_model}b.

The loss function in a \ac{PINN}, designed specifically for holographic reconstruction \cite{huang_self-supervised_2023}, can be expressed as:

\begin{equation}
\mathcal{L}_{\text{pinn}} = \sum_n \alpha_n \mathcal{L}_n (\tilde{I}, I_{\text{GT}}) + \lambda \mathcal{R}(\tilde{A}, \tilde{\phi})
\label{eq:pinn_loss}
\end{equation}
where $I_{\text{GT}}$ is the captured hologram, and $\tilde{I}$ is the hologram calculated by propagating the generated complex field ($\tilde{A}$, $\tilde{\phi}$) the correct distance $z$ back to the hologram plane. 
Thus, as opposed to the supervised loss in \cref{eq:supervised_loss}, which minimizes the difference between the estimated complex field and the presented ground truth, the \ac{PINN} model is encouraged to generate amplitude-phase profile pairs that, when propagated back to the hologram plane using the forward model of free-space propagation, could have created the measured hologram.

The importance of this distinction between the two loss functions cannot be overstated. 
In effect, the conventional supervised model is working blindly, with no assurance that the reconstruction it generates is physically possible given the hologram that was presented to it. 
In addition, there is a risk that what is learnt is biased towards a specific sample type or by the specific experimental setup used in the training data, hence the requirement for a large and diverse training set.

On the other hand, the outputs of the \ac{PINN} must adhere to the forward model and yield reconstructions which could have caused the measured holograms. 
By ensuring that the reconstructions adhere to the fundamental principles of propagation through freespace (through propagation using \ac{ASM}), the \ac{PINN} may be able to achieve more physically accurate results. 
In addition, since what is learnt by the network should, ideally, not change significantly with different sample types, as the same fundamental physical properties are maintained in the forward model, the generalization capabilities of the model should be enhanced.

\subsection{Holographic Reconstruction \ac{PINN}}

To reiterate the points made in the last subsection: a \ac{PINN}, capable of reconstructing holograms, is a generative model --- an image-toimage translator --- configured such that its outputs are encouraged to obey certain physical properties that can be modeled from the real world. In this case, the network is optimized to generate the complex object field which could have generated the intensity hologram that was shown to it. To nudge the weights of the network in the correct direction, in each training step of the learning of the network, its generated complex object field is propagated, via some numerical propagator, to the hologram plane. Therefore, it is learning which object field could, after propagation by the propagator, have been the object responsible for the captured hologram.

Now, regarding the design and architecture of the actual \ac{ML} model. Generative models and image-to-image translators can take on many forms; from the ever popular UNet architecture \cite{ronneberger_u-net_2015-2} to generative adversarial networks (\ac{GAN}s) \cite{goodfellow_generative_2014}, and hundreds more. Regardless of the specific architecture, the network backbone of the holographic \ac{PINN} must be adequately deep and complex to capture the nuanced relationship within both the holographic data and the associated transformation from intensity-only hologram to complex object. However, it is essential to strike a balance between model complexity and performance, such as to avoid unnecessary computational overhead. An overview of the design of the \ac{PINN} model, $M_{\text{pinn}}$, is given in Appendix F.

\subsubsection*{Inputs and Outputs}
For this implementation of a hologram-reconstructing \ac{PINN}, it was decided to employ the popular simplifying assumption that object amplitude and phase are wavelength independent ((\cref{eq:amplitude_independence,eq:phase_scaling})). As mentioned in Chapter 6, this is a popular assumption, allowing the use of the multi-wavelength data as phase diversity for more effective phase retrieval. Effectively, we are providing more information to the network than it has to generate. As such, the input to the \ac{PINN} will be a hologram, captured under illumination of three wavelengths; 450 nm, 530 nm, and 630 nm, such as to replicate the system in Chapter 6. The hologram images are of shape $512 \times 512 \times 3$, with 3 color channels and 512 pixels in both lateral directions. $512^2$ being the maximum hologram size to fit in the video random access memory (\ac{VRAM}) of the \ac{GPU} on which the \ac{PINN} is trained with a batch size of 3.

The output of the network is of size $512 \times 512 \times 2$, where the two channels denote the real and imaginary component of the reconstructed object field. These can then be combined to form the complex object field or its amplitude and phase profiles for use in the loss function.

\subsubsection*{Loss Function}
The effectiveness of learning the underlying physical phenomena hinges upon the loss function of the model. The loss function employed for this application consists of three individual terms:
\begin{equation}
\mathcal{L}_{\text{holopinn}} = \alpha \mathcal{L}_{\text{MSE}}(\tilde{I}, I_{\text{GT}}) + \beta \mathcal{L}_{\text{FD-MAE}}(\text{FFT}\{\tilde{I}\}, \text{FFT}\{I_{\text{GT}}\}) + \lambda \text{TV}(\tilde{C})
\label{eq:holopinn_loss}
\end{equation}

Where $\tilde{I}$ is the generated hologram following propagation, $I_{\text{GT}}$ is the input hologram, $\alpha$, $\beta$, and $\lambda$ are weightings of the three terms, \ac{FFT} denotes the \ac{2D} spatial Fourier transform, and $\tilde{C}$ is the generated complex object field. The three terms are:

\paragraph{mean squared error (mse):} This term quantifies the mean square error between the generated hologram and the ground truth hologram in the spatial domain. Minimizing this error encourages the network to produce reconstructions that produce holograms closely matching the ground truth.

\paragraph{fourier domain mean absolute error (fd-mae):} This term quantifies the mean absolute errors between the spatial frequencies present in the generated hologram and the ground

truth hologram, in order to preserve high-frequency information \cite{li_ftmixer_2024}.

\paragraph{complex total variation regularization (tv):} This regularization encourages the generation of smooth complex object fields by way of the total variation \cite{rudin_nonlinear_1992-1, fornasier_introduction_2010}, while still preserving details such as edges in the reconstruction.

\subsubsection*{Training Data}
A key advantage of the chosen self-supervised \ac{PINN} design lies in its data requirements, or lack thereof. Unlike many supervised learning approaches, the design does not require an enormous, diverse dataset of ground truth amplitude and phase image pairs captured from physical samples. Instead, its training relies entirely on a synthetic dataset generated numerically.

This offers several benefits. First, it eliminates any potential biases from specific sample characteristics, ensuring that learned principles are broadly applicable across all different types of specimen. For instance, in the literature, it is common to see reconstruction models excel at one sample type, while underperforming on others, e.g., blood samples and brain tissue. Secondly, since the synthetic data can be produced in virtually unlimited quantities, the physical constraints of sourcing large amounts of data are lifted.

For this application, the synthetic data is generated from images sourced from the DIV2K dataset \cite{agustsson_ntire_2017-1}; a repository of high-quality images of various objects and scenes (from mice to buildings and everything in between). These images are sampled randomly, and paired to create diverse amplitude and phase configurations. These amplitude and phase profiles are combined to form complex fields and propagated a distance $z$ by \ac{ASM}, at the three chosen wavelengths, to generate the ground truth holograms used for training. The random combinations of images as amplitude and phase profiles ensures that the dataset contains both high and low spatial fre-

quencies and a wide dynamic range in both amplitude and phase values.

\section{Physics Informed Reconstructions}
\label{sec:pinn_reconstructions}

The \ac{PINN} defined in \cref{sec:pinn_working_principle} is implemented in Python using the PyTorch framework. The training data consists of 5000 \ac{RGB} holograms, generated according to \cref{sec:pinn_working_principle}, and 200 reserved for validation.

\begin{figure}[t]
  \centering
  \captionsetup{format = plain}
  \includegraphics[width=\linewidth]{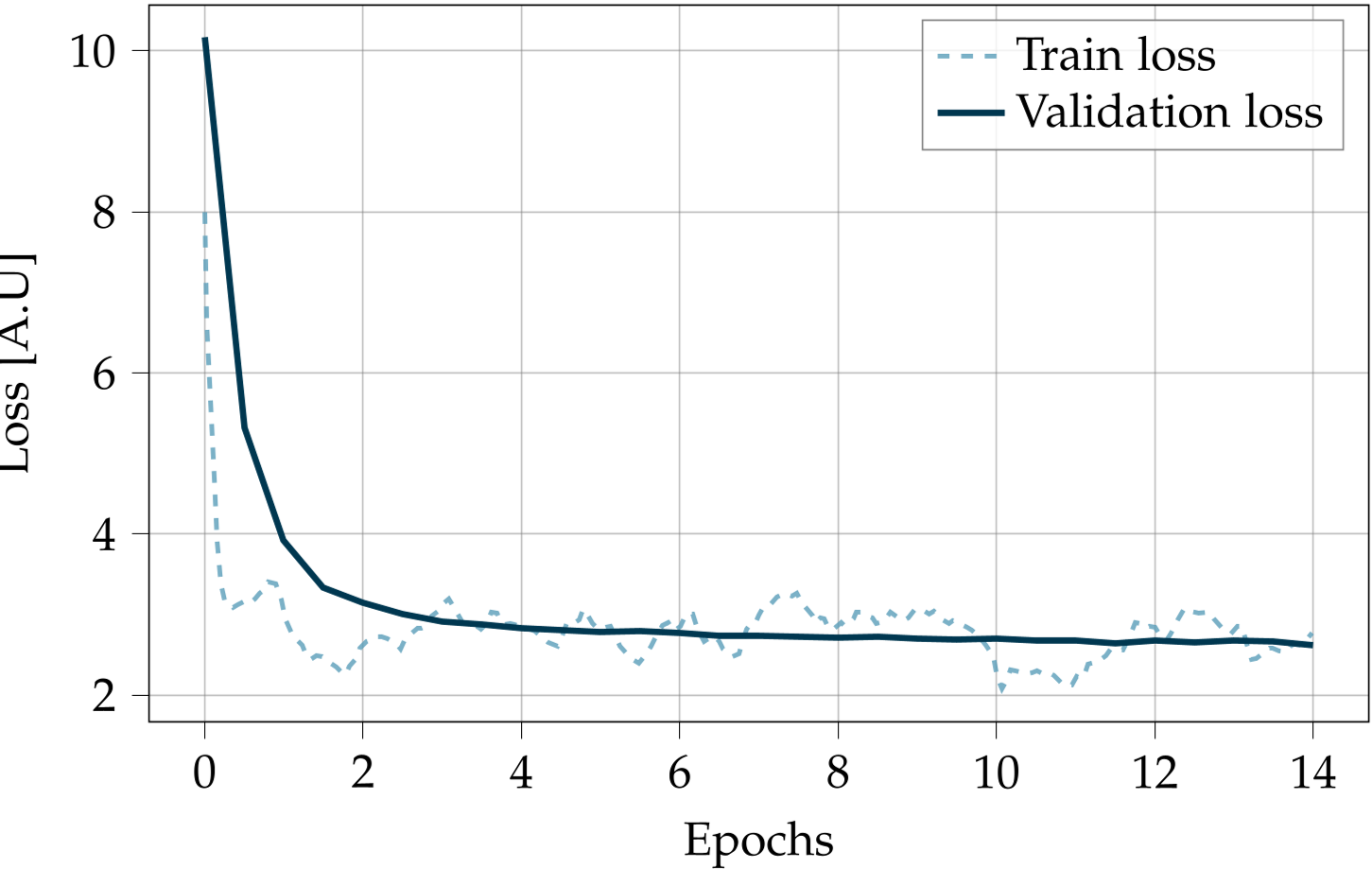}
  \caption{Training and validation loss for the training of the hologram-reconstructing \ac{PINN}. Loss plateaus after approximately 4 epochs.}
  \label{fig:pinn_training_loss}
\end{figure}

The final \ac{PINN} is trained for 14 epochs. In each epoch, the entire training dataset is passed through the network in batches of three holograms. 
The network outputs its best guess for the object reconstruction, and the result is evaluated in the loss function and the network weights are updated by gradient descent with the Adam optimizer \cite{kingma_adam_2017}. 
To facilitate convergence, an exponential learning rate scheduling is employed. By setting an initial learning rate $\text{lr}_0$, and a decay value $\gamma_{\text{lr}}$, the learning rate decreases exponentially each epoch. 
During training, both training and validation loss, as well as intermediate reconstructions, are logged. The losses are plotted in \cref{fig:pinn_training_loss}. 
The network quickly reaches a plateau after approximately 4 epochs, but does not show signs of overfitting.

\begin{figure}[t!]
  \centering
  \captionsetup{format = plain}
  \includegraphics[width=0.9\linewidth]{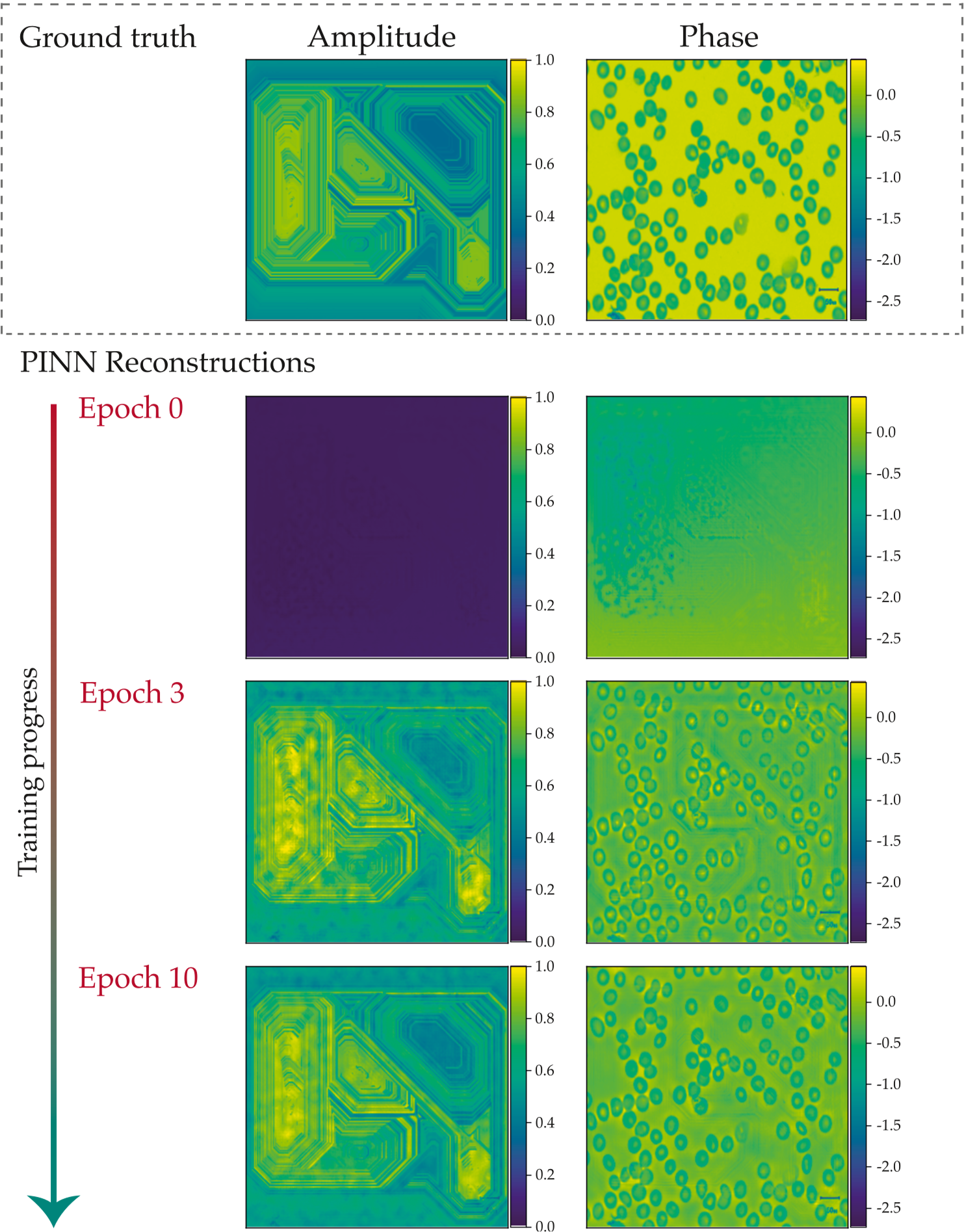}
  \caption{Training progress of the hologram-reconstructing \ac{PINN}. (Top row) The ground truth amplitude and phase of a sample the model has not seen, but attempts to replicate. As training progresses, the \ac{PINN} reconstructions appear closer to the ground truth. Note, the model is never shown the ground truth.}
  \label{fig:pinn_training_progress}
\end{figure}

In \cref{fig:pinn_training_progress}, intermediate reconstructions for one validation example at epochs 0, 3, and 10 are shown, as well as the ground truth amplitude and phase. 
Note, the ground truth amplitude and phase was not used for training, it is only shown here as a visual aid. 
As training progresses, the reconstructed profiles approach the ground truth, despite the network having never seem them. 
This validation example, as well as the rest of the training data, consists of samples with entirely different images used for the amplitude and phase profiles, hence the difficulty in reconstruction.

After the network has been stuck at the plateau for 10 epochs, training is stopped, and the network can be tested using experimentally captured holograms.

\subsection{Experimental Reconstructions}

\begin{figure}[t]
  \centering
  \captionsetup{format = plain}
  \includegraphics[width=\linewidth]{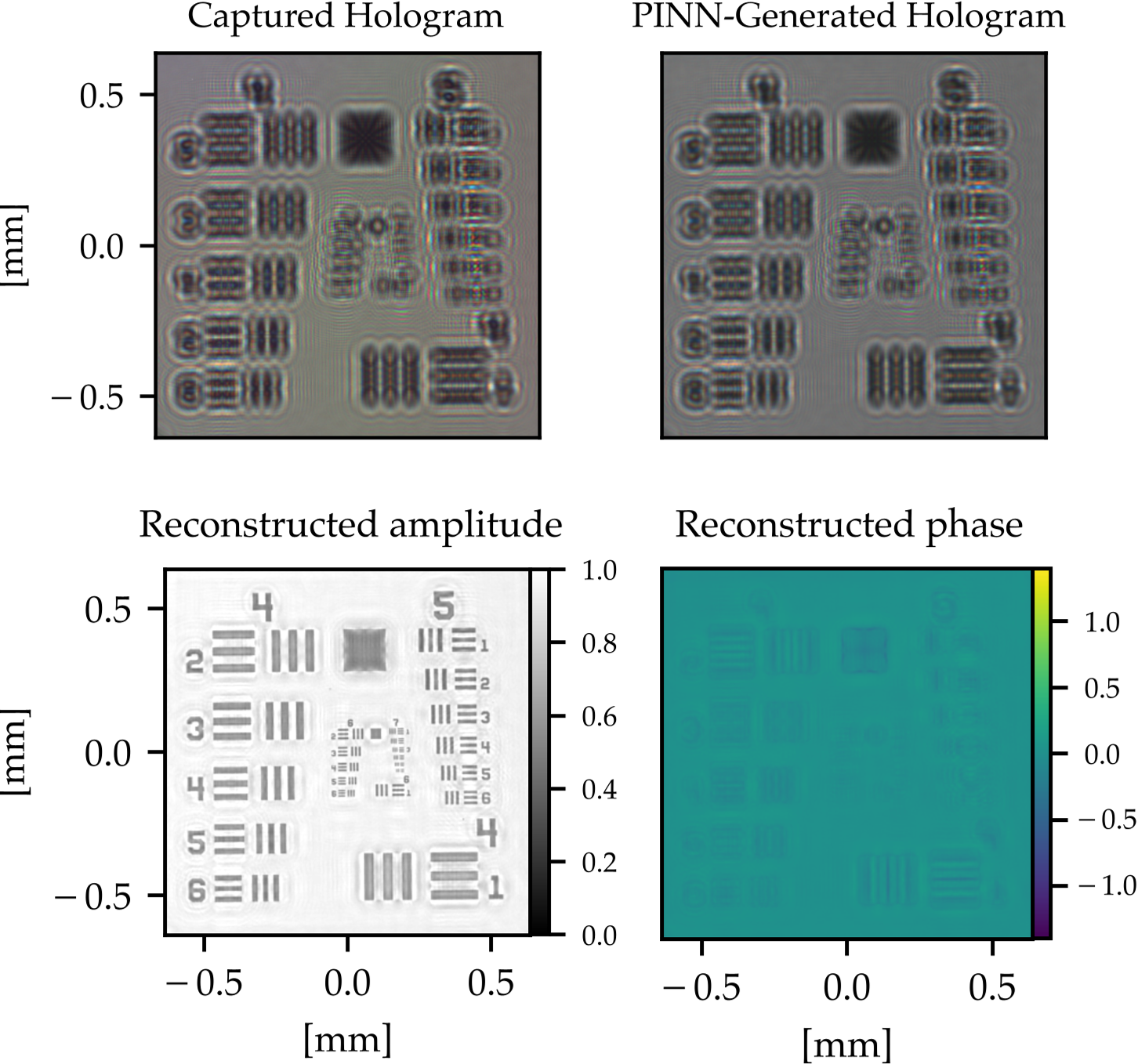}
  \caption{Experimentally captured hologram of amplitude resolution target tested on the \ac{PINN}. The captured hologram is input to the model, and the reconstructed amplitude and phase is output. Propagating the reconstructions back to the hologram yields the \ac{PINN}-generated hologram.}
  \label{fig:pinn_resolution_target}
\end{figure}

\begin{figure}[t]
  \centering
  \captionsetup{format = plain}
  \includegraphics[width=\linewidth]{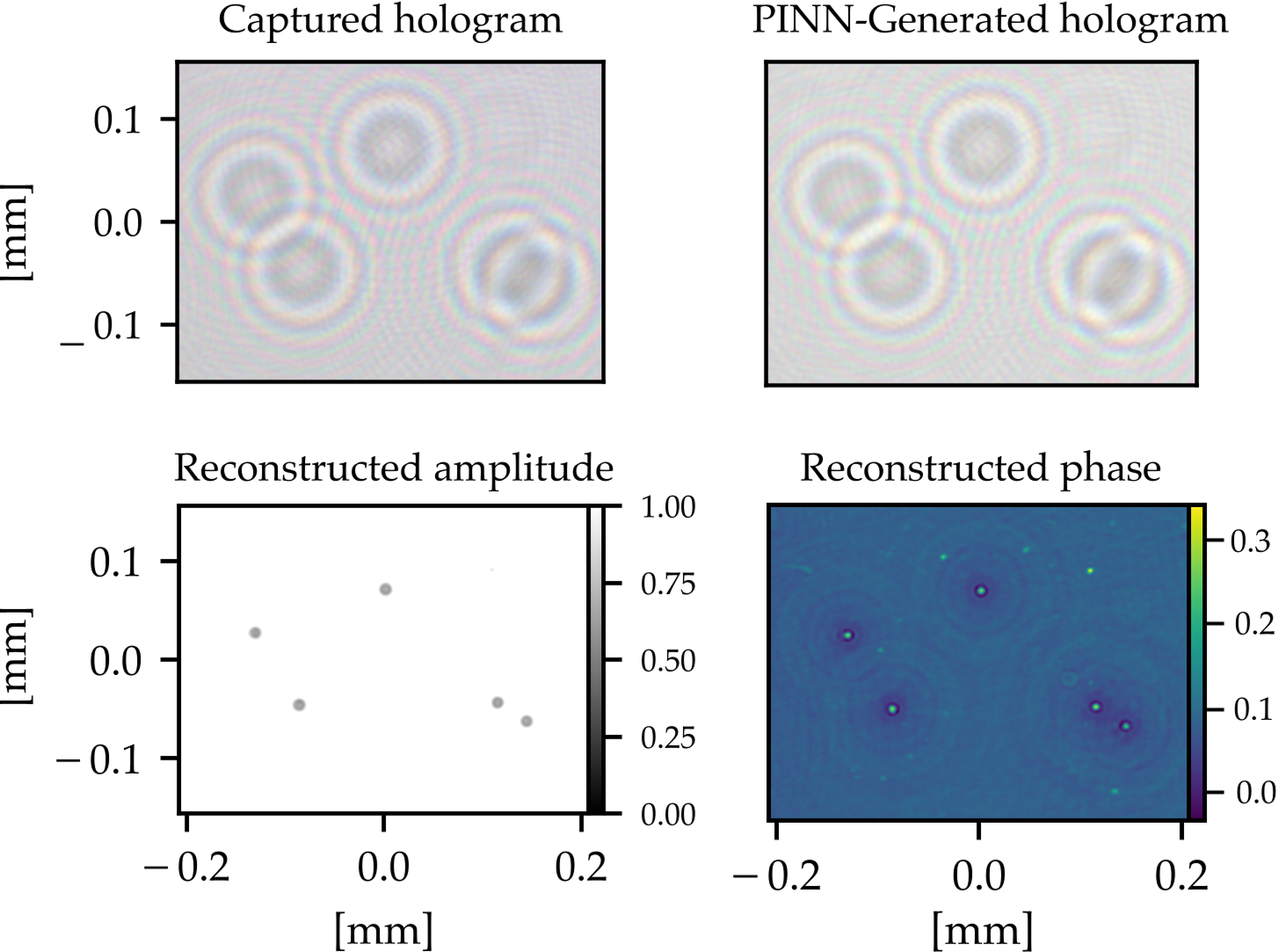}
  \caption{Experimentally captured hologram of 10 $\mu$m \ac{PS} beads tested on the \ac{PINN}. The captured hologram is input to the model, and the reconstructed amplitude and phase is output. Propagating the reconstructions back to the hologram yields the \ac{PINN}-generated hologram.}
  \label{fig:pinn_ps_beads}
\end{figure}

In \cref{fig:pinn_resolution_target,fig:pinn_ps_beads}, two \ac{RGB} holograms, captured under the multi-wavelength illumination, super-resolution approach described in \cref{sec:mw_experimental_setup}, and the result of their passing through the \ac{PINN} trained at their respective reconstruction distance, are shown. The results indicate that the \ac{PINN}, which has been trained on entirely synthetic data with limited spatial relation to the experimental objects, can generalize to experimentally captured holograms.

An amplitude resolution target is used, and shown in \cref{fig:pinn_resolution_target}. The reconstructed amplitude shows the target in focus, with minor artifacts surrounding the absorbing elements, resembling the twinimage. The reconstructed phase is nearly flat, as it should be, with minor deviations in the artifacts.

In \cref{fig:pinn_ps_beads}, the \ac{PINN} has been used to reconstruct a sample of blue-dyed \ac{PS} beads. The amplitude reconstruction shows virtually no signs of artifacts, and the beads are clearly seen. The phase profiles of the beads show some signs of noise in the background.

To improve further on these results, it should be considered to expand the depth of the \ac{PINN}, or research the use of alternative model architectures. Limited by computing power, the configuration explained in this chapter was not entirely successful in its reconstructions. While impressive that the experimental reconstructions show such a distinct semblance to these known objects, it is clear that further improvements can be made.

\section{Chapter Conclusion}
\label{sec:ch7_conclusion}

This chapter outlined the implementation of a physics-informed neural network capable of reconstructing experimental holograms. The model used a physically based loss function, in which the model is encouraged to output complex reconstructions which could, physically, have been the object responsible for the input hologram. This is done by propagating the generated complex reconstruction back to the hologram plane, and measuring the error between the input hologram and this synthesized hologram, both in the spatial and Fourier domain.

Importantly, this encourages the model to synthesize reconstructions that obey free-space propagation. Additionally, it enabled the training of the model to be performed on entirely synthetic data, alleviating the need for a large training set. The training data was generated by combining random images from the DIV2K dataset to form complex fields. These complex fields were then propagated to a hologram plane under simulated multi-wavelength illumination to form the training data.

The model was trained for 14 epochs, and the trained model was tested on experimentally captured holograms. The results indicate that the model has learned generalized rules of hologram recon-

struction. However, the results still contain remnants of twin-image noise and other artifacts. With further training using increased computational power and a deeper model, such that the model can better map the advanced relationships between holograms and their complex reconstructions, the model could perhaps outperform certain iterative reconstruction algorithms. However, the use of a deep learning model for image-to-image translation practically requires the use of hardware acceleration in order to take advantage of their superior inference rates. Great strides have been made in the area \ac{ML}-enabled mini computers with dedicated \ac{GPU}s \cite{noauthor_nvidias_nodate, noauthor_nvidia_nodate}. Such devices could facilitate on-device reconstruction using these complex models.

\part{HOLOTILE: RAPID AND SPECKLE-SUPPRESSED COMPUTER GENERATED HOLOGRAPHY}
\chapter{Computer-generated holography and HoloTile}
\label{ch:cgh_holotile}

Having explored the principle of holography through \ac{DHM} in Part I of this thesis, attention is now turned to the closely related field of computer-generated holography (\ac{CGH}). Both \ac{DHM} and \ac{CGH} utilize the wave nature of light and digital processing to manipulate or record optical information. \ac{DHM} captures the complex wavefront from a physical object on a digital sensor. \ac{CGH}, conversely, synthesizes this wavefront digitally, effectively creating a virtual object that exists only in the computer. Just as \ac{DHM} allows the digital reconstruction of a \ac{3D} image of a real object, \ac{CGH} enables the \ac{3D} projection of a virtual object into real space.

A computer generated hologram can be displayed on so-called spatial light modulators (\ac{SLM}s) or printed onto transparencies, effectively mimicking the original reconstruction process employed by Gabor, but with the ability to reconstruct arbitrary virtual objects. Because of this inherent duality and shared foundation in wave propagation, experience and knowledge gained in exploring \ac{CGH} is readily transferred to \ac{DHM}, and vice versa.

Just as there are unique advantages of using \ac{DHM} for imaging tasks, \ac{CGH} presents certain characteristics that makes it incredibly interesting as a complex light-delivery and light-shaping modality:

\paragraph{three-dimensional projection} The complex wavefront encoded in a hologram enables the reconstruction of \ac{3D} light distributions, effectively reproducing the properties of the

recorded or generated object. The computed holograms of \ac{CGH} allow for the reconstruction of entirely arbitrary \ac{3D} objects \cite{madsen_comparison_2022,shi_author_2021, yolalmaz_comprehensive_2022}, correctly reproducing important depth information.

\paragraph{projection efficiency} The complex wavefront generated by an object can be effectively encoded in the intensity variations of an interference pattern. Traditionally, holograms were produced by printing these patterns on transparencies, which reconstructed the object optically by modulating the amplitude of an incident light beam.

An alternative method is to encode the complex wavefront using phase variations. In this phase-only encoding approach, the wavefront is reconstructed by modulating the phase of the incident light rather than its amplitude. Since phase modulation does not attenuate the light, it enables the optical reconstruction to achieve nearly 100\% light efficiency.

\paragraph{aberration compensation} The complex wavefront encoded in a computer generated hologram is entirely user controlled through hologram generation algorithms. The generated holograms can be altered to correct for optical aberrations in the reconstruction path, e.g., scattering or lensing effects, by introducing compensating pre-aberrations.

Due to this flexibility and specialized superiority over conventional light-delivery methods, \ac{CGH} has found numerous uses. Examples include neuroscience and optogenetics \cite{papagiakoumou_scanless_2010, papagiakoumou_optical_2013, go_compact_2019}, three-dimensional displays for e.g., AR/VR\footnote{Augmented/virtual reality} \cite{gluckstad_holotile_2024-1, shi_towards_2021-1, yolalmaz_comprehensive_2022, shi_towards_2021-1, makey_breaking_2019, lee_high-contrast_2022}, laser material processing \cite{olsen_multibeam_2009}, optical micro-manipulation and trapping \cite{grier_holographic_2006-1, rodrigo_real-time_2004-1}, advanced light microscopy \cite{smith_programmable_2000}, and lately also \ac{3D} bio-fabrication \cite{madsen_axial_2025, gluckstad_holotile_2024-2, alvarez-castano_holographic_2024-2, alvarez-castano_holographic_2024-3, madsen_digital_2024}.

This chapter is based on the publications relating to HoloTile, a novel Fourier holography modality, capable of rapid, speckle-

suppressed hologram generation. By utilizing sub-hologram tiling and advanced \ac{PSF} shaping, HoloTile allows for a pseudo ``digital'' representation of holographic reconstructions, with output pixels that can both be resized, and amplitude modulated like conventional pixels. This solves a fundamental problem in conventional Fourier \ac{CGH}, in which overlapping \ac{PSF} contributions from spatial frequency components interfere to create a significant speckle patterns, the concept of which is shown in \cref{fig:cgh_speckle_formation}. The publications are reprinted in Appendices G and K to M, and each section will give a short summary of the contents and key takeaways from the relevant publication.

\begin{figure}[t]
  \centering
  \captionsetup{format = plain}
  \includegraphics[width=\linewidth]{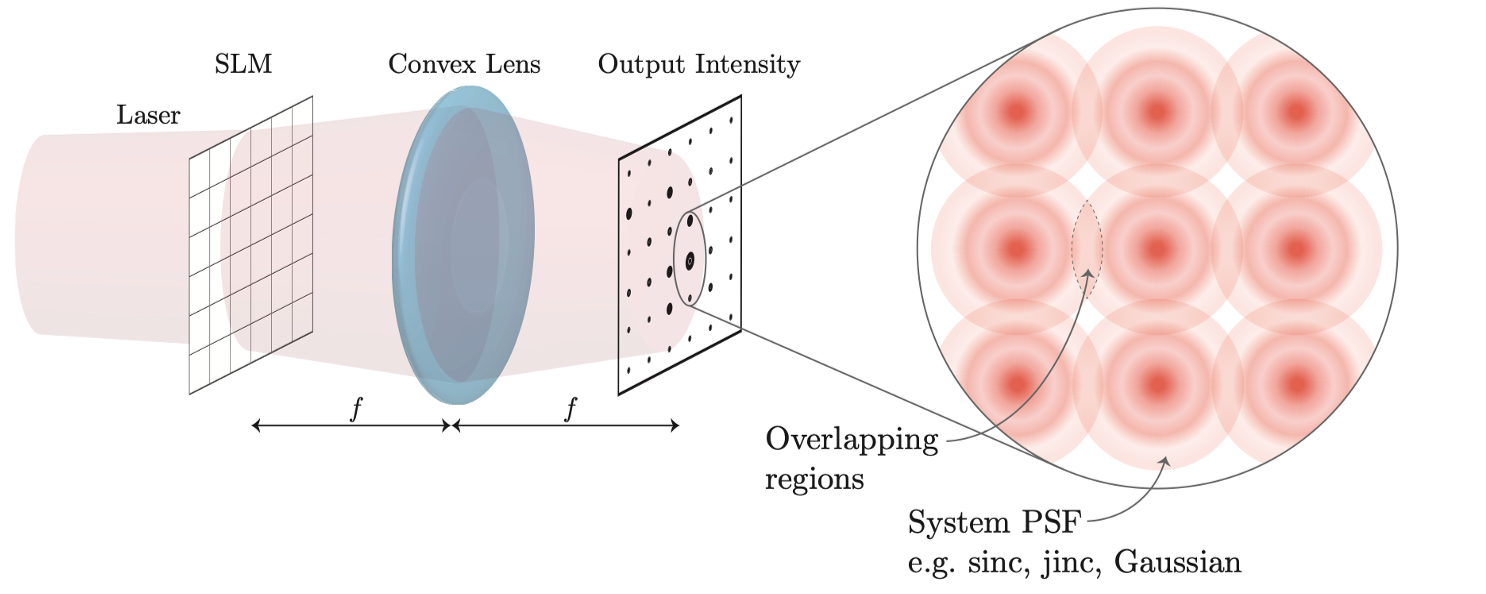}
  \caption{Formation of one source of speckle noise in \ac{CGH} reconstruction. The individual spatial frequency components (visualized here in a discrete grid) are each shaped as the system \ac{PSF} in the reconstruction plane. When the phase profile in this plane is unconstrained, these overlaps may lead to unwanted interference \cite{gluckstad_holotile_2024-3}.}
  \label{fig:cgh_speckle_formation}
\end{figure}

\section{Introduction to HoloTile}

The publication included in Appendix G - HoloTile: Rapid and Speckle-Suppressed Digital Holography by Matched Sub-Hologram Tiling and Point Spread Function Shaping \cite{madsen_holotile_2022} --- introduces HoloTile, a novel

Fourier \ac{CGH} technique aimed at real-time hologram generation while reducing the effect of the speckle noise generated by the overlapping system \ac{PSF}.

The HoloTile technique works by dividing the hologram generation algorithm in two distinct steps:

\begin{enumerate}
\item \textit{Sub-hologram calculation:} To isolate the spatial frequency components in a Fourier holographic reconstruction, the calculated hologram is repeated in a grid on the \ac{SLM} (a tiling operation of sub-holograms). In the reconstruction, this results in a representation of the target which is only defined in a grid of well-defined spots. The sub-hologram is calculated by any conventional hologram generation algorithm.

\item \textit{\ac{PSF}-shaping:} A hologram taking up the entire area of the \ac{SLM} is calculated analytically, responsible for shaping the spots in the reconstruction grid into square output pixels. This effectively eliminates the spatial overlap of the individual spatial frequency components, significantly reducing speckle noise.
\end{enumerate}

\begin{figure}[t]
  \centering
  \captionsetup{format = plain}
  \includegraphics[width=\linewidth]{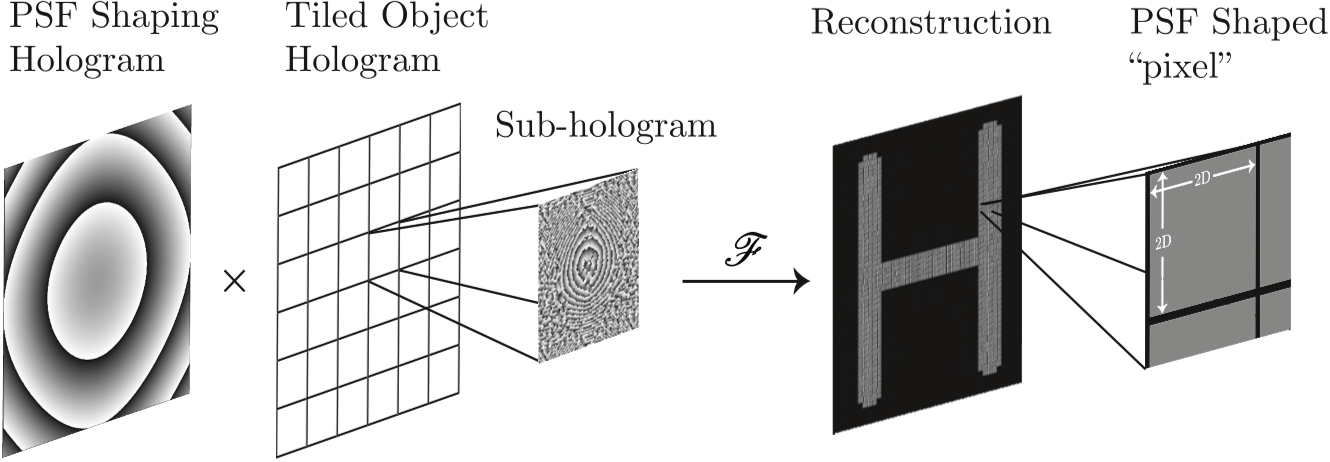}
  \caption{Calculation process of a HoloTile hologram. A \ac{PSF}-shaping hologram and a tiled object hologram are combined to facilitate the generation of square output pixels in the reconstruction plane \cite{madsen_holotile_2022}.}
  \label{fig:holotile_calculation}
\end{figure}

The two holograms; the tiled hologram and the \ac{PSF} shaping hologram, are then combined to form a final synthesized hologram, as seen in \cref{fig:holotile_calculation}.

\begin{figure}[t]
  \centering
  \captionsetup{format = plain}
  \includegraphics[width=.9\linewidth]{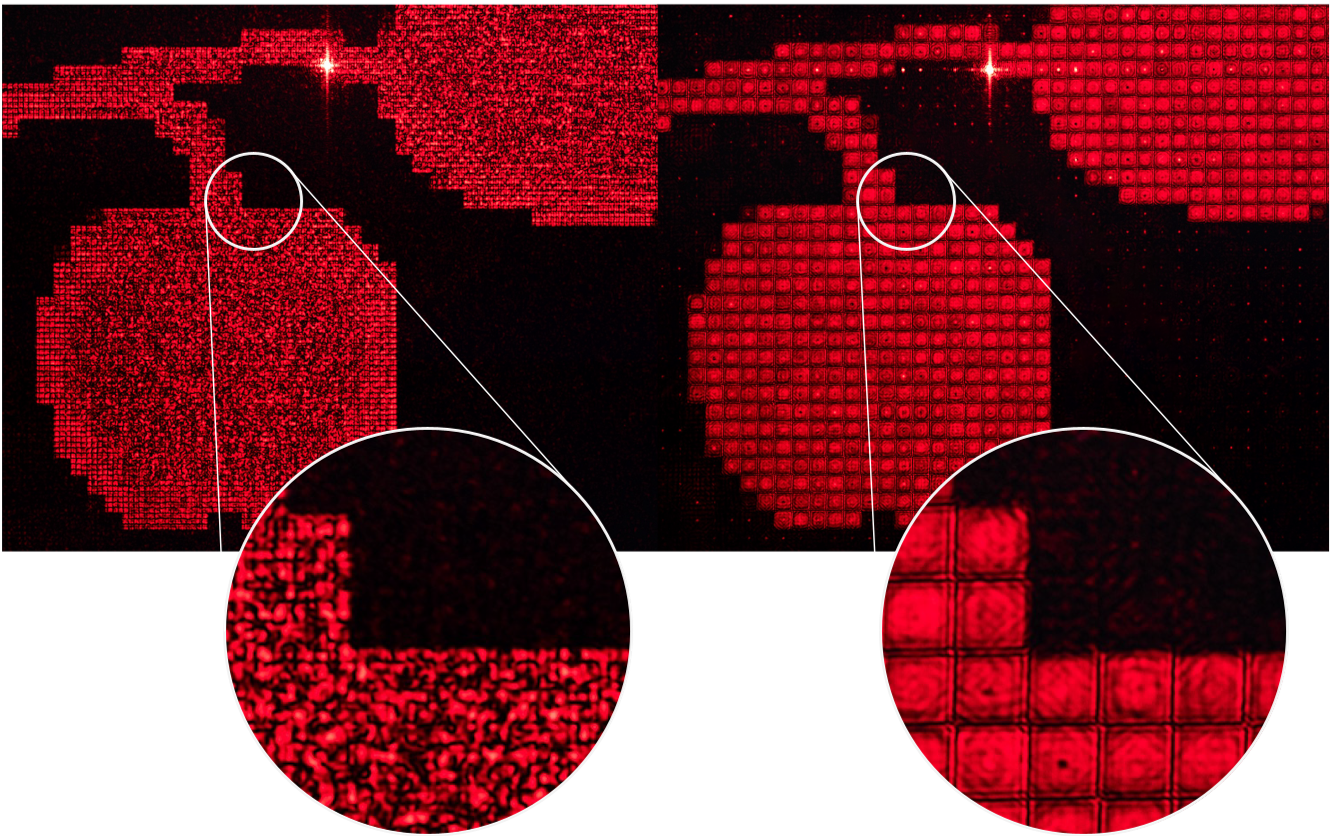}
  \caption{Comparison of homogeneity and output regularity between (Right) HoloTile and the (Left) Adaptive Weighted Gerchberg Saxton (AWGS) algorithm \cite{wu_adaptive_2021}. The HoloTile reconstruction shows limited cross-talk between output frequency components as opposed to the speckle noise seen in the AWGS reconstruction \cite{gluckstad_holotile_2024-3}.}
  \label{fig:holotile_vs_awgs}
\end{figure}

This decomposition of the displayed hologram serves several functions. First of all, the separation of the spatial frequencies, and subsequent ``pixelation'', of the reconstruction, ensures that there is virtually no overlap between the \ac{PSF} of adjacent spatial frequency components. This makes it nearly impossible for nearest-neighbor interference to occur, removing a significant source of the well-known noise of \ac{CGH} reconstruction. A comparison between the reconstructions of a conventionally calculated hologram and a HoloTile hologram is displayed in \cref{fig:holotile_vs_awgs}.

As such, the remaining dominant source of noise in the reconstruction is the sub-hologram generation algorithm itself, which can easily be substituted for new and better techniques. Future iterations may use \ac{ML} approaches for calculating the sub-hologram \cite{madsen_comparison_2022, shi_author_2021, yolalmaz_comprehensive_2022, shi_towards_2021-1}. Second of all, since the sub-holograms are all identical, and of smaller resolution than the \ac{SLM}, the tiled hologram can be calculated incredibly fast, while still taking advantage of the entire area of the \ac{SLM}. In Appendix G, a computational speed-up of approximately 15-25 times is reported, when compared to the calculation of single holograms using all pixels on the \ac{SLM}. However, in more recent implementations, a 100 times speed advantage has been observed.

\subsection{Generalized HoloTile}

\begin{figure}[t!]
  \centering
  \captionsetup{format = plain}
  \includegraphics[width=0.85\linewidth]{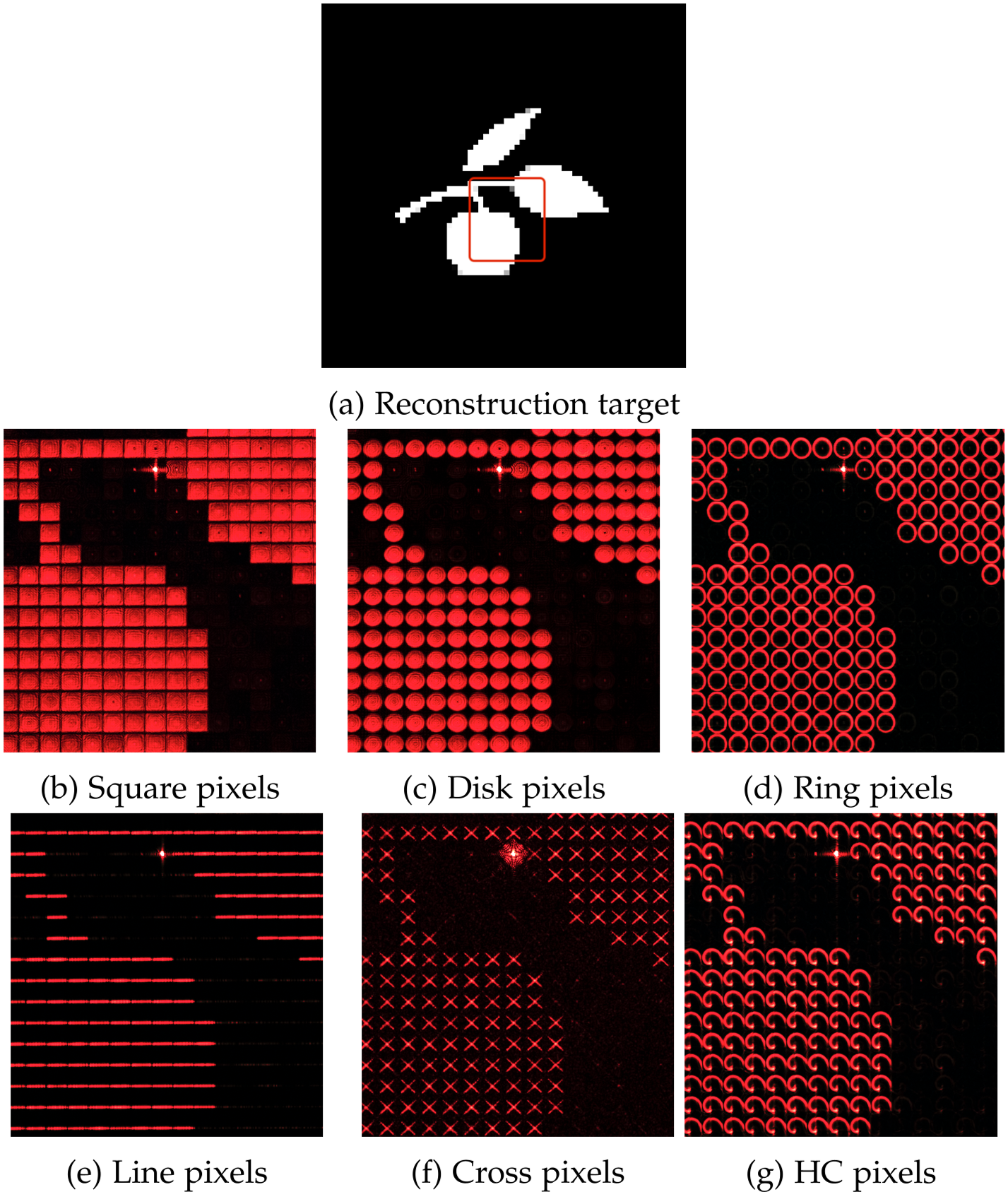}
  \caption{The \ac{PSF} shaping capabilities of HoloTile shown experimentally. For identical targets, the output pixels can be shaped near arbitrarily \cite{gluckstad_holotile_2024-3}.}
  \label{fig:holotile_psf_shapes}
\end{figure}

The HoloTile technique was expanded in the publication included\footnote{The arXiv version} in Appendix K --- HoloTile Light Engine: New Digital Holographic Modalities and Applications \cite{gluckstad_holotile_2024-1}. 
The publication introduces additional \ac{PSF} shaping modalities, as seen in \cref{fig:holotile_psf_shapes}, making it possible to not only represent square output pixels, but also disks, rings, lines, crosses, and even complex shapes such as the helico-conical (HC) \cite{alonzo_helico-conical_2005, overton_phase_2005} and axially extended Bessel beams. 
Since the \ac{PSF} shaping is independently calculated, any beam shape that can be engineered and encoded on the phase-only \ac{SLM} can be utilized as the output pixels shape, making it an immensely application-adaptive projection modality.

\begin{figure}[t]
  \centering
  \captionsetup{format = plain}
  \includegraphics[width=\linewidth]{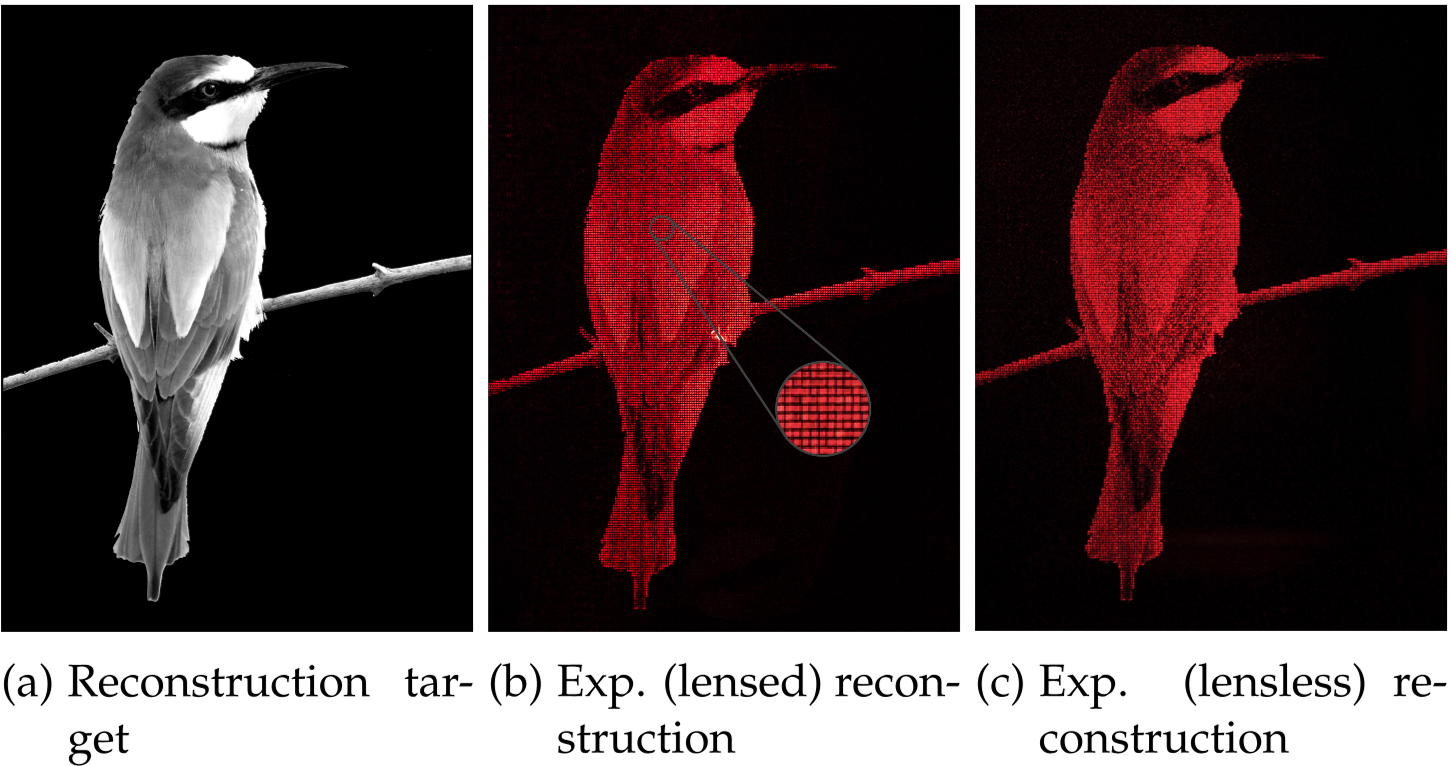}
  \caption{Target and experimental HoloTile reconstructions for lensed and lensless implementation \cite{gluckstad_holotile_2024-3}.}
  \label{fig:holotile_lensed_lensless}
\end{figure}

Additionally, lensless HoloTile is introduced. By encoding the Fourier transforming lens on the \ac{SLM}, it is possible to realize the specific advantages of HoloTile, while keeping cost and mechanical complexity to a minimum. 
With a setup requiring only a coherent light source and an \ac{SLM}, compact, ultra-high efficiency light projectors can be realized. In \cref{fig:holotile_lensed_lensless}, reconstructions of a high-resolution target using both the lensed and lensless configuration are shown.

A number of possible applications with this in mind are laid out, including optogenetics, particle trapping and manipulation, laser material processing, and volumetric additive manufacturing.

As an example, consider particle trapping and manipulation. Particles exhibiting certain properties such as low relative refractive index or strong absorbance are difficult to trap in conventional focused Gaussian beams as they are actively repelled from high light-intensity regions. To trap these particles, it is possible to use dark optical traps \cite{daria_dynamic_2004}. These consist of a region of low or no light intensity, surrounded by a barrier of high-intensity light. Employing the ring output pixel shape for this application would enable the trapping of hundreds, if not thousands, of particles concurrently. Additionally, the traps can be modulated both in size, intensity, and location at high frame rates.

\subsection{HoloTile: Axial}

The publication included in Appendix L --- Axial HoloTile: Extended Depth-of-Focus of Dynamic Holographic Light Projections \cite{madsen_axial_2025} --- extends HoloTile to three dimensions. 
The use of axicon phases as the \ac{PSF}-shaping hologram enables the parallel generation of axially extended Bessel beams. 

\begin{figure}[t!]
  \centering
  \captionsetup{format = plain}
  \includegraphics[width=0.95\linewidth]{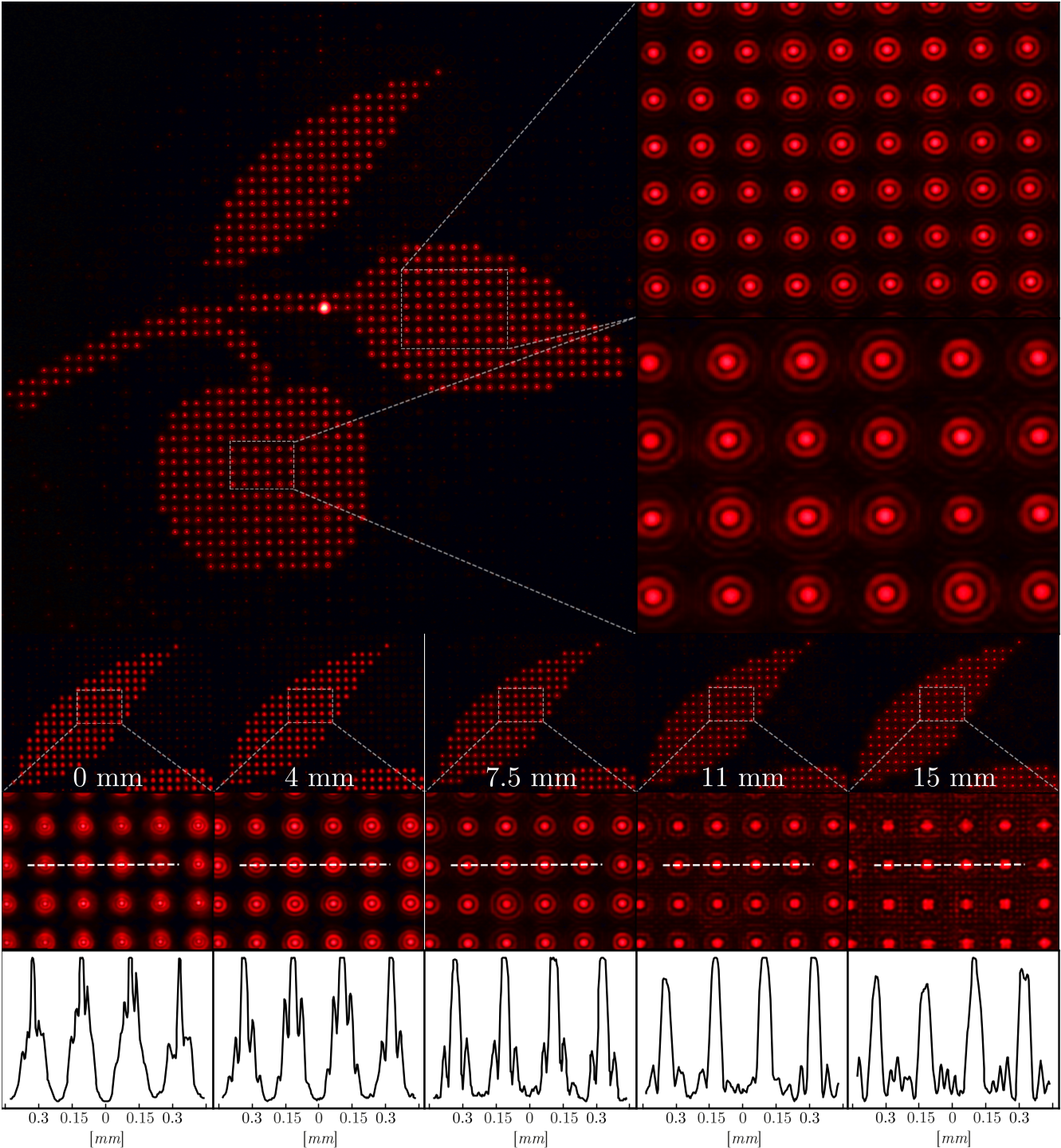}
  \caption{Axial propagation of the large reconstruction of the SDU logo. The Bessel beam shape is maintained over an axial distance of 15 mm \cite{madsen_axial_2025}}
  \label{fig:holotile_axial_propagation}
\end{figure}

These Bessel beams are automatically arranged in the pattern corresponding to the sub-hologram, thus allowing for the generation for numerous parallel propagating ``light pillars''. 
\cref{fig:holotile_axial_propagation} shows the propagation of these beams over a distance of 15 mm in the lensed configuration, however lensless is also possible and generates similar results. 
These reconfigurable Bessel beams could have serious implications for applications such as laser material processing, optical trapping, and especially tomographic volumetric additive manufacturing (\ac{T-VAM}) \cite{alvarez-castano_holographic_2024-3, kelly_volumetric_2019, loterie_high-resolution_2020-1}.

\ac{T-VAM} is a novel \ac{3D} fabrication technique, enabling the creation of \ac{3D} objects in seconds. Intuitively, the technique can be considered an inverse computed tomography (\ac{CT}) scanner. 
In \ac{CT} scanning, a \ac{3D} object is reconstructed in the computer by the repeated imaging by x-rays from multiple angles. 
The combined information from the angular-dependent images is enough to generate an accurate \ac{3D} representation of the scanned object. 
In \ac{T-VAM}, however, a physical \ac{3D} object is created by projecting angular dependent UV light patterns into a rotating vial of a photo-active curable medium. 
The accumulated UV dose solidifies the medium such that the \ac{3D} object is generated in the vial. 
This has the potential for printing in cell-laden hydrogel, allowing the creation of complex \ac{3D} tissue for applications in tissue engineering, regenerative medicine, drug testing, etc.

The introduction of \ac{CGH} in this field, here designated the name HoloVAM, could revolutionize its widespread use. 
The advantages of \ac{CGH}, and HoloTile in particular, makes it a promising candidate for improvement over the conventional \ac{T-VAM} light delivery methods. 
The Fourier holographic approach utilizing the entire \ac{SLM} area, in combination with phase modulation, means that a holographic light engine can be nearly 100\% light efficient. 
The conventional light engines, e.g., utilizing amplitude modulating digital micromirror devices (\ac{DMD}s) in imaging configurations, exhibit pattern dependent light efficiency. 
The projected patterns are often sparse, utilizing only a fraction of the \ac{DMD} pixels for actually exposing the active medium. 
It is therefore not uncommon to use large, high-power, multimode lasers to facilitate printing. 
With the light efficiency of HoloTile, printing can be performed with a laser not much more powerful than a laser pointer, making the system both more compact, safer to use, and more performant \cite{alvarez-castano_holographic_2024-3, gluckstad_holotile_2024, madsen_digital_2024}.

\begin{figure}[t]
  \centering
  \captionsetup{format = plain}
  \includegraphics[width=.9\linewidth]{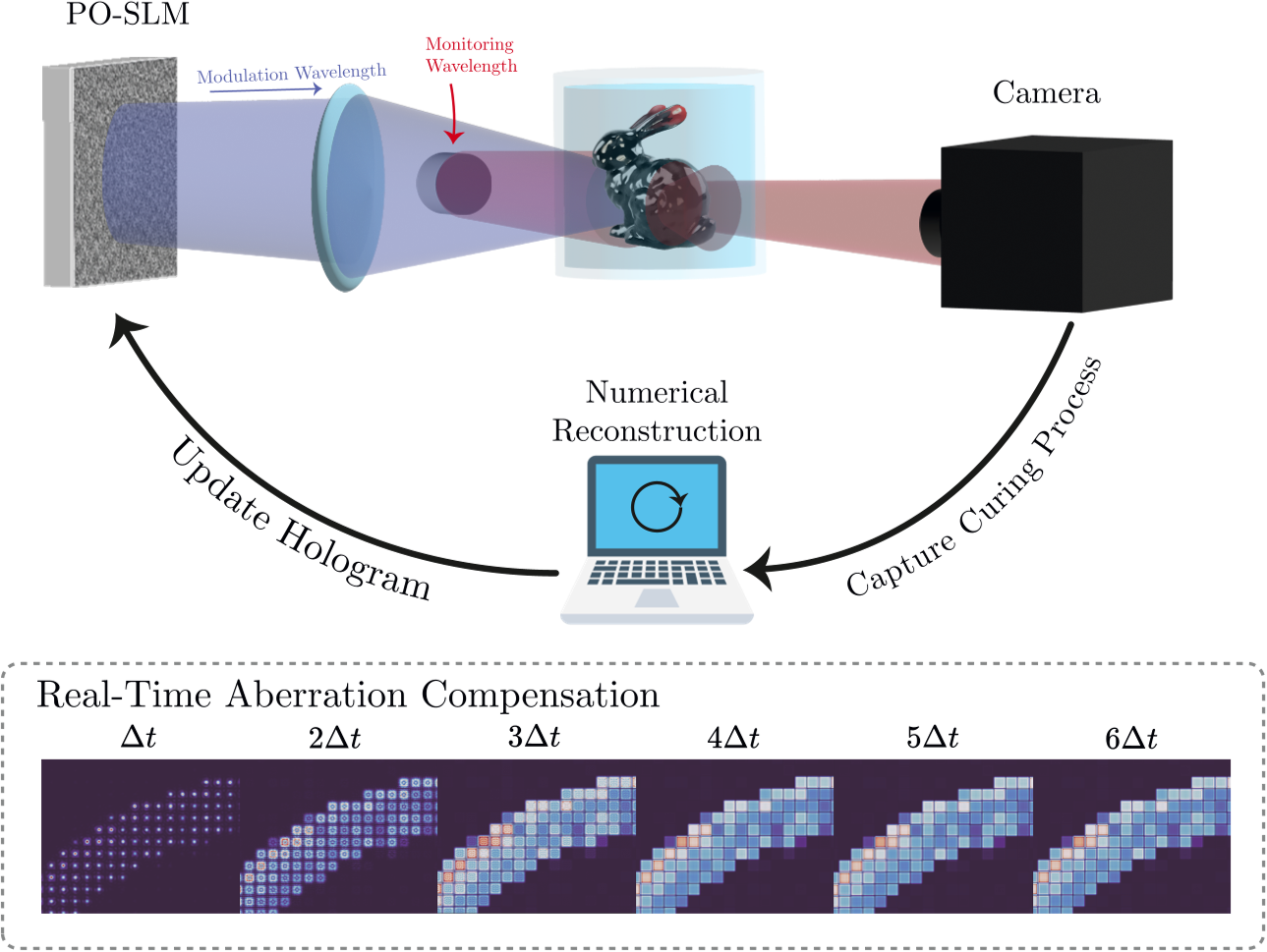}
  \caption{Illustration of the HoloVAM principle. A phase-only \ac{SLM} (PO-\ac{SLM}) displays calculated holograms which are reconstructed into the rotating vial. The fabrication process can be monitored and the projected reconstructions can be adapted to fit the current fabrication stage or compensate for aberrations \cite{gluckstad_holotile_2024-3}.}
  \label{fig:holovam_principle}
\end{figure}

In addition, the phase control of holography allows for explicit compensations of aberrations in the system such as lensing effects, refractive index mismatches, or even the scattering from a densely cell-laden hydrogel. 
In combination with its real-time hologram update rate, \ac{PSF} shaping could be employed to adaptively change the projected voxel shape while printing. 
A visual example of HoloVAM is given in \cref{fig:holovam_principle}.

\subsection{HoloTile: \ac{RGB}}

With the previous instantiations of HoloTile all showing monochromatic reconstructions, the publication included in Appendix M -- HoloTile \ac{RGB}: Ultra-fast Speckle-Free \ac{RGB} Computer Generated Holography \cite{madsen_holotile_2024-rgb} -- (currently available on arXiv \cite{alvarez-castano_holographic_2024-3}) brings HoloTile in color.

Color reconstruction in \ac{CGH} has conventionally mainly been attained via temporal multiplexing in order to reduce the effect of speckle noise, which is especially noticeable in color. 
By rapidly displaying numerous reconstructions, the human persistence of vision averages the contributions from the speckled reconstructions, creating a perceived cleaner image.

With the discrete output pixels and its superior speckle-suppression, HoloTile is a viable candidate for single-reconstruction \ac{CGH} in color display technologies. 
Utilizing an \ac{RGB} laser and a color compatible \ac{SLM} \cite{noauthor_gaea-2_2022}, HoloTile holograms can be generated for each wavelength, and superimposed in the reconstruction plane. 
Since only a single hologram is calculated per wavelength, HoloTile \ac{RGB}! (\ac{RGB}!) allows for real-time calculation and display of these speckle-reduced reconstructions, a feat that is virtually unheard of in \ac{CGH}, both for still images and for video purposes. 
Video rate color holography could enable the use of ultra-high efficiency dynamic light projectors for e.g., VR/AR, display projectors, etc.

\begin{figure}[t]
  \centering
  \captionsetup{format = plain}
  \includegraphics[width=.95\linewidth]{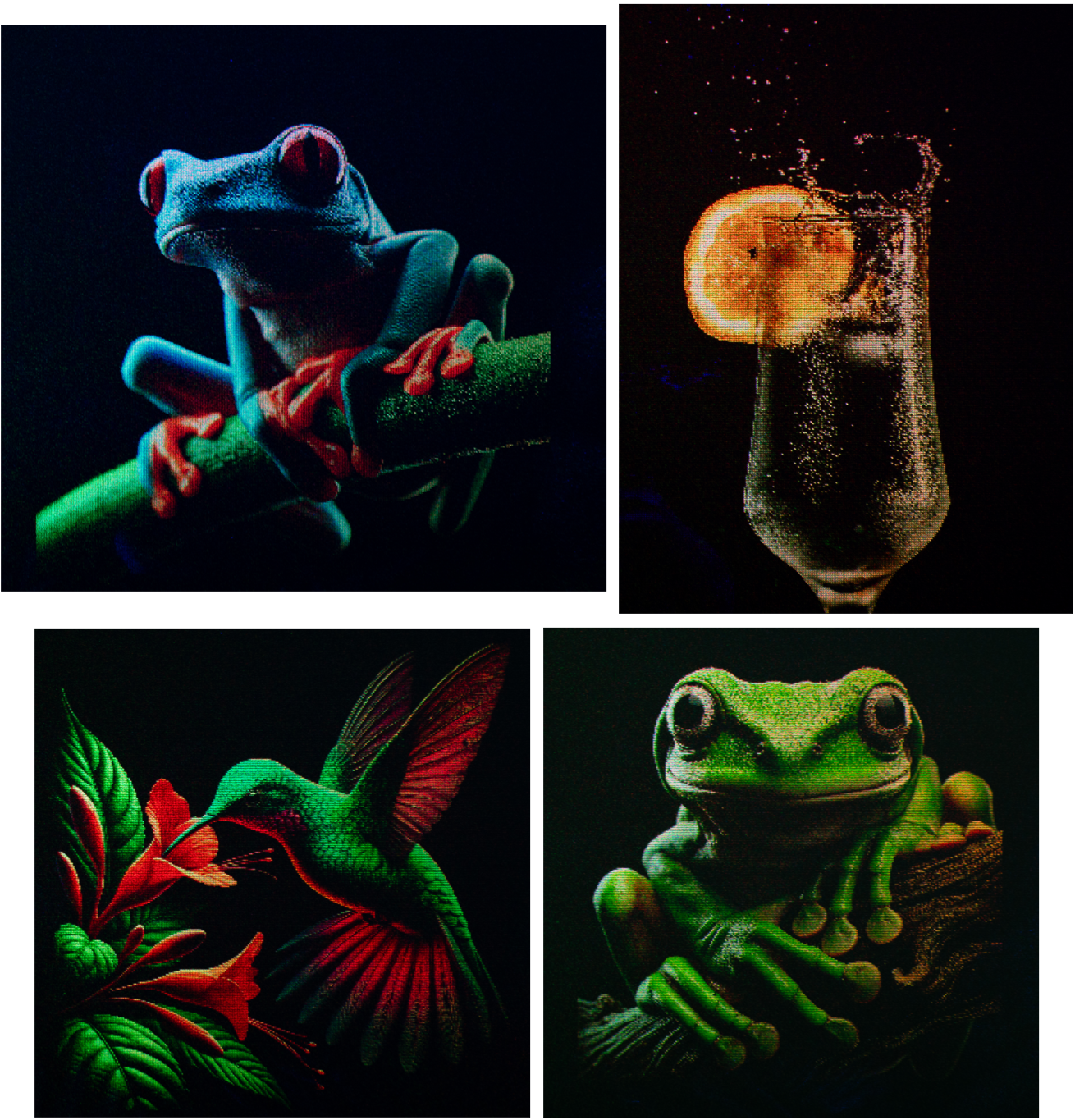}
  \caption{Lensed experimental \ac{RGB} reconstructions captured directly in camera \cite{madsen_holotile_2024-rgb}.}
  \label{fig:holotile_rgb_reconstruction}
\end{figure}

While the reconstructions displayed in \cref{fig:holotile_rgb_reconstruction} show color reconstructions for the lensed configuration, the lensless configuration produces similarly impressive results. A lensless holographic

color projector could be constructed extremely compact, while maintaining a bright projection image due to the high light-efficiency.

\chapter{Overall conclusion}
\label{ch:conclusion}

The primary aim of this industrial Ph.D. project was to explore the use of digital holographic microscopy (\ac{DHM}) for the specific task of performing a differential white blood cell count (\ac{dWBC}) in point-of-care (\ac{PoC}) devices suitable for acute care facilities, intensive care units (\ac{ICU}s), operating rooms, etc. A device constructed specifically for these environments must be optimized towards a simple, compact, and robust mechanical design, capable of rapid sample acquisition and processing.

To that end, the exploration began by constructing and testing an objective lens-based \ac{DHM} prototype, and associated 3-part differential white blood cell (\ac{WBC}) classifier. Conducting this initial investigation in a best-case scenario, in which the sample plane is optically magnified, allowed for the easier development and validation of the required algorithms, including hologram reconstruction, cell autofocus and localization, and the classifier itself. The prototype lens-based system and image classifying convolutional neural network (\ac{CNN}) achieved an overall 3-part classification accuracy of 89.6\%, classifying based on the reconstructed amplitude and phase planes of the cells, confirming the possibility of performing \ac{dWBC} using a \ac{DHM}-based system.

With these foundations laid and experiences gained, the development of a lensless prototype was carried out, thus simplifying the mechanical design. The removal of the objective lens also increased the field-of-view (\ac{FoV}), at the cost of spatial resolution, allowing for an increased rate of sample acquisition. Combining the algorithms developed for the lens-based prototype with optimizations made to the \ac{CNN} input representation and shape allowed for an increased overall 3-part classification accuracy of 92.65\%. The classification task was then challenged by introducing two additional cell types to form the full 5-part differential. The trained 5-part differential classifier demonstrated promising results with an overall classification accuracy of 89.44\%. However, notable misclassifications were observed for basophils, highlighting an area for future refinement. With the lensless prototype, the determination of additional diagnostic markers was also demonstrated. Specifically, the monocyte distribution width (\ac{MDW}), a promising biomarker for sepsis, was calculated from differentiated monocytes, showcasing the additional use of the system.

Both these prototypes were prioritized for acquisition speed and on-device reconstruction which, with the limited computational resources of a \ac{PoC} device, required the use of simple reconstruction via the angular spectrum method (\ac{ASM}). However, by alleviating these requirements, more sophisticated capture and reconstruction techniques were explored. In particular, a pixel super-resolution capture sequence and algorithm was implemented in order to provide more cell detail to a classifier. The super-resolution approach made possible a 4-8 times reduction in effective pixel size, effectively allowing the reconstruction of an approximate 46 mm² \ac{FoV} with sub-micron effective pixel sizes. Additionally, a multi-wavelength \ac{DHM} approach was explored as a way to regain cell color information for additional classifiable information. Lastly, a proof-of-principle self-supervised physics-informed neural network (\ac{PINN}), capable of reconstructing amplitude and phase planes from multi-wavelength holograms, was implemented and tested on experimental data. The \ac{PINN} relies on a physically based loss function, obviating the need for an extensive training set of ground truth reconstructions. The \ac{PINN} shows promising results, and should be explored further as a tool to increase reconstruction throughput.

In summary, this project served as the initial exploration phase for the possibility of integrating a \ac{DHM} system in a \ac{PoC} device for the purposes of adding \ac{dWBC} to the suite of health markers currently provided. This exploration was further aided by the investigation of \ac{CGH}, enabling an increased understanding of wavefront manipulation and digital holography. The contributions made here will hopefully spark further research of the application.

\backmatter

\appendix
\part{APPENDIX}
\chapter{White blood cell misclassification matrix}
\label{app:wbc_misclassification}

\begin{sidewaysfigure}[t]
  \centering
  \captionsetup{format = plain}
  \includegraphics[width=\linewidth]{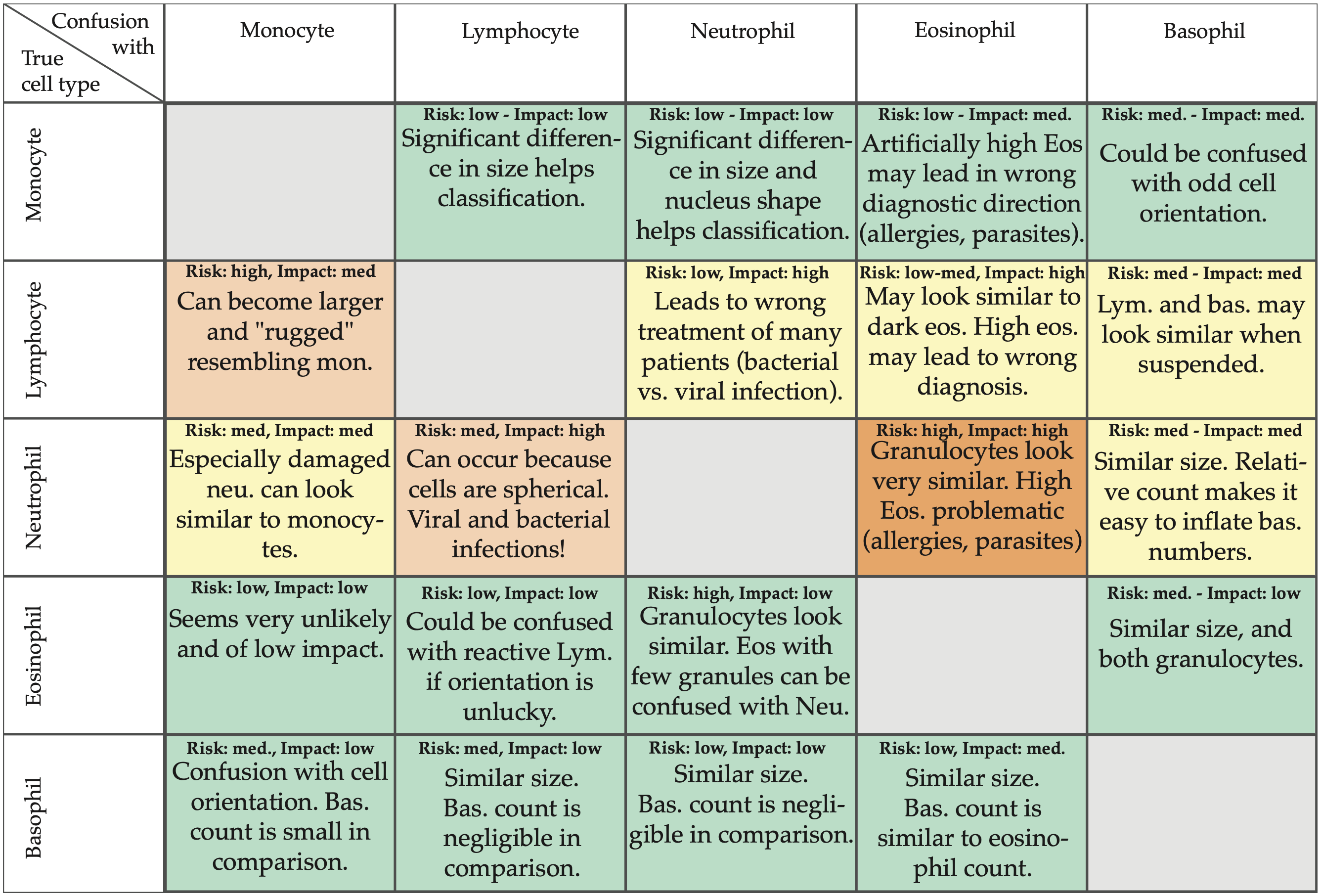}
  \caption{Misclassification matrix created and provided by Radiometer Medical ApS for documenting the chances and consequences of WBC misclassification.}
  \label{fig:wbc_misclassification_matrix}
\end{sidewaysfigure}
\chapter{Lateral localization of cells in reconstructed focal stack}
\label{app:lateral_localization}

The 2D cell-localization map, M is generated by the reconstruction stack consisting of complex-valued reconstruction planes. The algorithm starts out by computing the differences between consecutive reconstruction planes in the stack, capturing axial variations in both amplitude and phase. This results in two distinct stacks of axial differential maps, one for the amplitude changes and one for the phase shift changes. The two stacks are each condensed to single 2D differential maps by calculating the axial maximum through the stacks. These differential maps are then processed to enhance important features. Specifically, they are normalized, smoothed using a Gaussian filter to reduce noise, and then filtered by a Laplacian kernel to emphasize local peaks. The combined cell localization map, M, is generated by the sum of the squared amplitude and phase differential maps. The peaks in this combined map, exceeding a predefined threshold are identified as cell locations. The center $(i, j)$ pixel of the detected peaks are the reported cell locations. This technique, albeit confusing, leverages both amplitude and phase information in the complex-valued reconstruction planes to robustly detect regions with significant axial changes, indicative of the presence of cells.

\chapter{Holographic axial focus measures}
\label{app:focus_measures}

The $S_{\text{GradientSum}}$ metric indicates focus by the sum of gradient magnitudes of an image:
\begin{equation}
S_{\text{GradientSum}}(c) = \sum_{x,y} |\nabla c| = \sum_{x,y} \sqrt{\left(\frac{\partial c}{\partial x}\right)^2 + \left(\frac{\partial c}{\partial y}\right)^2}
\label{eq:gradient_sum}
\end{equation}

A higher sum of gradients indicates that there stronger edges in the image. The $S_{\text{GiniOfGradient}}$ metric indicates focus by calculating the Gini index of the gradient of an image:
\begin{equation}
S_{\text{GiniOfGradient}}(c) = \text{GI}(|\nabla c|)
\label{eq:gini_gradient}
\end{equation}
where $\text{GI}(\cdot)$ is the Gini index, a statistical measure of inequality. A higher Gini indicates greater inequality in the distribution of its index, which means that there are a few strong edges and many weak, or non-existent, edges. The $S_{\text{TamuraOfGradient}}$ metric indicates focus by the ratio of the standard deviation of the gradients in an image to the mean gradient \cite{zhang_robust_2018}:
\begin{equation}
S_{\text{TamuraOfGradient}}(c) = \text{TC}(|\nabla c|) = \frac{\sigma(|\nabla c|)}{\langle |\nabla c| \rangle}
\label{eq:tamura_gradient}
\end{equation}
where TC is the Tamura coefficient, $\sigma \cdot$ is the standard deviation and $\langle \cdot \rangle$ is the mean. A high Tamura coefficient indicates a wide spread of gradients, suggesting that there are several prominent edges.

\chapter{Spatial variations in filtered LED illumination wavelength}
\label{app:wavelength_variations}

It is commonly known that optical bandpass filter can exhibit a slight center wavelength change depending on the angle of incidence of the light to be filtered \cite{laser_components_gmbh_bandpass_2020}. This change in wavelength occurs due to the layered nature of the structure of the filter, and the tuning of these layers to a specific wavelength. In this case, the spatially filtered light exits the pinhole, and propagates in a divergent cone shape, thus introducing an angular spread of the rays incident on the bandpass filters. The capture and reconstruction of a hologram illuminated by a source where the illumination wavelength is significantly dependent on the radial distance from the center of the hologram should, naturally, be avoided. Large location dependent wavelength changes could result in a difference in focus across the hologram.

\begin{figure}[t]
  \centering
  \captionsetup{format = plain}
  \includegraphics[width=\linewidth]{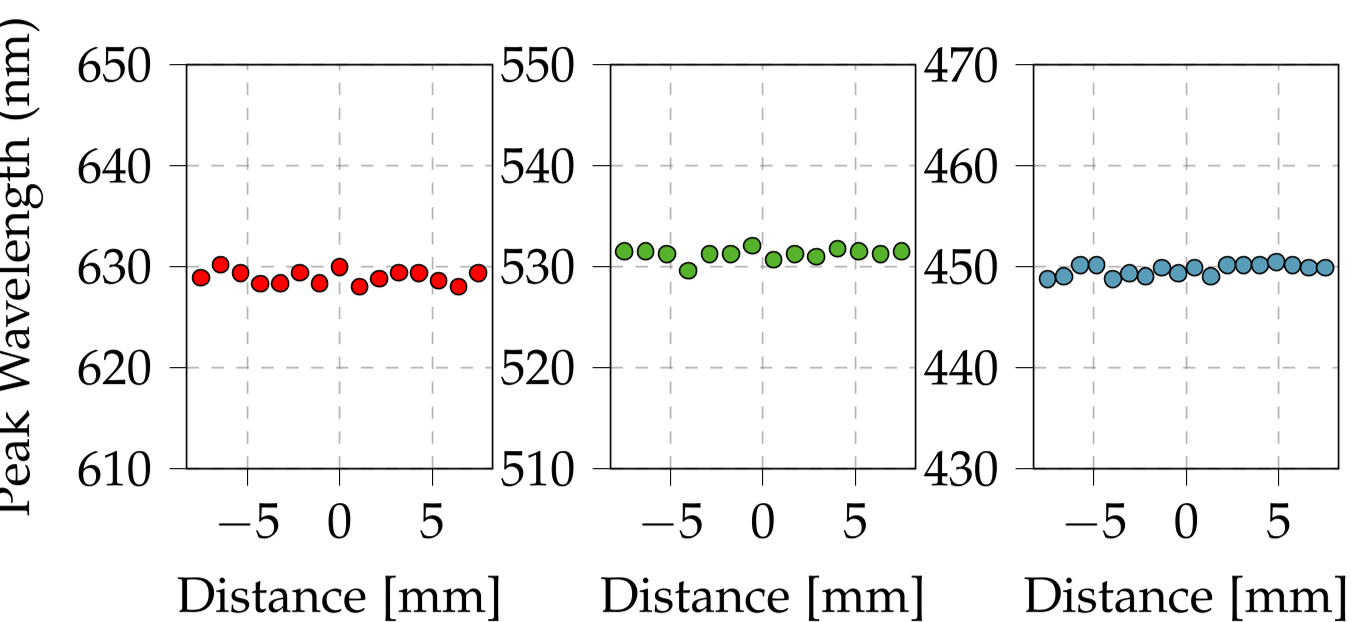}
  \caption{Wavelength measurements of the multi-wavelength illumination under filtering by the spectral bandpass filters as a function of the diagonal of the camera sensor.}
  \label{fig:wavelength_measurements}
\end{figure}

To ensure that the impact of the angular spread due to the pinhole does not affect the illumination of the holograms significantly, the camera sensor is removed in place of a spectrometer \cite{noauthor_avaspec-uls2048cl-evo_nodate}, and the wavelength change is measured. In order to observe any spectral variations in the illumination, the fiber tip of the spectrometer is mounted in a manual stage. Under illumination of the three selected wavelengths (bandpass filtered illumination from the LED-array), the spectrometer is scanned laterally across the largest dimension of the camera sensor area. Due to the cone-like angle spread of the incidence light, any wavelength change should be circularly symmetric around the center of the sweep. The results of the experiment are shown in \cref{fig:wavelength_measurements}, in which the peak wavelength at each location in the sweep is plotted, for each of the filtered wavelengths. Here it appears that no systematic change in wavelength occurs for the three wavelength bands, as the spectrometer is swept across the 15 mm of the camera sensor area.

\chapter{Super-resolution downsampling algorithm for Bayer patterns}
\label{app:bayer_downsampling}

\begin{algorithm}[htbp]
\caption{Downsampling pseudocode algorithm for Bayer pattern}
\label{alg:bayer_downsampling}

\KwIn{
$x$: High-resolution estimate $(N,N)$\;
$D$: Super-resolution magnification factor\;
channel $\in \{\texttt{r},\texttt{g1},\texttt{g2},\texttt{b}\}$\;
}
\KwOut{Downsampled estimate $u$ for the specified channel}

indexDict $\leftarrow \{\texttt{r}:0,\ \texttt{g1}:1,\ \texttt{g2}:2,\ \texttt{b}:3\}$\;
index $\leftarrow$ indexDict[channel]\;
row $\leftarrow (\text{index} \div 2)\times(D \div 2)$ \tcp*[r]{Row offset for subpixel}
col $\leftarrow (\text{index} \bmod 2)\times(D \div 2)$ \tcp*[r]{Column offset for subpixel}
$u \leftarrow$ empty array of size $(N/D, N/D)$\;

\For{$r \leftarrow$ row \KwTo $D + (D \div 2)$}{
    \For{$c \leftarrow$ col \KwTo $D + (D \div 2)$}{
        subpixels $\leftarrow x[r:N:D,\ c:N:D]$\; \tcp{Extract subpixels at intervals of $D$}
        $u \leftarrow u + \text{subpixels}$\;
    }
}

$u \leftarrow u/(D/2)^2$ \tcp*[r]{Normalize by the number of subpixels}
\Return $u$\;

\end{algorithm}

\chapter{Holographic PINN model design}
\label{app:pinn_design}

\begin{figure}[t]
  \centering
  \captionsetup{format = plain}
  \includegraphics[width=\textwidth]{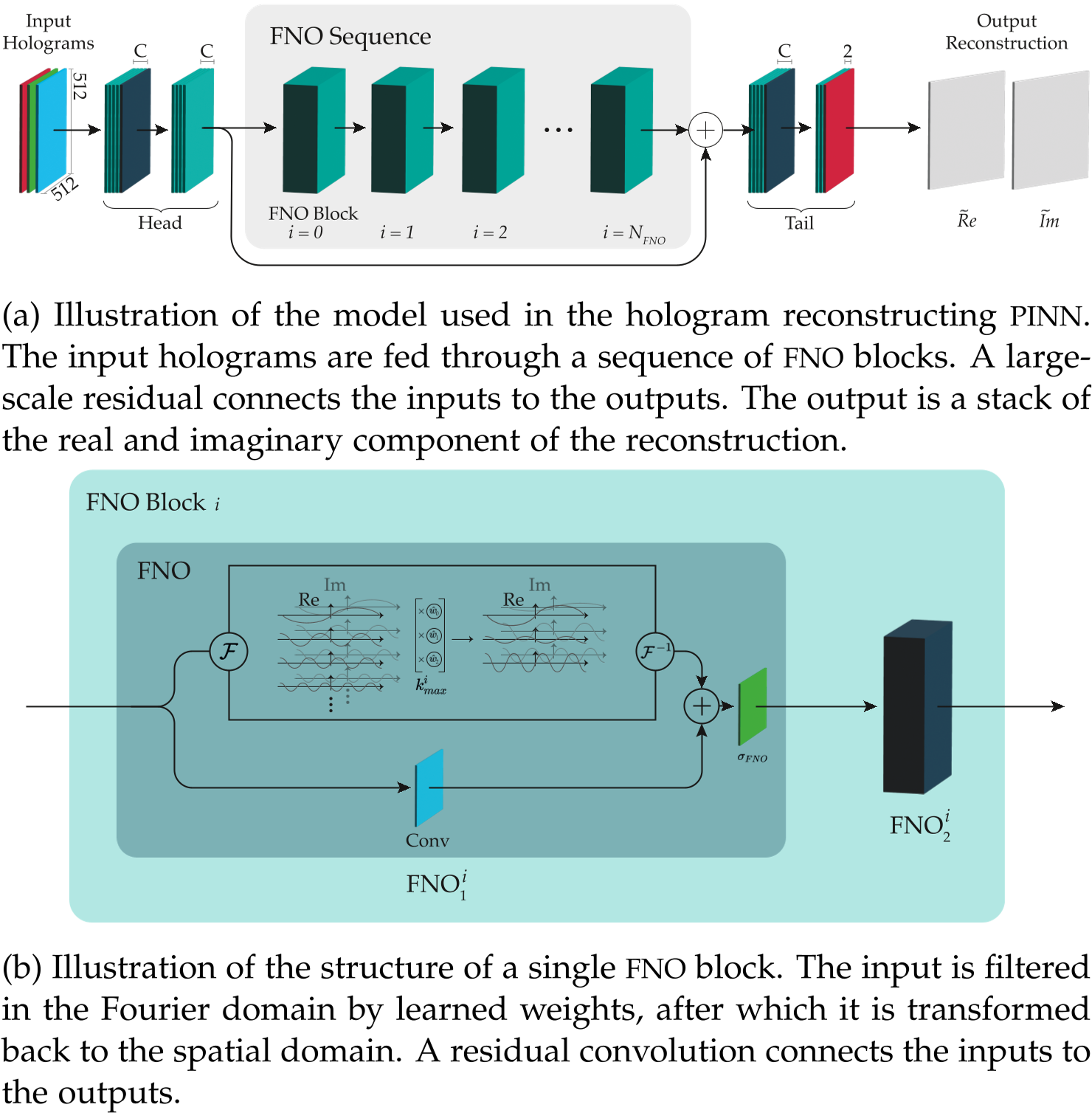}
  \caption{}
  \label{fig:pinn_architecture}
\end{figure}

A scalable generative network structure, as illustrated in \cref{fig:pinn_architecture}a, with an adaptive number of blocks is chosen. Specifically, the network consists of a sequence of fourier neural operator (FNO) blocks \cite{li_fourier_2021}, the structure of which will be detailed later. A head and tail are placed before and after the sequence of FNO blocks, respectively, each consisting of two convolution layers. There are responsible for converting between the input and output, and the deep network backbone. A large-scale residual connection adds the output of the head to the output of the sequence of FNO blocks. Residual connections have become standard in ML, due to the prominence of the ResNet architecture \cite{he_deep_2015-1}. They aid a deep network in converging faster, by allowing it to map the residual --- the difference between the input and the desired output --- instead of the full mapping. In addition, it has been observed that adding residual connections help avoid the problems of exploding and vanishing gradients \cite{he_deep_2015-1, huang_convergence_2022}.

The FNO blocks each consist of two FNO layers, as illustrated in \cref{fig:pinn_architecture}b, responsible for the bulk of the complex mapping between the input hologram and the output reconstructions, are based on the notion that both complex systems and patterns can be more efficiently represented in the Fourier domain, where both large- and small-scale interactions are captured as superpositions of oscillations. Formally, each FNO accepts some multi-dimensional input, typically of shape $(H \times W \times C)$, where C are the channels and H and W are the height and width in pixels, respectively. Each channel in the input is spatially Fourier transformed to yield a number of Fourier modes. A maximum number $k_{\text{max}}$ of Fourier modes\footnote{As sorted from the low-frequency end of the spectrum} are linearly transformed by element-wise multiplication with a learned weighting, such that the modes of importance can be emphasized, and vice-versa. The high-frequency information is discarded. The modified Fourier modes are then inversely Fourier transformed, and passed through an activation function $\sigma_{\text{FNO}}$. In addition, a residual connection containing a single convolutional layer connects the raw input to the FNO block to the output of the Fourier processing. This residual connection acts similarly to a bias term.

The maximum number of Fourier modes in each FNO block in the sequence is halved with each block, thus allowing an efficient transfer and maintaining of spatial features at multiple frequency scales through the depth of the network, similar to the information distilling in a conventional CNN by the repeated convolution and pooling operations.

The number of FNO blocks $N_{\text{FNO}}$, the number of initial Fourier modes $k_0^{\text{max}}$, and the number of channels C to use throughout the network are the hyperparameters of the network, and can be tuned to optimize performance while minimizing computational complexity.

\markboth{}{}
\printbibliography[heading=bibintoc]

\end{document}